\documentclass[twocolumn]{emulateapj}
\usepackage{aas_macros}
\usepackage{graphicx}
\usepackage{epstopdf}
\newcommand{\be}{\begin{equation}}
\newcommand{\ee}{\end{equation} }
\newcommand{\ba}{\begin{eqnarray}}
\newcommand{\ea}{\end{eqnarray}}

\newcommand{\bnabla}{\mbox{\boldmath$\nabla$}}

\newcommand{\bxi}{\mbox{\boldmath$\xi$}}
\newcommand{\nn}{\mbox{} \nonumber \\ \mbox{} }
\newcommand{\kB}{k_{\rm B}}

\begin{document}

\shorttitle{Non-thermal Gamma-Ray Emission from Delayed Pair Breakdown}
\shortauthors{Gill \& Thompson}
\title{Non-thermal Gamma-ray Emission from Delayed Pair Breakdown \\
       in a Magnetized and Photon-rich Outflow}
%{On The High energy prompt emission from gamma-ray bursts}
\author{Ramandeep Gill and Christopher Thompson}
\affil{Canadian Institute for Theoretical Astrophysics, 60 St. 
George St., Toronto, ON M5S 3H8, Canada}
\begin{abstract}
We consider delayed, volumetric heating in a magnetized
outflow that has broken out of a confining medium
and expanded to a high Lorentz factor ($\Gamma \sim 10^2-10^3$) 
and low optical depth to scattering ($\tau_{\rm T} \sim 10^{-3}-10^{-2}$).
The energy flux at breakout is dominated by the magnetic field, with a modest contribution 
from quasi-thermal gamma rays whose spectrum was calculated in Paper I.   We focus on the 
case of extreme baryon depletion in the magnetized material, but allow
for a separate baryonic component that is entrained from a confining medium.
Dissipation is driven by relativistic motion between these two components, which
develops once the photon compactness drops below $ 4\times 10^3(Y_e/0.5)^{-1}$.  
We first calculate the acceleration of the magnetized component
following breakout, showing that embedded MHD turbulence provides significant inertia,
the neglect of which leads to unrealistically high estimates of flow Lorentz factor.
After re-heating begins, the pair and photon distributions 
%using a simple prescription where heat is delivered at a constant 
%rate per unit logarithm of time. The pair and photon distributions 
are evolved self-consistently using a one-zone kinetic code that
incorporates an exact treatment of Compton scattering, pair production and 
annihilation, and Coulomb scattering.  Heating leads to a surge
in pair creation, and the scattering depth saturates at $\tau_{\rm T}
\sim 1$-4.  The plasma maintains a very low ratio of particle to 
magnetic pressure, and can support strong anisotropy in the charged
particle distribution, with cooling dominated by Compton scattering.
High-energy power-law spectra with photon indices in the range observed in GRBs
($-3 < \beta < -3/2$) are obtained by varying the ratio of 
heat input to the seed energy in quasi-thermal photons.  We contrast our results
with those for continuous heating across an expanding photosphere, and show
that the latter model produces soft-hard evolution that is inconsistent with
observations of GRBs.
\end{abstract}
\keywords{MHD --- plasmas --- radiative transfer --- scattering --- gamma rays: bursts}

\maketitle
%%%%%%%%%%%%%%%%%%%%%%% INTRODUCTION %%%%%%%%%%%%%%%%%%%%%%%%%%%%%%
\section{Introduction}

Most gamma-ray bursts (GRBs) appear to mark the birth of stellar-mass black holes
\citep{paczynski86,eichler89,woosley93}.  A magnetized jet extracting 
energy from a black hole ergosphere \citep{bz77} is strongly depleted in baryons, 
but the jet must propagate through a dense, confining medium.   As a result, the 
jet carries an intense thermal radiation field as well as a magnetic field.  

The central thesis of this paper is that non-thermal gamma-ray emission results from
the interaction between the thermal radiation field and a time-dependent magnetic field
\citep{thompson94,thompson06,meszaros97,spruit01,giannios06}.  Our focus is on the dynamics 
and dissipation of the jet after breakout, with a goal of accounting for the high-energy 
spectra of GRBs, and basic features of their pulse behavior.  

A major part of the problem involves understanding where dissipation is concentrated.
This is a significant theoretical challenge,
given that the jet maintains a very high energy density and compactness over
eight to ten decades in radius outward from the engine. 
Our approach is to divide the GRB emission process into
two major components:  dissipation before breakout, while the jet Lorentz factor is
still relatively low; and a second phase of dissipation that is delayed to a
large radius and -- importantly -- to a low scattering depth.   

This means that a magnetized GRB outflow characteristically develops {\it two} pair-dominated photospheres.
The radiation field advected by the jet is rich in electron-positron pairs close to the engine
\citep{goodman86,shemi90}, and a moderately large scattering depth in pairs can be
maintained by continued heating out to a considerable distance from the engine.  
When most of the jet energy flux is carried by the magnetic field at breakout, relaxation to
thermal equilibrium results in a flat spectrum below the spectral peak, as is
observed in GRBs (\citealt{tg13}, hereafter Paper I).  The spectral peak also sits in the observed range 
when the Lorentz factor inside breakout remains modest, $\Gamma \sim 1/\theta$.
Softer spectral peaks (which may correspond to X-ray flashes) result from jets
whose photospheres are dominated by electrons and ions (Paper I).  

A rapid transition to transparency after breakout allows the magnetofluid to 
accelerate outward, by a combination of radiation pressure and the Lorentz force \citep{russo13a,russo13b};
and helps to preserve a narrow peak in the spectrum.

Although the pairs can remain sub-relativistic during the first heating phase, they
become relativistic enough to upscatter thermal photons above the pair-creation threshold
during the second heating phase.   The resulting surge in pair creation leads to
a drop in mean particle energy while heating continues.  

We find that a broad, non-thermal, Comptonized 
spectrum is created.  There is a smooth connection to the thermal peak above a seed radiation compactness
$\ell_{\rm th} \sim 300$ and total compactness, including heat input, $\ell_{\rm tot} \sim 10^3$.  The required heating is
spatially distributed, and can easily be supplied by the damping of hydromagnetic turbulence.
This particular mechanism results in longitudinal heating of the embedded pairs along the background
magnetic field, with an enhancement of Compton emission over synchrotron
\citep{tb98,thompson06}.  In contrast with the approach taken by 
\cite{ghisellini99}, \cite{giannios06} and \cite{lazzati10}, the high-energy
spectrum is mainly the result of single scatterings of thermal photons by a gradually softening relativistic particle population,
not of multiple scattering by trans-relativistic pairs.  

The thermal radiation field also plays a more
central role in the outflow dynamics and emission than it does in the approach taken by
\cite{usov94} and \cite{lyutikov03} to strongly magnetized outflows.  Those authors assume that fireball radiation 
decouples early on from the magnetic field (forming, e.g. a soft thermal precursor), leaving frozen-in
electrons and pairs that emit synchrotron radiation after re-heating.

Regarding the trigger for delayed dissipation, we focus on the baryons that are embedded in the 
magnetized jet during breakout from the confining medium.  Some residual baryons are pulled
outward by the hyper-Eddington radiation flux, and decouple at a large radius where the jet compactness
drops below a well-defined level.  The magnetic field then is strongly distorted by the
differential motion of the baryons, which supplies enough energy to account for the non-thermal tails of GRBs.
The draining of baryons from the jet head also limits the Lorentz factor to $\Gamma \sim 1/\theta$ at breakout.

Magnetic reconnection remains a natural possibility in a magnetized jet, but pinning
down where it operates depends on understanding the time evolution of a dynamo
process in the engine.  The simplest version of a magnetically striped wind
\citep{drenkhahn02}, which is based on force-free models of pulsars,
is inconsistent with a black-hole driven jet.  A similiar difficulty arises in 
localizing the activity of internal shocks.

\subsection{Plan  of the Paper}

After some further introduction to the problem of
GRB prompt emission, in Section \ref{s:accel} we revisit the acceleration of 
a hot, magnetized jet that has become transparent to scattering.  We take into account the inertia
provided by MHD turbulence that is frozen into the expanding jet, which easily
dominates the inertia of the entrained pairs.  Section
\ref{s:reheat} outlines the effects of reheating in an optically thin, magnetized
jet on the electron and photon distributions.  We review the origin of a strongly anisotropic
particle distribution in Section \ref{s:synch}, and why the reabsorption of cyclo-synchrotron photons cannot
effectively isotropize the pairs during delayed reheating.

Direct kinetic calculations of the photon
and charged particle distributions are described in Section \ref{s:method}.
The results of these calculations are presented in Section \ref{s:results}, 
using as an initial condition the quasi-thermal GRB spectrum calculated in Paper I.
The calculation is repeated in an expanding medium in Section \ref{s:expand}. 

The residual effect of the regenerated $e^\pm$ shell
on the output spectrum is evaluated in Section \ref{s:rescatt} using the Monte Carlo approach
described separately in \citep{tg14b} (hereafter Paper III).  Scattering by an optically thick shell
is shown to have only a modest flattening effect on the low-energy spectrum,
in contradiction with recent claims in the literature.  We also test spectral models that 
invoke continuous heating starting at a modest scattering depth, and continuing across 
the photosphere.  This is shown to produce strong soft-hard evolution that strongly contradicts
the observed behavior.  

The implications of our results are summarized in Section \ref{s:discuss}.  
Appendix \ref{s:kinetic} gives further details of our kinetic code, and Appendix \ref{s:drag}
analyzes the different types of drag experienced by electrostatically heated particles in a magnetized plasma.  

In mathematical expressions we use the shorthand $X_n\times 10^n$ to describe a quantity 
X in cgs units.

\section{Challenges for a Model of \\ the Prompt Gamma-ray Emission}

Before presenting our spectral model, we set the stage by reviewing
several challenges to a theoretical understanding of the prompt emission of GRBs.

\subsection{Origin of the spectral peak in GRBs}
A common early approach to the GRB emission problem was to imagine that all parts of the
non-thermal spectrum originate in the same part of the outflow (e.g. \citealt{peer04,stern04,giannios05}).  Since the
{\it high-energy} part of the spectrum must originate at a high Lorentz factor, this
then implies that the spectral peak in the comoving frame is very low.  Some fine 
tuning is required to avoid pushing the spectral peak to either very
high or low values.  This is especially an issue in synchrotron-self-Compton emission models.

As has been noted by a number of authors, a fireball forming at the engine and then
diluted by adiabatic expansion is inconsistent with the spectral peaks of most
GRBs, producing a peak at too high an energy (e.g. \citealt{rm05}).  Continuing dissipation in a plasma of a very
high compactness naturally generates a spectral peak at $\hbar\omega_{\rm pk}' \sim 0.1 m_ec^2$ 
in the comoving frame through the exponential dependence of the pair density on temperature
\citep{thompson97,ghisellini99,eichler00}.  

Detailed calculations (Paper I) show that this result i) is
sensitive to the baryon loading, requiring a high magnetization $\sigma \gtrsim 10^5$ at jet breakout;
and ii) also depends on distributed heating that is consistent with the damping of bulk hydromagnetic distortions
of the jet fluid, but is probably not consistent with very localized heating by reconnection events.  
When heating is too fast, there is a rapid build-up of cold pairs which drive copious production of
soft photons and a hardening of the low-energy spectrum.

Agreement with the observed spectra of GRBs is obtained if the bulk Lorentz factor is $\Gamma_{\rm br} \sim
1/\theta \sim 3$-10 during this initial heating episode, as would be appropriate for breakout over an angular
width $\theta$:
\be
\hbar\omega_{\rm pk} \sim {4\over 3}\Gamma_{\rm br} \times 0.1~m_ec^2 \sim 200\,\left({\Gamma_{\rm br}\over 3}\right)\quad {\rm keV}.
\ee
In this approach, the low-energy part of the spectrum arises at a moderate radius, 
and is reprocessed to higher energies by delayed dissipation operating at a higher
Lorentz factor and a larger radius.  The origin of this delayed dissipation is one focus
of this paper.

Baryons can dominate the photospheric opacity at breakout
even when the magnetic energy still exceeds the baryon rest mass energy.
Therefore even modest amounts of baryon contamination can force a transition from GRB to X-ray
flash (Paper I).  The cyclo-synchrotron process was found to be the largest source of seed
photons in a plasma with $B^2/8\pi \gtrsim 0.1 P$; otherwise double-Compton emission tends to
dominate.  The dependence of spectral peak on the parameters of a baryon-dominated jet with 
a weak magnetization has been considered by \cite{beloborodov13}; and over an intermediate range of
magnetizations, with particular attention to cyclo-synchrotron emission, by \cite{vurm13}.  As 
these authors note, a baryon-dominated phase could still source the GRB spectral peak if the 
Lorentz factor were somewhat higher than argued for here, e.g. $\Gamma \gtrsim 30$.

\subsection{Role of Finite Scattering Depth}
The role of a scattering photosphere has played
a somewhat nebulous role in modelling the spectra of GRBs.  All emission mechanisms involving
rapidly accelerated, non-thermal particles naturally lead to large scattering depths in
$e^\pm$ pairs, if pushed to a large compactness \citep{guilbert83,meszaros00}.  Two drawbacks
here are that i) the multiplication of the pair density can cut off the high-energy spectrum;
and ii) will rapidly feed back on the energy
of a synchrotron or inverse-Compton peak (with $\omega_{\rm pk}$ scaling as $(n_{e^+}/n_p)^{-2}$
or $(n_{e^+}/n_p)^{-4}$, respectively).  

The second issue is a particularly serious one,
since very bright GRBs with low-energy spectral peaks are not observed.  Although the
strong sensitivity of peak energy on pair density can be partly mitigated by
introducing seed thermal photons \citep{peer06}, some fine tuning
is required to avoid the appearance of a cooling spectrum $F_\omega \propto \omega^{-1/2}$
above the spectral peak \citep{ghisellini99}.  In effect, the injected non-thermal particle spectrum
must extend downward to sub-relativistic energies.  If shocks are responsible, they must
be sub-relativistic and, in turn, cannot efficiently convert bulk kinetic energy to radiation 
\citep{beloborodov00}.

The pair density is naturally buffered to a moderate value $\sim 10$ in a thermal gas, and therefore
may play a role in the creation of the spectral peak \citep{thompson97,ghisellini99}.  Our calculations
show that the low-energy spectrum is consistent with that of a GRB if the outflow is strongly
magnetized (Paper I).  

Runaway pair creation can mediate the formation of a high-energy gamma ray tail to a seed
thermal spectrum, in a way that has only been briefly discussed in the
GRB literature \citep{thompson06}.   Starting at a {\it low} scattering depth, but still
high compactness, distributed heating of the plasma creates relativistic particles.  The
mean particle energy declines following a surge in pair creation, as the injected
energy is shared amongst a greater number of particles.  The inverse-Compton
image of the seed thermal peak then scans through a broad range of energies.

The calculations presented here focus on this mechanism.  We find that the created optical
depth to scattering never exceeds $\sim 1$-4, allowing the high-energy tail to connect
smoothly to the thermal peak.  The closest analog to this mechanism is synchrotron-self Compton
emission by continuously heated pairs \citep{stern04}:  in the absence of seed thermal
radiation the peak energy covers a broad range as the pair density develops.  

Continuous heating of the outflow (e.g. \citealt{drenkhahn02}) could, in principle, maintain 
a finite scattering depth in pairs over a very wide range of radius.  The continuously created pairs
are, however, very sensitive to any (temporary) shutoff in heating.
The pairs mostly annihilate after such a shutoff, and the outflow is rapidly accelerated outward,
leading to a freezeout of causal dissipative processes such as magnetic reconnection
\citep{russo13b}.  An additional argument against such an approach is provided by measurements of GRB
pulse evolution:  our Monte Carlo calculations (Section \ref{s:shell}, Paper III) show that the pulses emerging 
from such a continuously heated photosphere are {\it broader} at higher energies, in strong contrast
to the observed behavior.

\begin{figure}
\includegraphics[width=0.45\textwidth]{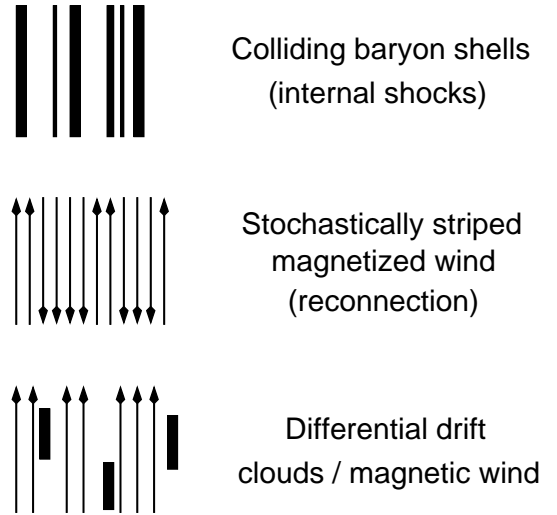}
\caption{Variable gamma-ray emission from a relativistic outflow depends on some type of irregularity.  
A schematic of proposed mechanisms.
1.  Differential motion of baryon shells.   2. Striping of a non-radial magnetic field.  In a black-hole
driven jet this field has a stochastic radial structure, imprinted by a dynamo process in the accreting material.  
3.  Differential motion of a magnetized jet with respect to baryon clouds that are swept up from an external medium.
This third mechanism is distinguished from the others by depending on an intense radiation field:  differential
motion re-emerges below a radiation compactness $\sim 10^3$.}
\vskip .1in
\label{fig:freeenergy}
\end{figure}

\subsection{Powering the High-energy Emission}
Energy can be stored in a GRB fireball in
the structure of the magnetic field \citep{thompson94,spruit01,zhang11,mckinney12} and in differential motion
of baryon shells \citep{rm94,kobayashi97,daigne98}.  

A third possibility (Figure \ref{fig:freeenergy}) involves the differential motion
of the magnetic field and baryons that are collected from the confining medium \citep{thompson06}.
Here we revisit the question of how these two components are accelerated,
and in Paper III examine again how much mass is entrained by the magnetized jet.
The entrained baryons are light enough to be accelerated outward beyond
breakout, but heavy enough to strongly disturb the magnetic field after 
both components have achieved relativistic expansion.

\subsection{Lorentz Factor Growth in the Outflow}  
The photons and pairs in a simple, baryon-free fireball expand ballistically
from the point of the `explosion'.  Here we consider what is, effectively, a radially
offset explosion (actually multiple such explosions) with a significant contribution to the 
energy flux from an entrained magnetic field, and a subdominant contribution from baryon
clumps at breakout.

The Lorentz factor profile of a magnetized jet while confined depends on the details
of the confining medium.  
A common -- but probably erroneous -- assumption is that the jet moves into nearly free
expansion once it leaves the vicinity of the engine.  Such a rapid spreading would lead
to rapid growth in Lorentz factor within a short distance outside the engine.  But in a GRB,
the neutron torus is itself the source of a trans-relativistic wind that is driven by neutrino annihilation 
heating on the torus surface (e.g. \citealt{dessart09}).  Global simulations of accreting 
black holes that include the driving effect of magnetorotational heating (but not
of such neutrino heating) show such an trans-relativistic sheath surrounding the relativistic
jet core \citep{sadowski13}.

After breakout from a confining medium, a magnetized and pair-loaded jet rapidly becomes
transparent and is accelerated outward by a combination of radiation pressure, and
the Lorentz force due to diverging magnetic flux surfaces \citep{russo13a,russo13b}.

\subsection{Radial Localization of the \\ High-energy Emission Process}
Although the temporal power spectrum of a GRB is broad, representing pulses of a range
of widths \citep{bss00}, there is little evidence for systematic evolution of 
the power spectrum within a typical GRB -- as might be expected if dissipation continued
over decades in radius.  The output of essentially all radiation processes depends on the plasma 
energy density and radiation compactness.  The width of the emitted pulses is also
sensitive to radius through the curvature delay of off-axis photons \citep{sari97}, and through the
changing size of dissipating zones, as limited by causal growth of inhomogeneities.

For these reasons, \cite{thompson06} argued that the high-energy gamma-ray emission
is triggered by a feedback process, and pointed to the interaction of the radiation
field with ambient baryons.  As the GRB outflow expands, the radiation field weakens
and its compactness drops.  The magnetic field is strongly perturbed by the differential
motion of the baryons, with the timing of this interaction being determined by a 
reduction in the photon compactness below a critical value.   

Two sources of the baryonic material can be considered:
an external medium that formed before the collapse to a black hole; and denser
material that is derived from the progenitor only after the collapse.  The first
is present to a significant degree only in collapsars, which emit powerful winds during
a Wolf-Rayet phase.  The second is present in both collapsars and binary neutron star
mergers, because the merger product releases a dense neutron-rich wind before the 
collapse to a black hole, which extends to at least $\sim 10^9$ cm from the engine
by the time the MHD jet is fully developed \citep{dessart09}.  

Our focus, here and in Paper III, is on the second channel, baryons that are entrained by an MHD jet
from a confining medium.  When the co-moving radiation compactness is above $\ell_{\rm th}
\sim (Y_e m_e/m_p)^{-1}\sim 4\times 10^3/Y_{e\,0.5}$, where $Y_e = 0.5Y_{e\,0.5}$ is the electron fraction
of the confining medium, baryons can be pushed outward by the intense radiation pressure.  
(Material derived from the surface of a Wolf-Rayet star 
typically has electron fraction $Y_e \sim 0.5$, whereas the neutron-rich outflow from the remnant of a binary
neutron star merger is more electron poor, $Y_e \lesssim 0.1$: \citealt{dessart09}.)  
Acceleration of an MHD fluid containing a light $e^\pm$ gas can continue down to a
much lower compactness, so that the baryons and magnetofluid develop a large differential
Lorentz factor when the seed thermal radiation compactness has dropped to $\ell_{\rm th} \sim 10^2$-$10^3$.

\subsection{Nature of the Engine}

In this situation, a rapidly rotating magnetar is disfavored for a few reasons.  First, the 
magnetized outflow is polluted by a neutron-rich wind from the hot neutron star 
surface \citep{duncan86}, and remains too dirty to support an ion magnetization
as high as $\sim 10^5$ until an interval $\gtrsim 10^2$ s has lapsed \citep{metzger11}.  Second, an
orbiting torus that would help to collimate a polar jet is not excluded by centrifugal forces
from the magnetar surface as it is from the horizon of a black hole.  Indeed it requires fine tuning 
to supply enough angular momentum to the magnetar to power a long GRB without creating such
a torus.  Third, fine collimation of the jet (half-opening angle $\theta_j \lesssim 0.1$ rad)
is required to puncture a CO core before it collapses (\citealt{lazzati09}, Paper III), which
is difficult to achieve with a quasi-spherical outflow from a neutron star.

Even though the black hole in a GRB engine
is surrounded by a very dense, neutron-rich torus, the baryon flux away from the horizon is easily
suppressed by the back-pressure of a dense photon-electron-positron gas.  Such a relativistic gas
is injected into the jet funnel by annihilating neutrinos emitted by the torus,
$\nu_e + \bar\nu_e \rightarrow e^+ + e^-$ \citep{eichler89,zalamea11}.

\section{Acceleration of an Optically Thin and Strongly Magnetized Shell}\label{s:accel}

We consider a transient, magnetized outflow, of duration $t_{\rm eng}$, that is sourced by the 
horizon of a hyper-accreting black hole.  
The outflow contains a thermal radiation field, with a flat spectrum below the peak that is generated 
during an intermediate stage of heating during breakout (Paper I).

The baryonic magnetization is very high, $\sigma_{\rm ion} = B^2/4\pi\rho_{\rm ion} c^2 \gtrsim 10^5$ 
in the frame of the engine.  The magnetic field $B$ is predominantly non-radial over a wide range of radius.  
The rest mass density $\rho_{\rm ion}$ here refers only to baryons advected out from the black hole
ergosphere.  This constraint on $\sigma_{\rm ion}$ derives from the requirement that the relativistic
component of the outflow is pair-dominated during breakout; otherwise the low-energy 
spectrum is harder and the peak softer (Paper I, see also \citealt{vurm13,beloborodov13}).
In order to power the high-energy emission of a GRB, some component of the outflow other than thermal
radiation carries much of the energy at breakout.  Given that the density of embedded pairs
is exponentially suppressed near breakout, the magnetic field can be viewed as a default choice.

The simplest case is a single pulse of activity of the central engine.  One
frequently encounters the idea that the engine may be sporadic, leading to radial structure 
in the outflow.  In Paper III, we explore the role of angular variations in producing the pulse
structure of GRBs, and the possibility that\footnote{The standard measure of the duration of the
prompt gamma-ray emission, encompassing 90\% of the fluence.}  $T_{90}$ is much longer than $t_{\rm eng}$ 
as measured at jet breakout.  

The outflow escapes a confining medium at a distance $R_{\rm br}$ from the engine,
where `br' labels breakout.  We refer to this ambient medium in a generalized sense, because
it can also be in bulk motion away from the engine (e.g. \citealt{rr02}).  Deconfinement of a relativistic, magnetized
fluid may even extend close to, or beyond, the transition
between `jet' (${\cal R}_{\rm br} < 1$) and `pancake' (${\cal R}_{\rm br} > 1$) geometries:
\be\label{eq:rbreak}
R_{\rm br} \equiv {\cal R}_{\rm br} \cdot 2\Gamma_{\rm br}^2 ct_{\rm eng}.
\ee
Here $\Gamma_{\rm br}$ represents the Lorentz factor of the baryonic material through which
the magnetized fluid is moving.  

A corrugation instability is triggered in a forward baryon
shell when it becomes geometrically thin, which is possible when the shell and the magnetofluid
behind it expand to ${\cal R}_{\rm br} > 1$.  Then, as is discussed further in Paper III,
the duration of the gamma-ray emission is dominated by the curvature delay across the shell:
\be
T_{90} \sim {\cal R}_{\rm br} t_{\rm eng}.
\ee

It should be noted that the numerical value of (\ref{eq:rbreak}), 
\be\label{eq:rbreak2}
R_{\rm br} \sim 2\times 10^{11}{\cal R}_{\rm br}
\left({\Gamma_{\rm br}\over 3}\right)^2\left({t_{\rm eng}\over{\rm s}}\right)~{\rm cm},
\ee
can exceed the radius of the pre-existing `envelope'.  For a long GRB, this may be the radius of the
Wolf-Rayet progenitor and, for a short GRB, the neutron-rich outflow that is emitted by a merged neutron
star binary,
\be
R_{\rm env} \sim  \left\{
\begin{array}{ll}
\lesssim R_\odot\quad & ({\rm WR}), \\
ct_{\rm eng}/3 \sim 1\times 10^9\,(t_{\rm eng}/0.1~{\rm s})~{\rm cm}\quad & ({\rm merger}).
\end{array}\right.
\ee

The Poynting and radiation energy fluxes at breakout are expressed in terms of the compactness,
\be
\ell_{\rm P,br} = {\sigma_T\over m_ec^2} {{B'}^2\over 8\pi} {R_{\rm br}\over \Gamma_{\rm br}};
\quad\quad \ell_{\rm th, br} = {\sigma_T\over m_ec^2} U_\gamma' {R_{\rm br}\over \Gamma_{\rm br}},
\ee
as defined in the comoving (primed) frame.  The apparent net energies carried by thermal radiation and 
magnetic Poynting flux are
\be
E_{\rm P,iso} = \Gamma^2 {B'}^2 r^2 ct_{\rm eng};  \quad\quad 
E_{\gamma,\rm iso} = {4\over 3}\Gamma^2 U_\gamma' 4\pi r^2 ct_{\rm eng}.
\ee
Then
\be
\ell_{\rm P,br} = {\sigma_T E_{\rm P,iso}\over 16\pi {\cal R}_{\rm br}\Gamma_{\rm br}^5 m_ec^4t_{\rm eng}^2};\quad\quad
\ell_{\rm th,br} = {3\over 2}{E_{\rm \gamma,iso}\over E_{\rm P\,iso}}\ell_{\rm P,br}.
\ee
Numerically, this works out to 
\be\label{eq:lbreak}
\ell_{\rm th,br} = {1\times 10^8\over ({\cal R}_{\rm br}/10)}\,\left({E_{\rm \gamma,iso}\over 10^{52}~{\rm erg}}\right)\,\left({\Gamma_{\rm br}\over 3}\right)^{-5}
\left({t_{\rm eng}\over {\rm s}}\right)^{-2}.
\ee

The radiation luminosity, normalized here at breakout, continues to grow as the magnetized component 
is accelerated outward by the Lorentz force (Section \ref{s:magaccel});
and after the embedded pairs are reheated (Sections \ref{s:reheat}-\ref{s:results}).

Although the breakout compactness increases in proportion to $E_{\rm iso}$, it also has a strong inverse 
dependence on $\Gamma_{\rm br}$.  If the angle-integrated burst energy is regulated by the binding energy
of the core \citep{tmr07}, which varies weakly with progenitor mass, and if breakout occurs in a causal 
manner on an angular scale $\delta\theta$, then $\Gamma_{\rm br}\,\delta\theta \sim 1$.  For a single pulse, 
one has $E_{\rm iso} (\delta\theta)^2 \sim E_{\rm iso}/\Gamma_{\rm br}^2 \sim$ const, and so 
\be\label{eq:elleiso}
\ell_{\rm br} \propto E_{\rm iso}^{-3/2}.
\ee
One sees that more luminous GRBs can be inferred to have a lower breakout compactness.

\subsection{Acceleration of Matter by Anisotropic Photon Pressure Outside Breakout}

Once the outflow becomes optically thin, the photon component self-collimates and defines a frame
in which entrained particles move relativistically.  In this section, we proceed first by neglecting
the Lorentz force and the inertia of the magnetic field. 

The net radiation force vanishes in a frame moving with Lorentz factor
\be\label{eq:gameq}
\Gamma_{\rm eq}(r) \simeq \Gamma_{\rm br}\left({r\over R_{\rm br}}\right).
\ee
Then the radial flow of the entrained electrons and positrons closely approximates $\Gamma \simeq \Gamma_{\rm eq}$
until $\ell_{\rm th}$ drops below unity.  Since 
\be\label{eq:lgam}
\ell_{\rm th}(r) = \ell_{\rm th,br}\left({r\over R_{\rm br}}\right)^{-1}\left({\Gamma\over \Gamma_{\rm br}}\right)^{-3},
\ee
one finds that $\Gamma$ saturates at
\ba\label{eq:gamsat}
\Gamma_{\rm sat} &\sim& \Gamma_{\rm br} \left(\ell_{\rm th,br}\right)^{1/4}\nn
               &=& 600\,{(f_{\rm th,br} E_{j,51})^{1/4}\over (t_{\rm eng}/{\rm s})^{1/2} (\Gamma_{\rm br}\theta_j)^{1/2}}
                   \left({\Gamma_{\rm br}/3\over {\cal R}_{\rm br}}\right)^{1/4}
\ea   
at a radius
\ba\label{eq:rsat}
R_{\rm sat} &=& {\Gamma_{\rm sat}\over\Gamma_{\rm br}}\,R_{\rm br}\nn
            &=& 5.4\times 10^{14}\,\left({{\cal R}_{\rm br} t_{\rm eng}\over 10~{\rm s}}\right) \left({\ell_{\rm th,br}\over 10^8}\right)^{1/4}
             \left({\Gamma_{\rm br}\over 3}\right)^2\quad {\rm cm}.\nn
\ea
Here, for illustration, we have re-written $E_{\rm \gamma,iso} = (2/\theta_j^2) f_{\rm th,br} E_j$,
where $E_j$ is the total (bi-axial) jet energy and a fraction $f_{\rm th,br}$ is
carried by thermal photons at breakout.

\subsection{Acceleration of a Very Strongly Magnetized Outflow with Frozen MHD Turbulence}\label{s:magaccel}

Now we take into account the Lorentz force acting on a magnetized outflow, and its interaction with the radiation force.
Both of these forces are calculated using the formalism of \cite{russo13b}.  The spreading of magnetic flux surfaces
outside breakout is incorporated with a simple causal prescription, and the radiation force and $\Gamma_{\rm eq}$
are calculated by taking moments of the radiation field in the small-angle approximation.

A non-radial magnetic field carried outward by a relativistic jet contributes negligible inertia beyond the fast magnetosonic surface,
which sits at Lorentz factor $\Gamma \simeq \sigma^{1/3}$.  The magnetization $\sigma$, as defined by the inertia of the embedded
pairs, is formally very high at breakout.  Then one must examine carefully other possible sources of inertia.

To illustrate how turbulence provides inertia, we first consider the expansion of a plane-symmetric,
magnetized slab into a vacuum.  We provide an analytic solution to the similarity problem posed by \cite{granot11}, here generalized
to include both cold matter and a background of Alfv\'en waves in the pre-expansion state.  Then we generalize the calculation of 
jet breakout with radiation pressure by \cite{russo13b} to include the effects of frozen MHD turbulence.   For the time being, 
we ignore any baryons derived from a confining medium.

\subsubsection{Self-similar Expansion of a Magnetized, \\ Turbulent Slab}

Consider a semi-infinite medium, initially filling $x < 0$ and containing a uniform magnetic field ${\bf B} = B_0 \hat z$ and 
perfectly conducting matter with proper density $\rho_0 \ll B_0^2/8\pi c^2$.  Superposed on this relativistic magnetofluid 
is a gas of Alfv\'en waves of energy density $U_t = \varepsilon_t B^2/8\pi$.   The usual magnetization parameter is 
$\sigma = B^2/4\pi\rho c^2$.  The effective magnetization, taking into account the inertia of the turbulence, is
\be\label{eq:sigeff}
\sigma_{\rm eff} = {B^2/4\pi\over U_t/2 + \Gamma\rho c^2} = {\sigma\over \varepsilon_t\sigma/4 + 1}.
\ee
There is an additional factor of 1/2 multiplying $U_t$ because the component of the magnetic field that is parallel to the direction of the
mean flow imparts a vanishing Lorentz force.

The medium begins to expand into a vacuum at $x > 0$ at time $t = 0$.
We follow the expansion with velocity ${\bf v} = v\hat x$ using the similarity coordinate $\chi = x/ct$,
\be\label{eq:sim}
B = B_0 \hat B\left({x\over ct}\right); \quad v = c\hat v\left({x\over ct}\right).
\ee
The $x-$component of the relativistic Euler equation is
\be\label{eq:euler}
\Gamma\rho\left[{\partial(\Gamma v)\over\partial t} + v{\partial(\Gamma v)\over\partial x}\right] = 
                 {1\over c}({\bf J}\times{\bf B})_x + {\bnabla\cdot{\bf E}\over 4\pi}E_x,
\ee
where ${\bf E}$, ${\bf B}$ denote electric and magnetic fields and ${\bf J}$ is the current density.
The Lorentz force has contributions from both the background laminar fluid and the turbulence,
\be
{1\over c}({\bf J}\times{\bf B})_x = -{B\over 4\pi}\left[{\partial B\over \partial x} + {1\over c}{\partial E_y\over\partial t}\right] + 
{1\over c}\langle B_{y,T} J_{z,T} - B_{z,T}J_{y,T}\rangle.
\ee
Here $\langle ...\rangle$ denotes a temporal and spatial average over quantities bilinear in the waves.
Only the turbulence contributes to the Coulomb force in this planar geometry.  Since $B_{z,T}' = E_{z,T}' = 0$ for Alfv\'en waves moving along the
$z$-magnetic field in the co-moving frame, the wave fields in the lab frame are
\ba\label{eq:transf}
B_{x,T} &=& B_{x,T}'; \quad B_{y,T} = \Gamma B_{y,T}'; \quad B_{z,T} = \Gamma{v\over c}E_{y,T}' \nn
E_{x,T} &=& E_{x,T}'; \quad E_{y,T} = \Gamma E_{y,T}'; \quad E_{z,T} = -\Gamma{v\over c}B_{y,T}'.\nn
\ea
The net contribution to the Lorentz force from large-scale $t$- and $x$-derivatives of these fields is
\ba\label{eq:lorturb}
F_t &\equiv& {1\over c}\left\langle ({\bf J}_t\times {\bf B}_t)_x\right\rangle + {1\over 4\pi}\left\langle {\partial E_{x,T}\over\partial x} E_{x,T}\right \rangle\nn
 &=& -{1\over 4\pi}\biggl\langle {1\over 2}{\partial\over\partial x}\left( B_{y,T}^2 + B_{z,T}^2 - E_{x,T}^2\right) \nn
         &&\quad\quad + B_{z,T}{1\over c}{\partial E_{y,T}\over\partial t} - B_{y,T}{1\over c}{\partial E_{z,T}\over\partial t}\biggr\rangle.
\ea

The mean magnetic field is imprinted in the fluid, and evolves according to 
\be\label{eq:induc}
{\partial B\over\partial t} + {\partial(vB)\over\partial x} = 0;  \quad B = B_0 {\Gamma\rho\over\rho_0}.
\ee
We also need an equation of state for the turbulent pressure in the comoving frame.  The pressure of
Alfv\'en waves in an isotropically expanding plasma evolves in the same way as photons, but here there
is no expansion parallel to the background field.  Then the adiabatic invariant is
\be\label{eq:scale}
(B_t')^2 \propto B' = {B\over\Gamma}.
\ee
Making use of this relation and equations (\ref{eq:transf}) and (\ref{eq:induc}), assuming equal contributions from 
the two polarization modes, and approximating $E_{x,T}' = B_{y,T}'$ (as appropriate for Alfv\'en waves in a very strongly
magnetized plasma), the turbulent Lorentz force (\ref{eq:lorturb}) simplifies to
\be
F_t = -{\varepsilon_t\over 4}\Gamma^3 {BB_0\over 4\pi c^2} \left({\partial v\over\partial t} + v{\partial v\over\partial x}\right).
\ee
The Euler equation becomes
\ba\label{eq:euler2}
&&\left[\Gamma^3\left(\rho_0 c^2 + {\varepsilon_t\over 4} {B_0^2\over 4\pi}\right) + {BB_0\over 4\pi}\right]{\partial v\over\partial t}\nn
&&+\left[\Gamma^3\left(\rho_0 c^2 + {\varepsilon_t\over 4} {B_0^2\over 4\pi}\right) - {BB_0\over 4\pi}\right]v{\partial v\over\partial x}
= -{1\over \Gamma^2}{B_0\over 4\pi}{\partial B\over\partial x}.\nn
\ea
We see that the turbulence produces a simple re-scaling
of the material energy density, corresponding to a magnetization (\ref{eq:sigeff}).  

In the remainder of this section, we provide a simple analytic solution, which can be applied to both the laminar 
and turbulent fluids.   Substituting the ansatz (\ref{eq:sim}) into equations 
(\ref{eq:euler}) and (\ref{eq:induc}) in combination with (\ref{eq:lorturb}) and (\ref{eq:scale}), and
replacing $\sigma_0$ with $\sigma_{\rm eff,0}$, gives
\ba\label{eq:reduced}
\left[\Gamma^3(\hat v - \chi) - \sigma_{\rm eff,0} \chi \hat B\right]{d\hat v\over d\chi} &=&
-\sigma_{\rm eff,0}(1-\chi\hat v){d\hat B\over d\chi};\nn
(\chi - \hat v){d\hat B\over d\chi} &=& \hat B {d\hat v\over d\chi}.
\ea
Combining these two equations gives a constraint on the evolved magnetic field,
\be\label{eq:integral}
\sigma_{\rm eff,0} \hat B =  \Gamma^3{(\chi -\hat v)^2\over 1-\chi^2}.
\ee

The inner boundary of the rarefaction wave is determined by setting $\hat B = 1$ and $\hat v = 0$, giving
\be\label{eq:inner}
x = -\left({\sigma_{\rm eff,0}\over 1 + \sigma_{\rm eff,0}}\right)^{1/2} ct.
\ee
This coincides with the position of a magnetosonic wave moving inward through the magnetofluid
and starting at $x=0$ at $t=0$.

The solution for the velocity field is obtained by differentiating equation (\ref{eq:integral})
with respect to the similarity variable $\chi$, and then substituting into the second of equations (\ref{eq:reduced}), giving
\be
{d\hat v\over d\chi} = {2(1-\hat v^2)\over 3(1-\chi^2)}.
\ee
This integrates to give 
\be
{1 + \hat v\over 1 - \hat v} = \left[(1+\sigma_{\rm eff,0})^{1/2} + \sigma_{\rm eff,0}^{1/2}\right]\left({1+\chi\over 1-\chi}\right)^{2/3}.
\ee
The coefficient has been determined by setting $\hat v = 0$ at the inner boundary (\ref{eq:inner}) of the rarefaction wave.
The outer boundary of the wave coincides with $\hat B =0$, corresponding to $\hat v = \chi$.  

The maximum Lorentz factor is reached at this boundary, and is found to be
\be
\Gamma_{\rm max} \simeq 2\sigma_{\rm eff,0}\quad (\sigma_{\rm eff,0} \gg 1).
\ee
This agrees with the laminar solution of \cite{granot11}, but with the important distinction that
$\sigma_{\rm eff,0}$ is rescaled downward from $\sigma_0$.  When the turbulent intensity $\varepsilon_t \gg 1/\sigma_0$,
as is almost certainly the case in the applications considered here, one has
\be
\sigma_{\rm eff,0} \simeq {1\over\varepsilon_t}; \quad\quad \Gamma_{\rm max} \simeq {2\over\varepsilon_t} \ll \sigma_0.
\ee

In the parts of the fluid which reach a high Lorentz factor, one finds
\be
\Gamma = \sigma_{\rm eff,0}^{1/3} \left[{1+\chi\over 2(1-\chi)}\right]^{1/3}\quad \chi \leq 1 - (4\sigma_{\rm eff,0})^{-2}.
\ee
Focusing on the thin, relativistic layer near the outer boundary ($\chi \simeq 1$), one 
finds for the magnetic and velocity fields,
\be
B = {B_0\over 2}\left(1 - {\Gamma\over 2\sigma_{\rm eff,0}}\right)^2;
\quad \rho = {\rho_0\over 2}\left(1 - {\Gamma\over 2\sigma_{\rm eff,0}}\right).
\ee

\subsubsection{Expansion of an Optically Thin, Turbulent Jet}

The jet material, now optically thin, accelerates outward by a combination of radiation pressure and the Lorentz force.
Then the radial causal distance $\sim r/\Gamma^2$ shrinks
in the background inertial frame.  The angular causal distance $\sim r/\Gamma$ also shrinks if the increase in $\Gamma$
is faster than linear \citep{tchek10,russo13b}.  Therefore MHD modes with wavelength $\sim r/\Gamma$, 
and especially those with a significant radial component, will become frozen into the flow, and only gradually be smoothed
out by expansion.

The frozen turbulence behaves like a relativistic fluid.  The enthalpy per scattering charge that is 
carried by the electromagnetic field is
\be
w_{\rm P} = {\hat B_p\cdot ({\bf E}\times {\bf B})\over 4\pi n_e}.
\ee
Here ${\bf B}$ and ${\bf E}$ are the magnetic and electric fields, and $n_e$ is the density
of scattering charges, all evaluated in the inertial frame.  $\hat B_p$ denotes
 the unit vector parallel to the poloidal magnetic field.  

We work in the approximations that i) the flow is radial, with small angular deviations leading to a large Lorentz force;
and ii) the background magnetic field is purely toroidal.\footnote{Here we can neglect the mean radial magnetic field
threading the jet, since it has expanded far beyond the speed-of-light cylinder of the engine.}  Then
\be
w_{\rm P} = {B_\phi^2+B_\theta^2\over 4\pi n_e} = \bar w_{\rm P} + w_t,
\ee
where
\be
\bar w_{\rm P} = {{\bar B}_\phi^2\over 4\pi n_e}
\ee
is the contribution from the mean flow, and 
\be
w_t = {(\delta B_\phi)^2 + (\delta B_\theta)^2\over 4\pi n_e}
\ee
from the frozen turbulence. 

We focus here on steady expansion, with a uniform rate of transfer of toroidal magnetic flux
along a poloidal flow line,
\be\label{eq:fluxtrans}
{\bar B_\phi \over n_e r\sin\theta} = {\rm const}.
\ee
To obtain the scaling of $w_t$ with radius, one notes that the wave field can be written
as 
\be
\delta {\bf B}' = {1\over r\sin\theta} {\partial \bxi\over\partial \phi} B_\phi'
\ee
where $\bxi$ is the Lagrangian displacement field of the magnetofluid and the prime denotes the
comoving frame.  The gradient scales as $r^{-1}$ under expansion, and $\xi^2 \propto 1/B_\phi'$,
hence for a nearly radial flow 
\be\label{eq:wt}
w_t \sim {(\Gamma \delta B')^2\over 4\pi n_e} \propto {\Gamma^2 B_\phi'\over r^2 n_e} \propto {\Gamma\over r}. \ee
Therefore $w_t$ evolves according to
\be\label{eq:ev1}
{1\over w_t}{dw_t\over dr} = {1\over \Gamma}{d\Gamma\over dr} - {1\over r}.
\ee
The turbulent energy per particle decays as $r^{-1}$ in the comoving frame, but may even grow slightly
in the inertial frame.

It is common to express the relative partitioning between rest energy and magnetic energy in terms of the magnetization,
\be
\sigma = {(\bar B_\phi)^2\over 4\pi n_e \mu c^2} = {\bar w_{\rm P}\over \mu c^2}.
\ee
This can be written in terms of the magnetic compactness,
\be
\ell_{\rm P}^{\rm lab} = {dL_{\rm P}/d\Omega\over dL_\gamma/d\Omega}\,\ell_{\rm th}^{\rm lab}
\ee
where 
\be
\ell_{\rm th}^{\rm lab} \sim {4\over 3}\Gamma^3 \ell_{\rm th} \equiv {\sigma_T\over \mu c^3 r} {dL_\gamma\over d\Omega}
\ee
is the photon compactness in the inertial frame.  Then (e.g. equation (14) of \citealt{russo13b}),
\be
\sigma = {\ell_{\rm P}^{\rm lab}\over 6\Gamma^2\tau_{\rm T}},
\ee
where the scattering depth is evaluated for a radial ray in the Thomson approximation,
\be\label{eq:tauT}
\tau_{\rm T}(r) = \int_r^\infty {\sigma_T n_e\over 2\Gamma^2} dr.
\ee

A minimal magnetization in a pair-dominated outflow is obtained by taking an inertia 
$\mu = 2m_p/Y_{e\,0.5}$.  Then at breakout ($\tau_{\rm T} \sim 3$) one has
\be\label{eq:sigbr}
\sigma = 4\times 10^6\,{\cal R}_{\rm br}^{-1}\left({E_{\rm P}\over 10^{52}~{\rm erg}}\right)\,\left({\Gamma\over 3}\right)^{-2}
\left({t_{\rm eng}\over {\rm s}}\right)^{-2}.
\ee

The response of the outflow to an imposed radial force (such as radiation pressure)
depends on $\sigma$ in a subtle way.  In a purely laminar outflow, 
the effective particle inertia is (e.g. \citealt{goldreich70,russo13b})
\be\label{eq:mueff}
\mu_{\rm eff} = m_e\left(1-{\sigma\over\Gamma^3}\right)\quad (\sigma, \Gamma \gg 1).
\ee
This would be negative at breakout in an outflow with magnetization (\ref{eq:sigbr}) and
Lorentz factor $\Gamma \sim 3-10$.   

At such large values of $\sigma$, the energy carried by the frozen turbulence dominates the 
kinetic energy of the entrained charges.  Then the effective magnetization is obtained by replacing
$\Gamma \mu c^2 \rightarrow w_t$,
\be
\sigma_t \sim \left({B_\phi\over \delta B}\right)^2 \Gamma.
\ee

We focus here on the case where the turbulent intensity at breakout is large enough to ensure
$\sigma_t < \Gamma^3$.  Then we can work with the total energy integral
\be\label{eq:enint}
w = \Gamma \mu c^2 + \bar w_{\rm P} + w_t + {\cal R},
\ee
where
\be
{\cal R} \simeq {1\over n_e \mu c^3 r^2}{dL_\gamma\over d\Omega} = {\ell_{\rm th}^{\rm lab} \over \sigma_T n_e r}
\ee
is the energy per particle that is carried by radiation, in units of $m_ec^2$.  The photons have energies around $0.1m_ec^2$
in the comoving frame (Paper I), so we focus here on Thomson scattering with
cross section $\sigma_T$.  Then from equation (32) of \cite{russo13b},
\be\label{eq:ev2}
{d{\cal R}\over dr} = {\sigma_T n_e\over 4\Gamma^2}\left[\left({\Gamma\over\Gamma_{\rm eq}}\right)^4-1\right] {\cal R}.
\ee
We work in the regime where
$w_t$, $\bar w_{\rm P}$, and ${\cal R}$ are all much larger than unity.  Then the kinetic term in (\ref{eq:enint}) can be
neglected, and making use of the scaling (\ref{eq:wt}), one finds that $dw/dr = 0$ gives
\be\label{eq:ev3}
{d\Gamma\over dr} = {\Gamma\over r} - {\Gamma\over w_t}\left({d{\cal R}\over dr} + {d\sigma\over dr}\right).
\ee

The change in magnetization is driven mainly by angular spreading of the magnetic field lines (\citealt{tchek10}, and references
therein).  Equation (\ref{eq:fluxtrans}) may be used to reference $\sigma$ to its value at breakout,
\be
\sigma = \sigma_{\rm br}{r\sin\theta\over (r\sin\theta)_{\rm br}}{\bar B_\phi\over \bar B_{\phi,\rm br}}, 
\ee
and taking the small-angle limit for the polar angle of a poloidal flux surface, $\theta = \delta\theta + \theta_{\rm br}$,
one finds
\be
\sigma = \sigma_{\rm br}{1+\delta\theta/\theta_{\rm br}\over 1 + d(\delta\theta)/d\theta_{\rm br}}
\ee
The change in $\sigma$ due to angular spreading can, in general, be of either sign.  We focus here on parts of
the outflow where $\sigma$ decreases with radius, corresponding to 
\be\label{eq:ev4}
{d\sigma\over dr} \sim - \sigma {|\beta_\theta|\over \theta_{\rm br} r}.
\ee
The angular velocity is allowed to grow only at a causal rate,
\be\label{eq:ev5}
{d|\beta_\theta|\over dr} \sim {1\over \Gamma r}.
\ee

Finally we must evaluate the change in the equilibrium frame of the radiation field.
One has $\Gamma_{\rm eq} \propto r$ in a freely expanding radiation field.    We follow the procedure of
\cite{russo13b} and take angular moments of the radiation field, 
\be
F_n = {1\over 2}\int d\mu (1-\mu)^n {1\over r^2}{dL_\gamma \over d\Omega}(\mu).
\ee
Then $\Gamma_{\rm eq}^2 = F_0/4F_1$, and by combining equations (30) and (31) of \cite{russo13b}, one finds
\be\label{eq:ev6}
\left({\Gamma_{\rm eq}\over r}\right)^{-1}{d\over dr}\left({\Gamma_{\rm eq}\over r}\right)
     = \left({\Gamma_{\rm eq}^2\over \Gamma^2} + {1\over 2}\right) {1\over {\cal R}}{d{\cal R}\over dr}.
\ee

A closed set of equations describing the acceleration of a hot, magnetized jet outside breakout is provided
by (\ref{eq:ev1}), (\ref{eq:ev2}), (\ref{eq:ev3}), (\ref{eq:ev4}), (\ref{eq:ev5}), and (\ref{eq:ev6}).
The value of $n_e$ at breakout is iterated to give a pre-determined value $\tau_{\rm T}(R_{\rm br})$.

\begin{figure}
\includegraphics[width=0.45\textwidth]{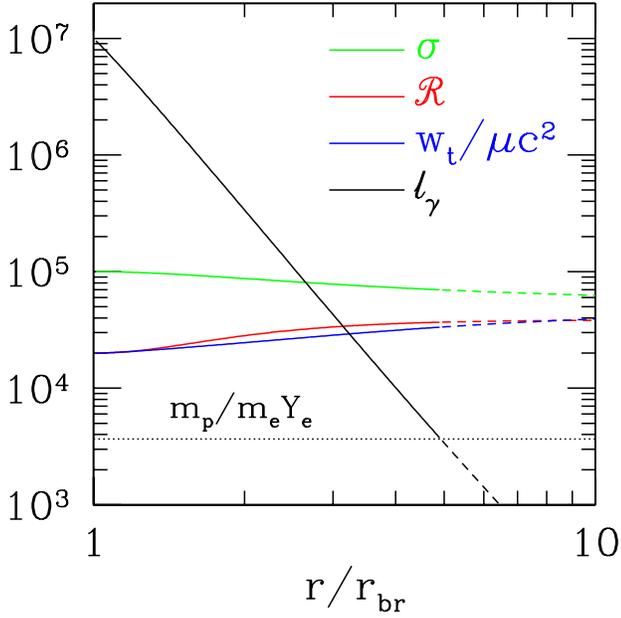}
%\vskip -0.5in
\caption{Various components of the energy of a magnetized jet as a function of distance
outside breakout from a confining medium.  At breakout, scattering depth $\tau(R_{\rm br} \rightarrow\infty) = 3$;
magnetization (Poynting energy/particle rest energy) $\sigma = 10^5$; and both the radiation energy ${\cal R}$
and turbulent energy $w_t/\mu c^2$ equal to $0.2\sigma$.  Curves are dashed outside the radius (\ref{eq:rbar}) at which
growth of the matter Lorentz factor stalls, but continuing acceleration of the magnetized jet is still possible.}
\vskip .2in
\label{fig:energy}
\end{figure}

\begin{figure}
\includegraphics[width=0.45\textwidth]{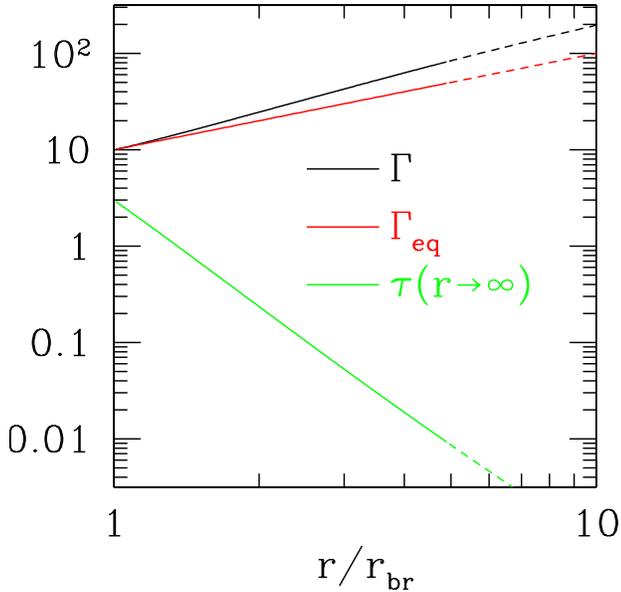}
%\vskip -0.5in
\caption{Lorentz factor $\Gamma$ of the outflow shown in Figure \ref{fig:energy}.  $\Gamma_{\rm eq}$
defines the frame in which the radiation force vanishes.  Electron scattering optical depth is measured from radius $r$
to infinity.}
\vskip .2in
\label{fig:gamma}
\end{figure}

The profile of magnetization, turbulent energy, and radiative energy per particle is shown in Figure
\ref{fig:energy}, for an outflow with optical depth $\tau_{\rm res} = 3$ at breakout,
Lorentz factor $\Gamma = 10$, and jet opening angle $\theta = \Gamma^{-1}$.  The growth of Lorentz factor,
shown in Figure \ref{fig:gamma}, is somewhat faster than linear ($d\ln\Gamma/d\ln r \simeq 1.4$),
and slightly outstrips the growth of $\Gamma_{\rm eq}$.


\vfil\eject
%%%%%%%%%%%%%%%%%%%%%%% PHYSICAL MODEL %%%%%%%%%%%%%%%%%%%%%%%%%%%%
\section{Distributed Heating in a Medium of Low Initial Optical Depth}\label{s:reheat}

\begin{figure}
\includegraphics[width=0.45\textwidth]{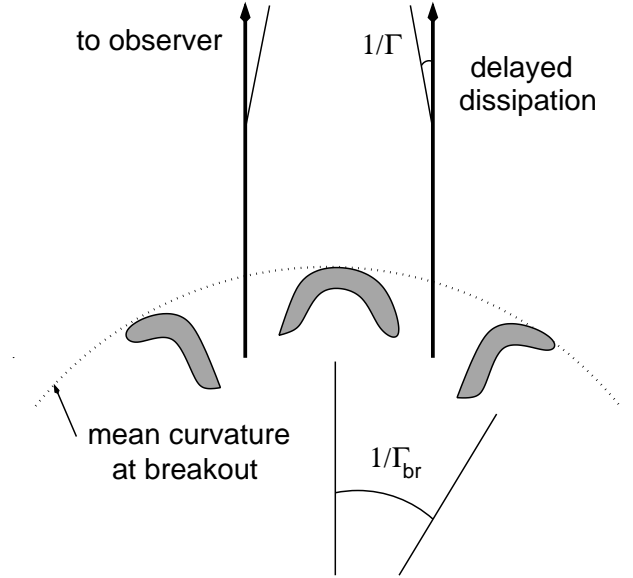}
\caption{Shocked material derived from the confining medium experiences a corrugation instability at breakout, once the contact
discontinuity accelerates to $\Gamma_c \sim 1/\theta_j$.  Magnetized material escaping through holes in the corrugated shell
deviates from purely radial flow by an angle $\delta\theta \sim 1/\Gamma_c$.  Causal contact is then lost across an angle $\sim \theta_j$
as the magnetized material accelerates beyond breakout.  When dissipation resumes at a larger radius (\ref{eq:rbar}), overlapping gamma ray
pulses can result from causally separated events.}
%Overlapping pulses which appear not to disturb each other are observed in some 
%GRBs (e.g. \citealt{pendleton97}).}
\label{fig:breakout}
\vskip .2in
\end{figure}

We now illustrate how a high-energy, non-thermal spectrum is generated by pair breakdown
in a magnetized plasma containing a thermal photon seed.  Detailed kinetic calculations are described
in Sections \ref{s:method}, \ref{s:results}, and \ref{s:expand}.

\subsection{Delayed Decoupling between Baryons and the Relativistic Components of the Outflow}\label{s:decouple}

A hyperluminous, magnetized jet that breaks through a cloud of baryonic material can entrain a certain
mass of baryons in spite of the corrugation instability that the baryons suffer at the jet head.
The radiation field entrains baryons if they are thin enough to cool radiatively on the dynamical
time, with a scattering depth
\be\label{eq:taucool}
\tau_{\rm T,cool} \lesssim \left({3\ell_{\rm P,br}\over \Gamma_{\rm br}} {m_eY_e\over m_p}\right)^{1/2}.
\ee
Here $\ell_{\rm P,br} = (\sigma_T/m_ec^2) (R_{\rm br}/\Gamma_{\rm br}) {B'}^2/8\pi$ is the compactness
of the magnetized outflow, measured in the comoving frame at breakout.
Then the rest-mass luminosity at breakout can be related to the Poynting luminosity in a straightforward way 
(equation (24) of Paper III),
\be\label{eq:fbar}
{\langle dL_{\rm rest}/d\Omega \rangle\over dL_P/d\Omega}\biggr|_{\rm br} =
           {f_{\rm cover}\cdot R_{\rm br}^2 \Sigma_b c^2 \over t_{\rm eng}\, dL_P/d\Omega}
  = {3f_{\rm cover}{\cal R}_{\rm br}\over 2\Gamma_{\rm br}^2\tau_{\rm T,cool}}.
\ee
Here $f_{\rm cover}$ is the angular covering factor of the residual shells of baryons, and ${\cal R}_{\rm br}$ 
is the dimensionless breakout radius (\ref{eq:rbreak}).   Because the compactness at breakout (\ref{eq:lbreak}) 
is still very large, this thin baryonic material remains optically thick, $f_{\rm cover}\tau_{\rm T,cool} \gg 1$.

As the jet Lorentz factor grows, there is a contraction in the angular distance
over which a causal disturbance can propagate,
\be
r{d\theta\over dr} \sim \Gamma^{-1}.
\ee
This means that parts of the outflow containing clumps of entrained baryons, and those which do not, can 
become causally separated.  Clumps of a typical size and separation $\delta\theta_{\rm br}$ at breakout
will lose contact when the outflow has expanded to a Lorentz factor $\Gamma > 1/\delta\theta_{\rm br}$.

Once the (comoving) radiative compactness drops below
\be\label{eq:lbar}
\ell_{\rm th} \sim {m_p\over Y_em_e},
\ee
the parts of the jet containing baryons are no longer accelerated outward by the anisotropic photon pressure.  

The limiting Lorentz factor depends on the radial flow profile. 
Here we choose a simple power-law expansion law, $\Gamma = \Gamma_{\rm br} (r/R_{\rm br})^\delta$.
Then the compactness and scattering depth decrease as $\ell_{\rm th} = \ell_{\rm th,br} (r/R_{\rm br})^{-1-3\delta}$
and $\tau_{\rm T} = \tau_{\rm T,br} (r/R_{\rm br})^{-1-2\delta}$.  The compactness reaches the limiting value
(\ref{eq:lbar}) at the radius
\be\label{eq:rbar}
R_{\rm sat,ei} \sim \left({m_e Y_e\over m_p} \ell_{\rm th,br}\right)^{-1/(1+3\delta)} R_{\rm br},
\ee
where the Lorentz factor has grown to 
\be\label{eq:gamsatei}
\Gamma_{\rm sat,ei} = \left({Y_e m_e\over m_p}\right)^{1/4}\Gamma_{\rm sat} = 0.13\,Y_{e\,0.5}^{1/4}
\Gamma_{\rm sat}
\ee
in the simplest case of linear expansion ($\delta = 1$).

The flow develops strong inhomogeneities in Lorentz factor beyond the radius (\ref{eq:rbar}), 
because the baryon-free parts of the jet continue to accelerate outward.  Disturbances from
the parts of the flow with Lorentz factor (\ref{eq:gamsatei}) propagate inward to the faster
components at a rate
\be
\delta\theta(r) \sim {1\over \Gamma_{\rm sat,ei}}\ln\left({r\over R_{\rm sat,ei}}\right)\quad(r>R_{\rm sat,ei}).
\ee
In this way, the energy source for the high-energy part of the photon
spectrum is {\it dynamically generated} outside the inner thermalization zone described in Paper I.

\subsection{Available Energy}

An important constraint on the reheating mechanism that generates the high-energy tail
of a GRB is that the tail emission typically carries comparable energy to the part of the
spectrum at or below the thermal peak.  For example, our kinetic calculations 
show that the photon index is $\sim -2.3$ when the injected energy is about twice the
seed thermal photon energy.  The kinetic energy of the embedded baryons, as driven by
the outward thermal photon flux inside the reheating zone, is limited by multiple scattering
to 
\be
{dL_k\over d\Omega} = \Gamma {dL_{\rm rest}\over d\Omega}
\lesssim {dL_{\rm th}\over d\Omega}.
\ee

After decoupling of the matter from the photons, the magnetofluid continues to accelerate outward,
so that its Lorentz factor exceeds the saturation Lorentz factor (\ref{eq:gamsat}) of the baryons,
\be
\Gamma_{\rm mag}  \sim {r\over R_{\rm sat,ei}}\Gamma_{\rm sat,ei}.
\ee
In the frame of the magnetofluid, the baryons move with a Lorentz factor
\be
\Gamma_{\rm ei}' = {1\over 2}\left({\Gamma_{\rm sat,ei}\over\Gamma_{\rm mag}} + 
                                   {\Gamma_{\rm mag}\over\Gamma_{\rm sat,ei}}\right)
\ee
and their density is $\rho_{\rm ion}' = (\Gamma_{\rm ei}'/\Gamma_{\rm sat,ei}) \rho_{\rm ion}$.
Hence the energy available to heat the embedded pairs is
\ba
{U_{\rm heat}'\over U'_{\rm th}} &\sim& {\Gamma_{\rm ei}'\rho_{\rm ion}' c^2\over U_{\rm th}'}
\sim {1\over 4}\left({\Gamma_{\rm mag}^2\over \Gamma_{\rm sat,ei}^2}  + 1\right)^2 
{dL_k/d\Omega\over dL_{\rm th}/d\Omega}  \nn
&\lesssim& {1\over 4}\left[\left({r\over R_{\rm sat,ei}}\right)^2 + 1\right]^2.
\ea

This bound is saturated only if outflow carries enough baryonic material.
To check this, we start with the ratio (\ref{eq:fbar}) of rest luminosity 
to Poynting luminosity at breakout, substitute expression (\ref{eq:taucool}) 
for the scattering depth and (\ref{eq:gamsatei}) for the limiting baryon 
Lorentz factor, to get
\ba\label{eq:maxbar}
{dL_k/d\Omega\over dL_{\rm th}/d\Omega} &\leq& \Gamma_{\rm sat,ei}{dL_{\rm rest}/d\Omega \over dL_{\rm th}/d\Omega} \nn
&=& 1.2\;{f_{\rm cover} \over Y_{e\,0.5}^{1/4} }
\,{{\cal R}_{\rm br}/10\over (\Gamma_{\rm br}/3)^{1/2}}
\,{(\ell_{\rm P,br}/10^9)^{1/2}\over (\ell_{\rm th,br}/10^8)^{3/4}}.\nn
\ea
The result depends weakly on the compactness at breakout (as $\sim \ell^{-1/4}$)
as well as on the thermalization efficiency $\ell_{\rm th}/\ell_P$.  A combination of covering factor and shell expansion factor corresponding to $f_{\rm cover}
{\cal R}_{\rm br} \gtrsim 10$ is required for efficient heating.  Note that we have normalized the engine lifetime in the breakout compactness
(\ref{eq:lbreak}) to $t_{\rm eng} \sim 1$ s, when then implies an expansion factor ${\cal R}_{\rm br} \sim 10$ in a burst with
$T_{\rm 90} \sim 10$ s.  

We also note that increasing the isotropic energy tends to a {\it reduce} the outbreak compactness (equation (\ref{eq:elleiso})), which then raises the right-hand-side
of equation (\ref{eq:maxbar}).  An evaluation of $f_{\rm cover}$ requires a numerical simulation along the lines of 
\citep{jiang13}, but including relativistic hydrodynamics and a magnetic field.

\subsection{Optical Depth of the Frozen Pairs}

The optical depth of the residual pairs in the magnetized jet has an important
influence on the radiative signature of delayed heating.  During jet breakout,
annihilation freezes out and the scattering depth (\ref{eq:tauT}) evolves mainly by expansion 
below $\tau_{\rm T}^\pm(R_{\rm br}) \sim 3$,
\be\label{eq:tauscale}
\tau_{\rm T}^\pm(r) \simeq \tau_{\rm T}^\pm(R_{\rm br})\left({r\over R_{\rm br}}\right)^{-1}
                        \left({\Gamma\over \Gamma_{\rm br}}\right)^{-2}.
\ee
At the radius (\ref{eq:rbar}) the depth through the frozen pairs has decreased to
\be\label{eq:taubar}
\tau_{\rm T}^\pm(R_{\rm sat,ei}) \sim 
\tau_{\rm T,br}\left({m_e Y_e\over m_p} \ell_{\rm th,br}\right)^{-(1+2\delta)/(1+3\delta)}.
\ee
This works out to 
\be\label{eq:tauheat}
\tau_{\rm T}^\pm(R_{\rm sat,ei}) \sim 
1.4\times 10^{-3}\,\left({\tau_{\rm T,br}\over 3}\right)\left({\ell_{\rm th,br}\over 10^8}\right)^{-3/4}
\ee
in the case of linear acceleration ($\delta = 1$).

\subsection{Pair Breakdown and Inverse Compton Spectrum}

We describe the heating by a volumetric input
$dU_{\rm heat}/dt$ that extends over a total time $t_{\rm tot} \ll r/\Gamma c$,
\be\label{eq:heatprof}
{dU_{\rm heat}\over dt} = {\Delta U_{\rm heat} \over t_{\rm tot}}  \left({t\over t_{\rm tot}}\right)^{-\alpha_h}.
\ee
In this section all quantities are evaluated in the comoving frame.
The net heat released can be expressed in terms of a total compactness $\ell_{\rm heat}$,
which accumulates from some initial time $t_0 < t_{\rm tot}$ according to
\begin{equation}\label{eq:lprofile}
{\ell_{\rm heat}(t)\over \ell_{\rm heat}} = \left\{
\begin{array}{ll}
%\begin{cases}
\ln(t/t_0)/\ln(t_{\rm tot}/t_0), & \alpha_h = 1 \\
(1-(t/t_0)^{1-\alpha_h})/(1-(t_{\rm tot}/t_0)^{1-\alpha_h}), & \alpha_h \neq 1
\end{array}\right.
%\end{cases}
\end{equation}

In the initial state of the plasma considered here ($\tau_{\rm T} \sim 10^{-3}-10^{-2}$, the advected pairs quickly become relativistic 
after the onset of heating.  They begin to upscatter the advected thermal
photons to energies exceeding $m_e c^2$ in the comoving frame.  

At a high radiative compactness (we consider an initial thermal compactness $\ell_{\rm th} \sim 300-10^3$), particle 
heating and cooling are in near balance and adiabatic losses can be neglected.  Then
\be
{4\over 3}(\gamma_e^2-1)n_e\sigma_T U_\gamma c = {dU_{\rm heat}\over dt},
\ee
where $\gamma_e$ is the Lorentz factor of the pairs in the comoving frame. 
In the kinetic calculation described below,
we find a nearly mono-energetic distribution while the particles are relativistic.  During the first stages of 
heating, $U_\gamma \sim U_{\rm th}$, the seed thermal photon energy density.  So 
\be\label{eq:gamsq}
{4\over 3}(\gamma_e^2-1) =  {1\over \tau_{\rm T}} {\Delta U_{\rm heat}\over U_{\rm th}}\left({t\over t_{\rm tot}}\right)^{-\alpha_h}
                         = {t_{\rm tot}\over \tau_{\rm T} \ell_{\rm th}} {d\ell_{\rm heat}\over dt},
\ee
where $\tau_{\rm T} \equiv n_e\sigma_T ct_{\rm tot}$ in this section.

The inverse-Compton image of the thermal peak during this first stage of
heating sits at a comoving energy $\hbar\omega_{\rm IC,pk} \sim {4\over 3}(\gamma_e^2~-~1)\hbar\omega_{\rm pk}$.  
The initial seed peak is produced during a very compact thermalization phase before breakout (Paper I),
with an energy $\hbar\omega_{\rm pk} \sim 0.1 m_ec^2$.  Following this episode, the radiation is trapped
by a forward baryon shell for an expansion factor as long as $\sim {\cal R}_{\rm br}$ while the Lorentz factor
of the shell remains approximately constant.  The corresponding adiabatic diluation of the peak energy in the
comoving frame is in the range
\be
1 \gtrsim f_{\rm ad} \gtrsim 0.2\left({{\cal R}_{\rm br}\over 10}\right)^{-2/3}.
\ee
After breakout, we focus on the simplest case of linear growth of $\Gamma$, corresponding to 
$\omega_{\rm pk} \propto r^{-1}$.  

Combining these effects, the comoving peak energy drops to
\ba
\hbar\omega_{\rm pk} &\sim& 
0.1~m_ec^2\,f_{\rm ad}\left({r\over R_{\rm br}}\right)^{-1} \nn
&\sim& 0.008~m_ec^2\,f_{\rm ad}\left({\ell_{\rm th,br}\over 10^8}\right)^{-1/4}
\left({r\over R_{\rm sat,ei}}\right)^{-1}
\ea
beyond the radius (\ref{eq:rbar}) where baryons and magnetic field begin to moving differentially.  Heating at a particular place in the magnetofluid will
typically be delayed beyond this transition, depending on the proximity of
baryonic material.
Substituting equations (\ref{eq:tauscale}) and (\ref{eq:taubar}) into (\ref{eq:gamsq}) gives
the scaling $\hbar\omega_{\rm IC,pk} \propto (r/R_{\rm sat,ei})^2$, and
\be\label{eq:omic}
{\hbar\omega_{\rm IC,pk}\over m_ec^2} \sim 3\,f_{\rm ad}\,{\Delta U_{\rm heat}/U_{\rm th}\over (t/t_{\rm tot})^{\alpha_h}}
\left({\ell_{\rm th,br}\over 10^8}\right)^{1/2}\left({r\over R_{\rm sat,ei}}\right)^2.
\ee
This works out to $\hbar\omega_{\rm IC,pk} \sim 10\,f_{\rm ad}\,m_ec^2$, since we
are considering moderately relativistic differential motion between the baryon-loaded
and baryon-free parts of the outflow ($\Delta U_{\rm heat} \gtrsim U_{\rm th}$ and 
$r \sim 2R_{\rm sat,ei}$).

As photons accumulate above the pair-creation threshold, $\tau_{\rm T}$ begins to rise, reducing
the equilibrium particle energy and eventually shutting off the source of pair-creating photons.  
Pair creation continues due to the accumulation of hard photons, and $\omega_{\rm IC,pk}$ drops
toward the seed peak.  A compactness $\ell_{\rm tot} = \ell_{\rm th} + \ell_{\rm heat} \sim 10^3$ is large enough
to ensure that $\tau_{\rm T} > 1$ at the end of heating.  Then the plasma enters a sub-relativistic state and, 
as we demonstrate in Section (\ref{s:results}), the Compton upscattered peak merges smoothly with the
seed thermal peak.  

A first estimate of the high-energy spectral index is then obtained by relating the drop
in $\omega_{\rm IC,pk}$ to the rise in optical depth.  The energy spectrum can be
written
\be
\hbar \omega_{\rm IC}^2 {dn_\gamma\over d\omega_{\rm IC}}
\sim \hbar \omega_{\rm IC}^2 {n_{\gamma,\rm pk}\,\sigma_T n_e c \over d\omega_{\rm IC}/dt},
\ee
where
\be
{1\over\omega_{\rm IC}}{d\omega_{\rm IC}\over dt} = {1\over \gamma_e^2}{d\gamma_e^2\over dt}
         = -\left({\alpha_h\over t} + {1\over n_e}{dn_e\over dt}\right).
\ee
Substituting equation (\ref{eq:gamsq}) gives
\be
\hbar \omega_{\rm IC}^2 {dn_\gamma\over d\omega_{\rm IC}}
\sim {\Delta U_{\rm heat}\over \alpha_h + d\ln n_e/d\ln t}\,\left({t\over t_{\rm tot}}\right)^{1-\alpha_h}.
\ee
For example, if $dn_e/dt \sim {\rm const}$, then $\omega_{\rm IC,pk} \propto t^{-1}$
and $\hbar \omega_{\rm IC}^2 dn_\gamma/d\omega_{\rm IC} \propto \omega_{\rm IC}^{-\alpha_h+1}$.

So far we have neglected the effects of expansion.  These are included in Section \ref{s:expand}
after we first examine the non-expanding case in some detail.  It should be noted that,
in the kinetic calculations described here, most of the non-thermal tail emerges on 
a modest fraction of $t_{\rm tot}$, which itself may be a fraction of the flow time $r/\Gamma c$.

\section{Relative Importance of Inverse Compton and Synchrotron Cooling}\label{s:synch}

Our focus here is on the simplest mechanism of distributed heating, by decaying turbulence.   We are
invoking essentially the same heating mechanism during the two main stages of spectral evolution.
The qualitative difference in output spectrum (thermal vs. non-thermal) is mainly a consequence of
the large drop in scattering depth after jet outbreak, as we explained in Section \ref{s:reheat}.

During the first stage of heating, leading to the formation of the spectral peak and the low-energy
slope, there is an indirect argument (Paper I) in favor of this mechanism over more sporadic and localized bursts of 
heating, such as might be
mediated by magnetic reconnection.  When the optical depth is high, the pairs remain subrelativistic
if the heating is smooth, and the low-energy spectrum that emerges is much flatter than Planckian and
comparable to that observed in a GRB.   Localized and intense heating generates a higher pair density,
which in turn pushes the Compton parameter much higher than in a thermal plasma, and allows the low-energy
spectrum to push closer to Rayleigh-Jeans.

Decaying Alfv\'enic turbulence only heats the embedded $e^\pm$ along the magnetic field.
At a high wavenumber, sheared Alfv\'en waves have the dispersion relation \footnote{This dispersion relation is modified when
the particles have a relativistic dispersion along the magnetic field.   Electron-supported modes such as whistlers
and kinetic Alfv\'en waves are not present in the pair plasma because of its charge symmetry.  Even if ions supply most of the positive charge,
the dispersion relation remains equivalent to (\ref{eq:alfdisp}) at $k_\perp \sim \omega_{Pe}/c \gg k_\parallel$, as long as the magnetic energy
dominates the rest energy of the electrons, $B^2/8\pi \gg n_e m_ec^2$.}
\be\label{eq:alfdisp}
\omega^2(k_\parallel) = {c^2\over \sqrt{1 + c^2k_\perp^2/\omega_{Pe}^2}}.
\ee
Here $k_\perp, k_\parallel$ are the components of the wavevector perpendicular and parallel to the magnetic field,
and $\omega_{Pe}$ is the electron plasma frequency.  The waves therefore Landau damp on the longitudinal motion of 
the $e^\pm$ where $k_\perp \sim \omega_{Pe}c$, 
even while the wave frequency remains orders of magnitude below the electron cyclotron frequency $\omega_{ce} = eB/m_ec$.  

The waves also become charge-starved during the first stages of reheating, in the sense that the fluctuating current
density $\delta J \sim (c/4\pi) k_\perp \delta B$ exceeds the maximum conduction current $n_e e c$ at lower wavenumbers
than those associated with Landau damping:  see \cite{thompson06} and Section \ref{s:outst}.

Energy can also be transferred to small scales through an independent cascade that is mediated by the fast mode.  In a strongly
magnetized plasma, the fast mode closely approximates a vacuum electromagnetic wave, with isotropic dispersion relation
$\omega \simeq ck$.   The mode then carries a weak electric current compared with a sheared Alfv\'en mode of similar amplitude
and wavenumber $k_\perp \sim k$.  The fast waves damp rapidly at a frequency $\omega \sim kc \sim \omega_{Pe}$,
via conversion to Langmuir waves, e.g. $f + f \rightarrow \ell + \ell$.   In a relativistic plasma with magnetization
$\sigma \gg 1$, the limiting fast wave frequency is then $\omega \simeq kc \sim \sigma^{-1/2}\omega_{ce} \ll \omega_{\rm ce}$.
Once again, parallel heating of the embedded pairs is the main effect.  

There is a close correspondence between fast turbulence 
and acoustic turbulence, for which the spectral energy density scales as $k U_k \sim k^{-1/2}$.  
Coincidentally this is the same scaling as has been obtained in most recent simulations of Alfv\'enic turbulence
\citep{maron01,boldyrev06}.  Hence fast waves and Alfv\'en could have comparable amplitudes at fixed $|k|$. 

\subsection{Particle Heating in a Pair Plasma with \\  Very Low $\beta = 8\pi P/B^2$}\label{s:heat}

The jet magnetization can be defined in two different ways, depending on whether the photon pressure is
included in the material pressure.  As regards the bulk dynamics of the jet, the photons are tied
to the magnetic field inside breakout and contribute effectively to the plasma inertia; but outside
breakout the two components have to be considered separately (Section \ref{s:accel}).

The magnetization also influences plasma instabilities, such as firehose and cyclotron
modes, that modify the particle distribution and operate on very short timescales.
Here the photons can be ignored.  The comoving plasma parameter is very small,
\be
\beta \equiv {8\pi n_e\kB T_e\over B^2} \sim {\tau_{\rm T}\over \ell_{\rm P}}\left({T_e\over m_ec^2}\right).
\ee
Using the scalings (\ref{eq:lgam}) for the compactness and (\ref{eq:tauscale}) for the scattering depth,
one has 
\be
\beta(r) \sim 1\times 10^{-10}\,{\Gamma(r)\over \Gamma_{\rm br}}
\left({\tau_{\rm T,br}\over 3}\right)\left({\ell_{\rm P,br}\over 10^9}\right)^{-1}
\left({3T_e\over 0.1m_ec^2}\right)
\ee
before the onset of reheating.

The pair pressure is somewhat higher during reheating,
\be
\beta \sim {8\pi n_e \gamma_e m_ec^2\over B^2} \sim {1\over \gamma_e\ell_{\rm P}}\left({\Delta U_{\rm heat}\over U_{\rm th}}\right)^{1/2}.
\ee
Here $\ell_{\rm th} \lesssim m_p/(Y_em_e) \sim 4\times 10^3Y_{e\,0.5}$, with $\ell_{\rm P} > \ell_{\rm th}$ to power the
high-energy tail of a GRB, so $\beta \sim 10^{-3}/\gamma_e$.

We conclude that $\beta$ is small enough during reheating that particle isotropization by cyclotron
and firehose instabilities appears to be ineffective.  

\subsection{Pitch Angle Excitation}

Even when the magnetic energy density dominates that stored in thermal radiation, $e^\pm$ which
experience strong parallel heating will cool mainly by Compton scattering the radiation field.
Once again we work in the frame in which the bulk plasma is at rest.

Synchrotron emission depends on excitation of the gyromotion.  Coulomb scattering between
relativistic electrons and positrons is negligible at low $\tau_{\rm T}$.  Supposing that relativistic
motion along the magnetic field is sustained by an electrostatic force, we can ask whether the equilibrium perpendicular
temperature of the $e^\pm$ differs from the ambient color temperature $T_c$ of low-energy photons.  

In the presence
of an isotropic, Rayleigh-Jeans spectrum of soft photons, a particle moving with speed $\beta_{e\,\parallel} c$ along ${\bf B}$
sees blackbody radiation of a temperature $T_c' \sim 2\gamma_{e\,\parallel} T_c$ in a cone of solid angle $\Delta\Omega'
\sim \pi/\gamma_{e\,\parallel}^2$ in the rest frame of its guiding center.  Here $\gamma_{e\,\parallel} =
(1-\beta_{e\,\parallel}^2)^{-1/2}$. Since the particle re-radiates isotropically, it's equilibrium temperature is 
\be
T_\perp \sim {T_c\over 2\gamma_{e\,\parallel}}.
\ee

More effective gyro-heating is provided by multiple Compton scatterings of thermal photons, which operates 
in a diffusive manner.  The gyrational momentum accumulated over a comoving time $\delta t$ is  \citep{thompson06}
\be
p_\perp^2 \sim {4\over 5}\gamma_e^2\biggl\langle\left({\hbar\omega\over c}\right)^2\biggr\rangle\cdot n_\gamma \sigma_T c \delta t,
\ee
and the corresponding pitch angle is
\be
{p_\perp\over \gamma_e m_ec} 
\;\sim\; \ell_\gamma^{1/2}\left({E_{\rm pk}\over m_ec^2}\right)^{1/2}\,\left({c\delta t\over r/\Gamma}\right)^{1/2},
\ee
where $E_{\rm pk} \simeq 3T_\gamma$ is the spectral peak energy and $\ell_\gamma$ the radiation compactness.

The gyrational state of the particles can, in some circumstances, maintain an equilibrium 
between synchrotron damping and excitation by Compton scattering.  Self-absorption effects are important when the gyrational motion 
is mildly relativistic, $p_\perp \gtrsim m_ec$, and so we allow for a suppression of synchrotron emission by a factor
$f_{\rm sa} < 1$.  Synchrotron cooling damps only the perpendicular momentum of a particle,
while leaving its gyrational rest frame unaltered.   In this regime, $\gamma_e = \gamma_{e\,\parallel} (p_\perp/m_ec)$ and the synchrotron power is
\be\label{eq:synchpower}
\gamma_{e\,\parallel} c {dp_\perp\over dt} = -2f_{\rm sa}\left({p_\perp\over m_ec}\right)^2 \sigma_T {B^2\over 8\pi} c
\quad\quad (p_\perp \gg m_ec).
\ee
A self-consistent equilibrium 
\be
{p_\perp\over m_ec} = \gamma_e^{3/4}\left({1\over 3f_{\rm sa}} {E_{\rm pk}\over m_ec^2}{U_\gamma\over B^2/8\pi}\right)^{1/4}
\ee
is obtained as long as $p_\perp/m_ec \ll \gamma_e$.
The synchrotron power (\ref{eq:synchpower}) can be compared with the Compton power $(dE/dt)_{\rm IC} = (4\gamma_e^2/3) \sigma_T U_\gamma c$,
\be
{(dE/dt)_{\rm synch}\over (dE/dt)_{\rm IC}} = {2f_{\rm sa}^{1/2}\over (3\gamma_e)^{1/2}}
\left({E_{\rm pk}\over m_ec^2}\right)^{1/2}\left({U_\gamma\over B^2/8\pi}\right)^{-1/2}.
\ee
In the reheating zone, the comoving plasma parameters are $U_\gamma \lesssim B^2/8\pi$ and $E_{\rm pk} \sim 10^{-3} m_ec^2$,
and so synchrotron losses are subdominant but not entirely negligible.

Low-frequency photons that are absorbed by exciting the gyrational motion of relativistic pairs contribute negligbly
to the parallel drag force (Appendix \ref{s:drag}).

%%%%%%%%%%%%%%%%%%%%% NUMERICAL METHOD %%%%%%%%%%%%%%%%%%%%%%%%%%%%%
\section{Numerical Method}\label{s:method}
We now describe our approach to calculating the evolution of $e^\pm$
immersed in a thermal photon gas and subjected to continuous
heating.  Here and in Appendix \ref{s:kinetic} we use a simplified notation,
with $n_\pm(p)$ representing the number density of $e^\pm$ pairs per 
dimensionless momentum $p$ (in units of $m_ec$), and $n_\gamma(x)$ 
the number density of photons per dimensionless energy 
$x = \hbar\omega/m_e c^2$. 

The state of the photo-pair plasma is 
governed by the following two integro-differential equations 
\citep[e.g.][]{PeerWaxman2005,BelmontMalzacMarcowith2008,
VurmPoutanen2009}
\ba\label{eq:particle}
\partial_tn_\pm(p) &=& \dot{n}_{\pm,\text{cs}}(p)+\dot{n}_{\pm,\text{pp}}(p)+
\dot{n}_{\pm,\text{pa}}(p) \nn 
&&+\dot{n}_{\pm,\text{coul}}(p)+\dot{n}_{\pm,\text{heat}}(p)+
\dot{n}_{\pm,\rm exp}(p) 
\ea
\be\label{eq:photon}
\partial_tn_\gamma(x) = \dot{n}_{\gamma,\text{cs}}(x)+
\dot{n}_{\gamma,\text{pp}}(x)+\dot{n}_{\gamma,\text{pa}}(x)
+\dot{n}_{\gamma,\rm exp}(x)
\ee
Here allowance is made for expansion, in the final term on the right-hand
side, but there is no escape of photons and pairs.   We describe the solutions
in static and expanding boxes in Sections \ref{s:results} and 
\ref{s:expand_results}, respectively.
The two distributions interact with each other and themselves via the 
processes of Compton scattering (cs, $\gamma + e^\pm \rightarrow \gamma + e^\pm$), 
pair production (pp, $\gamma + \gamma \rightarrow e^+ + e^-$), pair annihilation 
(pa, $e^+ + e^- \rightarrow \gamma + \gamma$), and Coulomb scattering (coul,
$e^\pm + e^\pm \rightarrow e^\pm + e^\pm$).  

The initial
state is assumed to contain cold thermal pairs with a compactness $\ell_{e,0}$ 
along with soft thermal photons with a compactness $\ell_{\rm th}$.
Heating of the pair gas is represented by the term $\dot{n}_{\pm,\rm heat}$ in equation (\ref{eq:particle}),
with a cumulative compactness $\ell_{\text{heat}}$ injected by the end of the
simulation.  

The formation of a flat
low-frequency spectrum by thermal Comptonization of cyclotron photons inside jet breakout
has already been examined in Paper I.  In the present calculations we take this `thermal GRB'
spectrum 
\begin{equation}\label{eq:grbthermal}
\frac{dn_\gamma}{d\ln x} \propto
\left\{
\begin{array}{ll}
K x_{\text{pk}}^3\exp(-x_{\rm pk}/\theta_\gamma), & x<x_{\text{pk}}\\
K x^3\exp(-x/\theta_\gamma), & x\geq x_{\text{pk}}
\end{array}
\right.
\label{eq:softdist}
\end{equation}
as the input.  The temperature $\theta_\gamma = \kB T_\gamma/m_ec^2$ is a fit parameter, and during
the initial heating phase differs slightly from the pair temperature.
Equation (\ref{eq:grbthermal}) smoothly matches a flat low-energy spectrum,
$dn_\gamma/d\ln x \propto x^0$ onto a Wien spectrum at energies above
$x_{\rm pk} = 3\theta_\gamma$.

The system of interest has an elongated $e^\pm$ momentum distribution in the direction of the magnetic field
(Section \ref{s:synch}, \citealt{thompson06}), but also a low to moderate scattering 
depth and significant inhomogeneity.  The inhomogeneity is required to induce 
heating: our particular model involves large-amplitude distortions of the magnetic 
field by embedded baryon clouds (Paper III).  The combination of 
inhomogeneity with a long mean free path for photons leads to some isotropization of the interactions.

The elongation of the $e^\pm$ distribution implies a suppression of cyclo-synchrotron emission
(Section \ref{s:synch}), which is difficult to handle quantitatively.  Here we simply shut off synchrotron
processes, and focus on Compton scattering of the seed thermal photons.  
Because the photon occupation number is everywhere small
in the energy range calculated, we also neglect stimulated effects in Compton scattering.

By neglecting the anisotropy of the $e^\pm$, we also somewhat underestimate the rate of
pair creation:  photons that are Compton scattered by relativistic particles moving
parallel to ${\bf B}$ are themselves beamed along the magnetic field, so that counterstreaming
gamma rays have an enhanced center-of-momentum energy.

\subsection{Details of Time Evolution}

The evolution of the photon-pair plasma is divided into regimes
of small and large energy exchange, as defined (e.g.) by the
fractional energy shift of a photon after scattering.
The Fokker-Planck (F-P) equations are used in the regime of small energy 
exchange, which for isotropic and homogeneous distributions can 
be written as \citep[e.g.][]{NayakshinMelia1998}
\be\label{eq:fpeq1}
\partial_t n_\pm^{\text{F-P}}(p) =
\partial_p\left[\frac{\gamma_e}{p}A_\pm n_\pm\right]+
\frac{1}{2}\partial_p\left[\frac{\gamma_e}{p}\partial_p
\left\{\frac{\gamma_e}{p}D_\pm n_\pm\right\}\right],
\ee
\be\label{eq:fpeq2}
\partial_tn_\gamma^{\text{F-P}}(x) = \partial_x(A_\gamma n_\gamma)+
\frac{1}{2}\partial^2_x(D_\gamma n_\gamma).
\ee
The coefficients $A_\pm(p,t)$, $A_\gamma(x,t)$ represent the average 
rate of change of particle or photon energy (due e.g. to secular cooling
processes or expansion), and $D_\pm(p,t)$, $D_\gamma(x,t)$ are the corresponding diffusivities.
Each of these terms receives contributions from Compton and Coulomb scattering,
and the advection terms from expansion:
\be
\{A,D\}_\pm = \{A,D\}_{\pm,\text{cs}} + \{A,D\}_{\pm,\text{coul}} + 
A_{\pm,\text{heat}} + A_{\pm,\rm exp}
\ee
\be
\{A,D\}_\gamma = \{A,D\}_{\gamma,\text{cs}} + A_{\gamma,\rm exp}.
\ee
The F-P approach to Compton scattering is essentially equivalent to 
the standard Kompane'ets formalism (but with the stimulated term here neglected)
and is essential for treating Coulomb collisions between thermal pairs. 
% is valid for describing Compton scattering of photons 
%and pairs in a phase space ($x_0,p_0$) where the small energy 
%condition is satisfied. It is particularly useful for treating Coulomb 
%collisions among pairs and plasma heating. 
%Thus, the two coefficients 
%represent a sum over the different radiative process

The remaining interactions, including Compton scattering of soft 
photons by energetic particles (where photons receive a large energy 
boost, $\omega \rightarrow 4\gamma_e^2\omega_0$) are described by exact collision integrals 
over the two distributions.  These integrals, and the division in energy and
momentum space between the two time-evolution methods, is described in Appendix 
\ref{s:kinetic}.  The full distribution functions are evolved by integro-differential 
equations combining both approaches:
\begin{equation}
\partial_t\{n_\pm, n_\gamma\} = 
\partial_t\{n_\pm, n_\gamma\}^{\text{F-P}} + 
\partial_t\{n_\pm, n_\gamma\}^{\text{col}}.
\end{equation}

We choose to solve the coupled equations 
(\ref{eq:fpeq1}), (\ref{eq:fpeq2})  using the \citet{ChangCooper1970} 
fully-implicit finite difference scheme, due to its robustness and 
guarantee of yielding positive spectra. This scheme is only accurate 
to first-order both in space and time, but much more stable than any 
of the higher order schemes \citep[e.g.][]{ParkPetrosian1996}. We use 
logarithmic grids for the particle distribution spanning five orders 
of magnitude ($p = 10^{-3} - 10^2$) in momentum and the photon 
distribution spanning eight orders ($x = 10^{-6} - 10^2$) in energy, 
with a grid size of 256 points for both distributions. 

The 
Chang-Cooper scheme conserves particle number exactly by imposing the 
condition of vanishing flux at the grid boundaries. On the other hand, 
greater care must be taken in the numerical accuracy of the collision integrals,
so as to avoid significant non-conservation of particle number.
This is especially true for Compton scattering, where a double integral 
is carried out. To this end, we use adaptive quadrature routines to calculate the integrals 
exactly, rather than using methods such as the composite Simpson's rule, 
which is sufficient for all other interactions included in the 
simulation. This allows us to conserve particle number and energy to 
better than 1\% in all the simulations presented in this study. 

One advantage of using implicit schemes is that they are free from the 
Courant condition, which depends on the size of the smallest bin for 
logarithmic grids, on how large the time step can be. However, to 
ensure convergence to the right solution, the time step should be of 
the order of the fastest cooling time to accurately track diffusion of 
particles in energy space. Among all of the radiative processes 
considered here, particles cool predominantly by Compton scattering 
and the timescale of which scales with compactness of the photon field 
and particle energy:
\begin{equation}
t_{\rm cs}\sim\frac{t}{\gamma_e \ell_\gamma}.
\end{equation}
This can be problematic for simulating highly compact plasmas, but
for the values of $\ell_\gamma$ ($\sim 10^2 - 10^4$) used in this study, 
each run takes between a few minutes and few days on multiple 
processors. 
%The most time consuming operation turned out to be the 
%precise calculation of the double integrals for Compton scattering.

%%%%%%%%%%%%%%%%% CONTINUOUS HEATING IN A STATIC MEDIUM %%%%%%%%%%%%%%%
\section{Distributed Heating in a Static Medium}\label{s:results}

We can use the static approximation when the heating episode is brief compared
with the flow time. The effect of scattering  after the heating turns off can
be evaluated with the Monte Carlo approach of Paper III, assuming a cold, 
frozen pair flow beyond the heating layer.

Our simulations start with the thermal GRB photon spectrum (\ref{eq:grbthermal})
and an initial photon compactness $\ell_{\rm th}$ that is determined by fixing
i) the total compactness 
\be
\ell_{\text{tot}} = \ell_{e,0} + \ell_{\rm th} + \ell_{\text{heat}}
\ee
that has accumulated at the end of heating; and ii)
the initial scattering depth $\tau_{T,0} = \sigma_T n_{e,0} ct_{\rm tot}$.  

We consider both uniform heating, $\alpha_h = 0$
in equation (\ref{eq:heatprof}), as well as heating distributed logarithmically 
over time ($\alpha_h = 1$).  Heating therefore starts at a finite
time $t_0 = 0.1t_{\text{tot}}$ and stops at time $t_{\rm tot}$.

The initial temperature of the pairs is set equal to $\theta_\gamma$ in equation 
(\ref{eq:grbthermal}), corresponding to a mean initial energy $\langle\gamma_{e,0}\rangle \simeq 
1 + (3/2)\theta_\gamma$.  The particle compactness is 
\begin{equation}\label{eq:le0}
\ell_{e,0} = \sigma_Tct_{\text{tot}}U_{e,0} = \langle\gamma_{e,0}\rangle
\tau_{\rm T,0}.
\end{equation}

%fixed budget of the total compactness $l_{\text{tot}} = l_{e,0} + 
%l_{\text{th}} + l_{\text{heat}}$, that is to be shared among the 
%injected pair and photon distributions at $t=t_0$ and the heat 
%injected continuously until $t=t_{\text{tot}}$. 
%The initial pair 
%compactness is set by the optical depth of the pair gas $\tau_{\rm T,0} = 
%n_{e,0}\sigma_Tct_{\text{tot}}$, where $n_e \equiv n_+ + n_-$, and 
%the average energy of the pairs $\langle\gamma_0\rangle$. 

%For a non-relativistic Maxwellian distribution 
%\begin{equation}
%n_\pm(p) = n_\pm\sqrt{\frac{2}{\pi}}\theta_\pm^{3/2}p^2\exp
%\left(-\frac{1}{2}\frac{p^2}{\theta_\pm}\right)
%\end{equation}
%$\langle\gamma\rangle = 1+\frac{3}{2}\theta_\pm$, where $\theta_\pm = k_BT_\pm/m_ec^2$ is the 
%dimensionless temperature of the pair gas, and $k_B$ is the Boltzmann 
%constant. The dimensionless energy density of the pair gas (in units 
%of $m_ec^2$) $U_{e,0} = \langle\gamma_0\rangle n_{e,0}$ yields the 
%compactness

%A quasi-thermal distribution with a flat spectrum below the peak was 
%shown to arise naturally in the evolution of a strongly magnetized, 
%highly compact, expanding, dilute pair fireball that is being 
%continuously heated \citep{tg13}. 

It is convenient to write the initial photon field compactness in terms of the total 
heating compactness, 
\be\label{eq:fth}
\ell_{\rm th} = f_{\rm th} \ell_{\text{heat}},
\ee
so that
\begin{equation}
\sigma_Tct_{\text{tot}}\int \frac{dn_\gamma}{d\ln x} dx = 
\frac{f_{\text{th}}(\ell_{\text{tot}}-\ell_{e,0})}{1+f_{\text{th}}}.
\end{equation}

%%%%%%%%%%%%%%%%%% HEATING OF THE PAIR GAS %%%%%%%%%%%%%%%%%%%%%%%%%%%
\subsection{Heating of the pair gas}
We consider a simple heating model where energy is injected into the 
pair plasma with the time distribution (\ref{eq:lprofile}).
%\begin{equation}
%\frac{dl_{\text{heat}}(t)}{dt} \propto t^{-\alpha_h}
%\label{eq:dlhdt}
%\end{equation}
%Integration of this expression over time yields the heating profile
%\begin{equation}
%l_{\text{heat}}(t) = 
%\left\{
%\begin{array}{ll}
%l_{\text{heat}}\left(\ln \frac{t_{\text{tot}}}{t_0}\right)^{-1}\ln 
%\frac{t}{t_0}, & \alpha_h = 1 \\
%\frac{1-(t/t_0)^{1-\alpha_h}}{1-(t_{\text{tot}}/t_0)^{1-\alpha_h}}
%l_{\text{heat}}, & \alpha_h \neq 1
%\end{array}
%\right.
%\end{equation}
In this study, we only consider two cases: $\alpha_h = 0$ (constant rate of heating)
and $\alpha_h = 1$ (cumulative heat input grows logarithmically with time).

The continuous fashion in which energy is delivered to the pairs is 
described by the advective term in the F-P equation. The rate of change 
of the average energy of the pairs due to heating is
\begin{equation}
\frac{1}{2}\frac{dU_{\text{heat}}}{dt} = \int\gamma_e~\partial_p
\left[\frac{\gamma_e}{p}A_{\pm,\text{heat}}n_\pm(p)\right]dp
\end{equation}
The factor of $1/2$ takes into account that heat is deposited equally
into electrons and positrons.   From here it is easy to show that
\begin{equation}
A_{\pm,\text{heat}} = \frac{1}{2}\left[\left\{\frac{\gamma_e^2}{p}
n_\pm(p)\right\}_{p_{\text{min}}}^{p_{\text{max}}}-n_\pm\right]^{-1}
\frac{dU_{\text{heat}}}{dt}
\end{equation}

%%%%%%%%%%%%%%%%% TEMPORAL EVOLUTION OF PAIRS %%%%%%%%%%%%%%%%%%%%%%%%
\subsection{Time evolution of the pair distribution}
As heat is injected into the pair gas, the average energy of the pairs 
begins to rise.   The pairs are also radiatively 
cooled by Compton scattering soft photons, and very quickly the
steady state energy (\ref{eq:gamsq}) is reached.   The maximum
energy achieved by the pairs is found by setting $\tau_{\rm T} = 
\tau_{\rm T,0}$.
%\begin{equation}
%\langle\gamma\rangle_{\text{max}}^2 \approx 
%1+\frac{3t_{\text{tot}}}{4\tau_{T,0}\ell_{\text{th}}}
%\frac{d\ell_{\text{heat}}(t_0)}{dt}
%\label{eq:gmax}
%\end{equation}
The equilibrium
energy changes adiabatically as soft photons are scattered over the
pair creation threshold, as determined by
\begin{equation}
x_1x_2 \geq \frac{2}{1-\hat k_1\cdot\hat k_2}.
\end{equation}
Here $\hat k_i$ is the unit vector along the wavevector of photon $i$.
The addition of fresh pairs lowers the heating rate per particle.

%The rise in the average 
%energy of the pairs continues until the pairs are able to scatter 
%softer photons over the pair creation threshold. Subsequently, the 
%addition of fresh pairs increases the number density while lowering 
%the average energy per particle. This feedback of pair creation 
%briefly establishes a steady state where $d\langle\gamma\rangle/dt$ 
%vanishes. The steady state equation can be written as the following
%From equation (\ref{eq:dlhdt}), the heating time is very short compared 
%to $t_{\text{tot}}$, thus the stationary state is achieved almost at 
%$t\approx t_0$. Then, under the assumption that $\tau\approx\tau_0$ 
%and $l_\gamma\approx l_{\text{th}}$, it can be shown that 
%%%%%%%%%%%%%% Figure %%%%%%%%%%%%%%%%%%%%%%%%%
\begin{figure}
\includegraphics[width=0.48\textwidth]{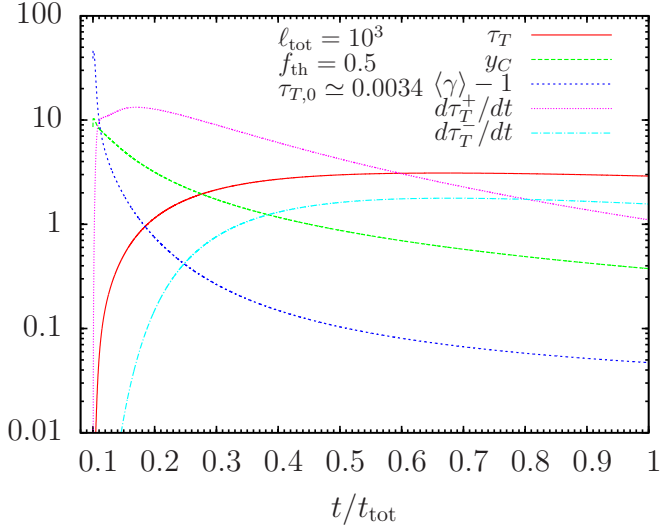}
\caption{State of the pair plasma as a function of time.  Total
compactness $\ell_{\text{tot}}=10^3$, $f_{\text{th}}=0.5$, and heating
profile $\alpha_h=1$.
Initial scattering depth $\tau_{T,0} = 0.0034$ and seed thermal
peak energy $x_{\rm pk} = 6\times 10^{-4}$ correspond to maximum IC photon
energy of peak photons $x_{\rm IC,max} = 4\gamma_e^2 x_{\rm pk} = 5$. 
Plotted quantities:  scattering depth (red solid); $y_{\rm C} = (4/3)(\gamma_e^2-1) \tau_{\rm T}$ 
(green dashed); average kinetic energy of pairs (blue wide dot);
rates of pair production (purple dot) and annihilation (cyan dot-dashed).}
\label{fig:timeprofile}
\vskip .2in
\end{figure}
%%%%%%%%%%%%%%%%%%%%%%%%%%%%%%%%%%%%%%%%%%%%%%%
The evolving state of the pair plasma is shown in
Figure \ref{fig:timeprofile}, for total compactness $\ell_{\text{tot}}=10^3$,
$f_{\rm th}=0.5$, and seed temperature $\theta_\gamma = 2\times 10^{-4}$.  
Heating declines as $t^{-1}$ from an initial time $t_0 = 0.1\,t_{\rm tot}$; the
initial optical depth $\tau_{\rm T,0} = 0.0034$ corresponds to a mean
inverse-Compton energy $\langle x_{\rm IC}\rangle = 4\gamma_e^2\theta_\gamma = 1.6$ 
for photons drawn from the thermal peak, $x_{\rm pk} = 3\theta_\gamma$,
and a maximum $x_{\rm IC,max} = 5$.   We recall that for photons of initial energy $x_0$,
\be\label{eq:xicmax}
x_{\rm IC,max} = 4\gamma_e^2 x_0\quad\quad(\gamma_e \gg 1).
\ee

The optical depth of the pairs continues to rise until the pair annihilation rate,
\begin{equation}
\frac{d\tau_{\rm T}^-}{dt} = \frac{3}{16}\sigma_Tct_{\text{tot}}\tau_{\rm T}^2,
\end{equation}
catches up with the pair production rate.  Eventually pair production 
declines due to a depletion in hard photons, which are no longer generated.

%This pair cascade 
%injects secondary pairs that now share the total kinetic energy with 
%the primary population, which consequently lowers the average kinetic 
%energy per particle. The pairs rapidly become non-relativistic after 
%the initial burst in energy. 

\begin{figure}
\includegraphics[width=0.48\textwidth]{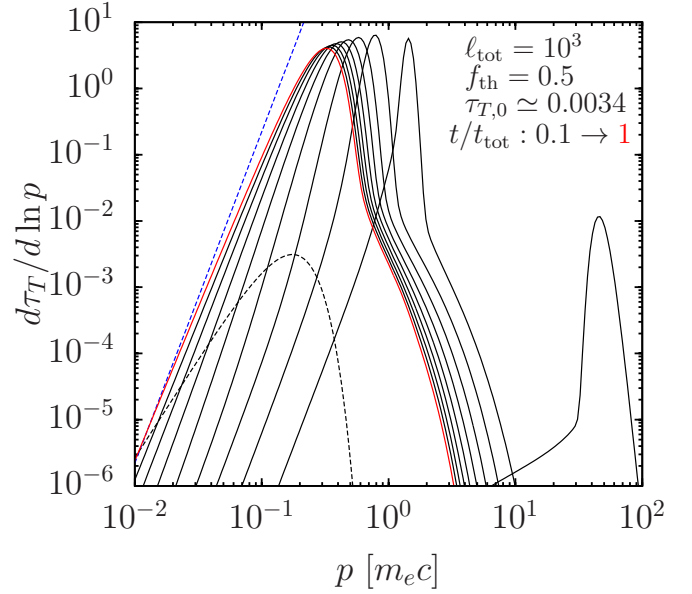}
\caption{Time evolution of the pair distribution in the simulation
of Figure \ref{fig:timeprofile}.
Black dashed line: initial distribution, non-relativistic Maxwellian with
$\theta_\pm = 0.01$.   Solid black curves:  evolving particle distribution
with peak trending from right to left as the pair density grows.  Curves separated
by $\Delta t/t_{\rm tot} = 0.1$, starting from $t_0 = 0.1t_{\rm tot}$. 
Blue dashed line: $n_\pm(p)\propto p^4$ (see text for details).}
\vskip .2in
\label{fig:partdist}
\end{figure}

Over the course of the simulation, the pair distribution, 
shown in Figure \ref{fig:partdist}, remains peaked around 
some $\langle\gamma_e\rangle$ that is set by the balance 
between heating and Compton cooling.  The Compton cooling rate of high-energy 
particles is $\propto p^2$, so heating dominates at the low-energy end.   
When Coulomb collisions are the dominant source of energy 
exchange between $e^\pm$, their distribution tends to a Maxwellian. 
Here strong heating and Compton cooling drive the distribution
away from complete thermalization, as is evident in 
Figure \ref{fig:partdist}.  

Consider in particular the low-momentum slope of the distribution, which is harder than a Maxwellian.
Although Coulomb collisions are most effective 
at low $p$, in this case the timescale to establish a 
thermal distribution is everywhere much longer than that of heating.  
%The Coulomb energy loss rate to colder pairs is approximately \citep[e.g][]{NayakshinMelia1998}
%\begin{equation}
%\dot{\gamma}_{\text{coul}} \approx -\frac{3}{2}\frac{\tau_{\rm T}}
%{t_{\text{tot}}}\frac{\ln\Lambda}{\beta\langle\gamma\rangle}
%\end{equation}
%where $\ln\Lambda\sim20$ is the Coulomb logarithm.  Then 
%\begin{equation}
%\frac{t_{\text{coul}}}{t_{\text{heat}}} = 
%\frac{2}{3}\frac{t_{\text{tot}}}{\ln\Lambda}\frac{d\ell_{\text{heat}}}{dt}
%\frac{\beta\langle\gamma\rangle}{\tau_{\rm T}^2} \sim \frac{1}{3}\ell_{\text{heat}}
%\gg 1.
%\label{eq:heat_coul_ratio}
%\end{equation}
%Here we take a constant heating rate, $\langle p\rangle\sim1$, and 
%a scattering depth of order unity. 
%the $e^\pm$ cannot completely thermalize, and the 
The low-$p$ slope can be inferred from the continuity equation
\begin{equation}
\frac{d}{dp}\left(\frac{\gamma_e}{p}n_\pm(p)\dot{\gamma_e}\right)
= \dot{n}_{\pm,\text{pp}}(p)
\end{equation}
which can be integrated over momentum to give
\begin{equation}\label{eq:modist}
n_\pm(p) = -\dot{\gamma_e}^{-1}\frac{p}{\gamma_e}\int_p^\infty 
dp~n_{\pm,\text{pp}}(p)
\end{equation}
Here $\dot{\gamma_e}\propto p^0$ is dominated by the heating term, and we 
find that $\dot{n}_{\pm,\text{pp}}\propto p^2$ at low energies from our 
numerical simulations. Then equation (\ref{eq:modist}) implies
$n_\pm(p) \propto p^4$ at low momenta.
%%%%%%%%%%%%%% Figure %%%%%%%%%%%%%%%%%%%%%%%%%
%%%%%%%%%%%%%%%%%%%%%%%%%%%%%%%%%%%%%%%%%%%%%%%
%%%%%%%%%%%%%%%% TEMPORAL EVOLUTION OF PHOTON SPECTRUM %%%%%%%%%%%%%%%%
\subsection{Time evolution of the photon spectrum}
\begin{figure}
\includegraphics[width=0.48\textwidth]{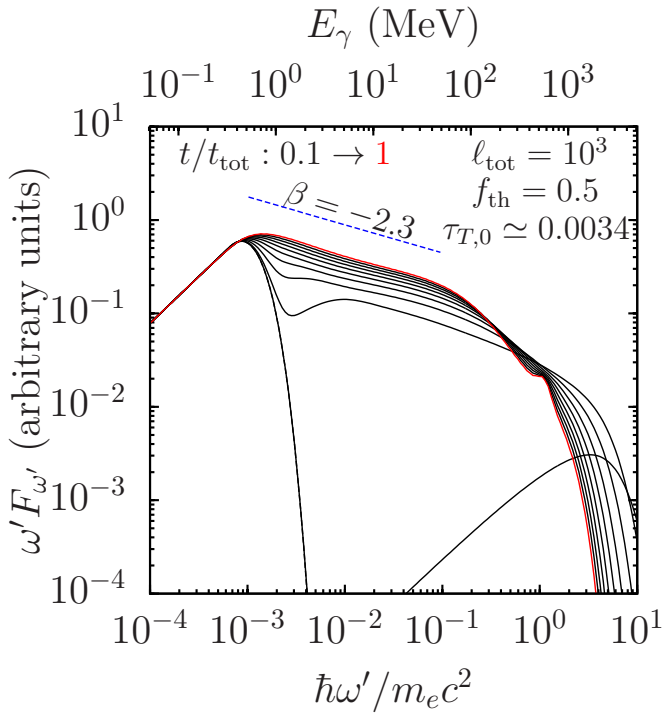}
\caption{Time evolution of the photon spectrum in the comoving 
frame of the burst ejecta for $\ell_{\text{tot}}=10^3$, 
$f_{\text{th}} = 0.5$, $\alpha_h=1$, and $\tau_0\simeq0.0034$. The curves 
correspond 
to the state of the photon-pair plasma at different times with 
$\Delta t/t_{\rm tot} = 0.1$ starting from $t=0.1t_{\rm tot}$ with 
compactness $\ell = \ell_{e,0}+\ell_{\text{th}}$. 
The horizontal axis at the top corresponds to the comoving energy 
boosted by a factor $\Gamma = 1000$ in the observer's frame. The blue 
dashed line segment indicates the high energy photon spectral index.}
\label{fig:ltot_1e3_fh_0.5_spec}
\vskip .2in
\end{figure}

%The heated pairs radiatively cool by upscattering soft photons
%to an average energy given by equation (\ref{eq:gamsq}).
% , where
%the average energy of the upscattered photon is
%\begin{equation}
%\langle x\rangle = \frac{4}{3}\gamma^2x_0
%\label{eq:xmax}
%\end{equation}
%and $x_0$ is the energy of the incoming photon. 
%For the soft photon distribution given in equation (\ref{eq:softdist}), one has
%\be
%\langle x_{\rm IC}\rangle \simeq 4\gamma_e^2\,\theta_\gamma.
%\ee
We plot the evolution 
of the photon distribution in the comoving frame of the burst ejecta 
in Figure \ref{fig:ltot_1e3_fh_0.5_spec}, in the case $\theta_\gamma \simeq 2\times10^{-4}$. 
The curves correspond to the states of the pairs shown in Figure 
\ref{fig:partdist}. As the particle energy declines,
the Compton upscattered peak shifts to a lower energy until it merges
smoothly with the seed thermal peak.

A higher rate of pair annihilation is found in the latter half of the simulation, 
as the $e^\pm$ become subrelativistic, which results in the 
formation of an annihilation feature at $x\simeq 1$.   A change in spectral slope
coincides with the line feature, and is driven by the annihilation of photons
of energy exceeding $m_ec^2$ with photons of a lower energy.

The final comoving photon spectrum looks remarkably like that observed 
for the majority of GRBs, with a typical low energy photon spectral 
index $\alpha \sim -1$ below the peak, and high energy index 
$\beta\sim-2.3$ above it.
 %%%%%%%%%%%%%% Figure %%%%%%%%%%%%%%%%%%%%%%%%%
%%%%%%%%%%%%%%%%%%%%%%%%%%%%%%%%%%%%%%%%%%%%%%%
%%%%%%%%%%%%%%% HEATING COMPACTNESS %%%%%%%%%%%%%%%%%%%%%%%%%%%%%%%%
%%%%%%%%%%% Figure - LTOT 1E3 FH COMPARE 1/T HEAT %%%%%%%%%%%%%%
\begin{figure*}
\epsscale{1.17}
\plottwo{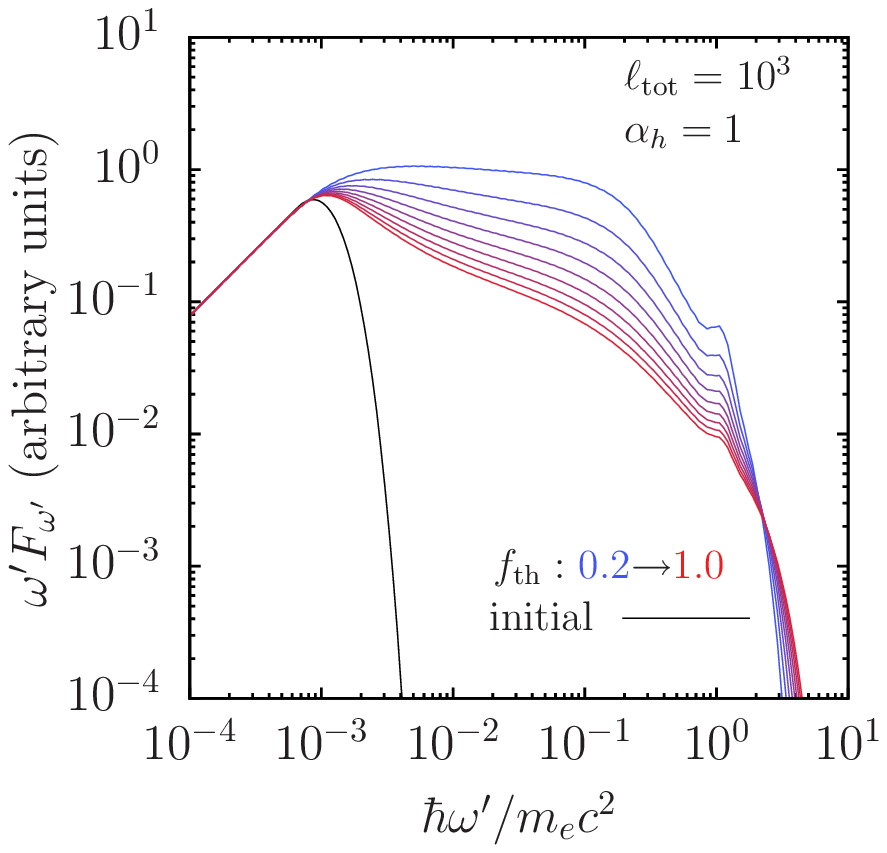}{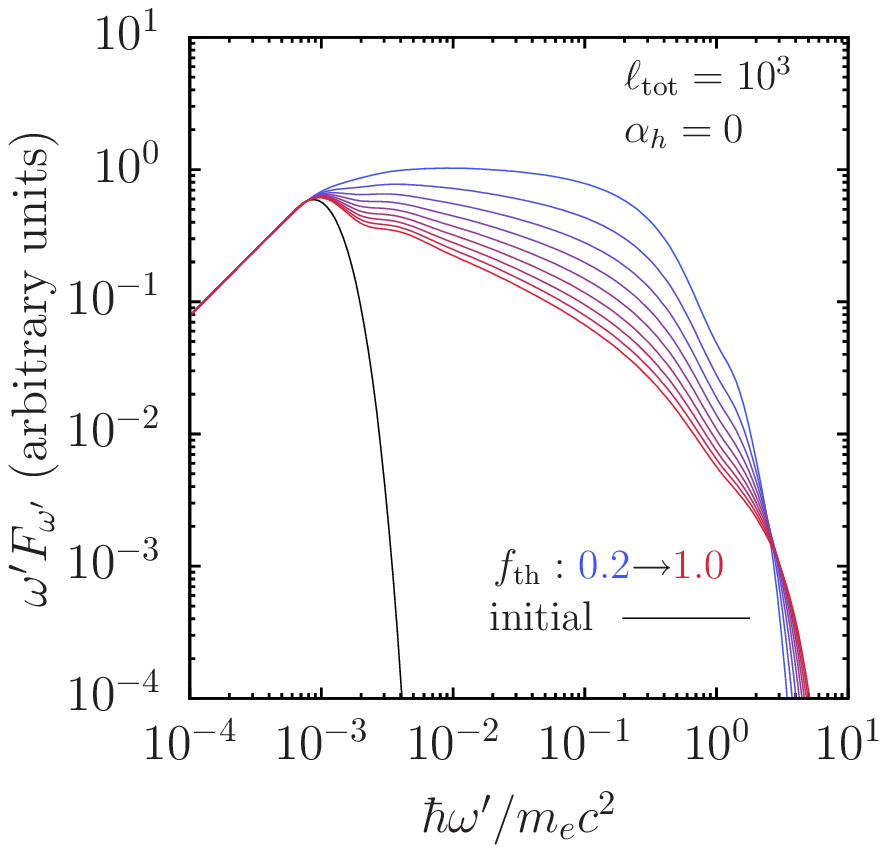}
\caption{Comparison of photon spectra (comoving frame) for total 
compactness $\ell_{\rm th} + \ell_{\rm heat} = 10^3$ and various
relative proportions $f_{\text{th}} = \ell_{\text{th}}/\ell_{\text{heat}}$ 
of seed thermal energy and added heat. 
Larger heating rates produce harder spectra.  Initial optical depth 
$\tau_{T,0}$ corresponds to $x_{\rm IC,max} = 5$ for photons draw from the
thermal peak.  GRB-thermal seed spectrum (equation (\ref{eq:grbthermal}))
with temperature $\theta_\gamma = 2\times10^{-4}$.  Left panel:  
$\alpha_h = 1$;  right panel: $\alpha_h=0$.}
\vskip .2in
\label{fig:fhcompare}
\end{figure*}
%\centering
%\subfigure[$l_{\text{tot}} = 10^3$, 
%$\alpha_h=1$]{%
%\includegraphics[width=0.48\textwidth]{ltot_1e3_fh_compare}
%%%%%%%% Figure - LTOT 1E3 FH COMPARE - CONST HEAT %%%%%%%%%%%%
%\subfigure[$l_{\text{tot}} = 10^3$, 
%$\alpha_h=0$]{%
%\label{fig:fhcompare_const}%
%\includegraphics[width=0.48\textwidth]{ltot_1e3_fh_compare_const_heat}
%}
%%%%%%%%%%%%%%% Figure - LTOT 1E3 FH COMPARE %%%%%%%%%%%%%%%%%%%
%%%%%%%%%%%%%%% TAU0 = 10*TAU0 %%%%%%%%%%%%%%%%%%%%%%%%%%%%%%%%%
\begin{figure*}
\epsscale{1.17}
\plottwo{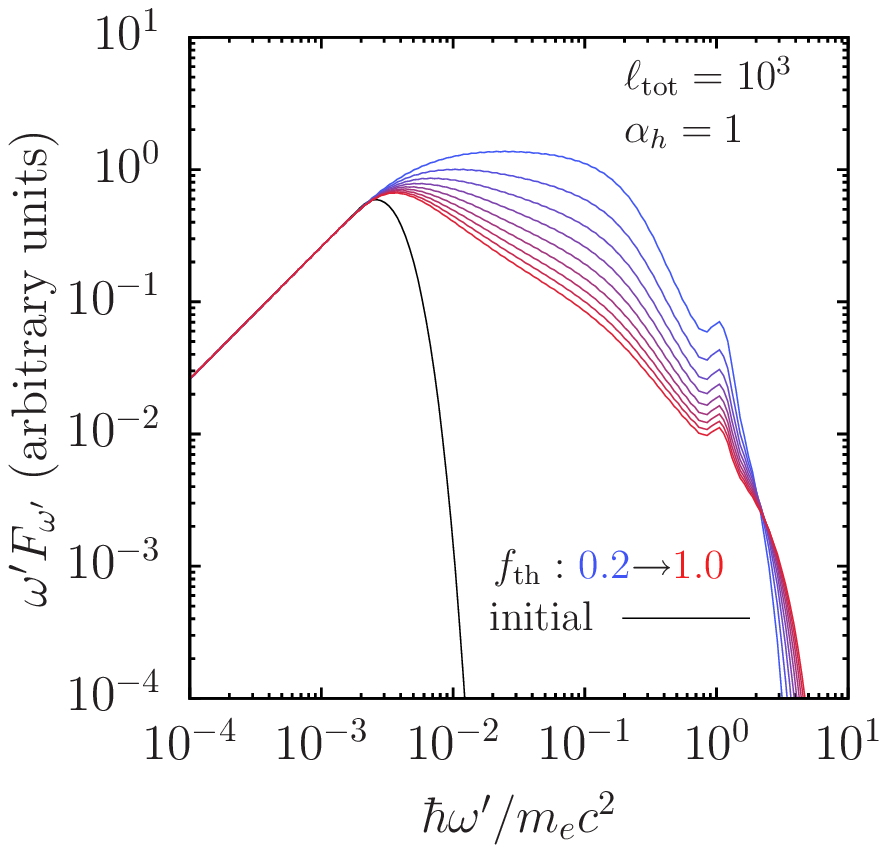}{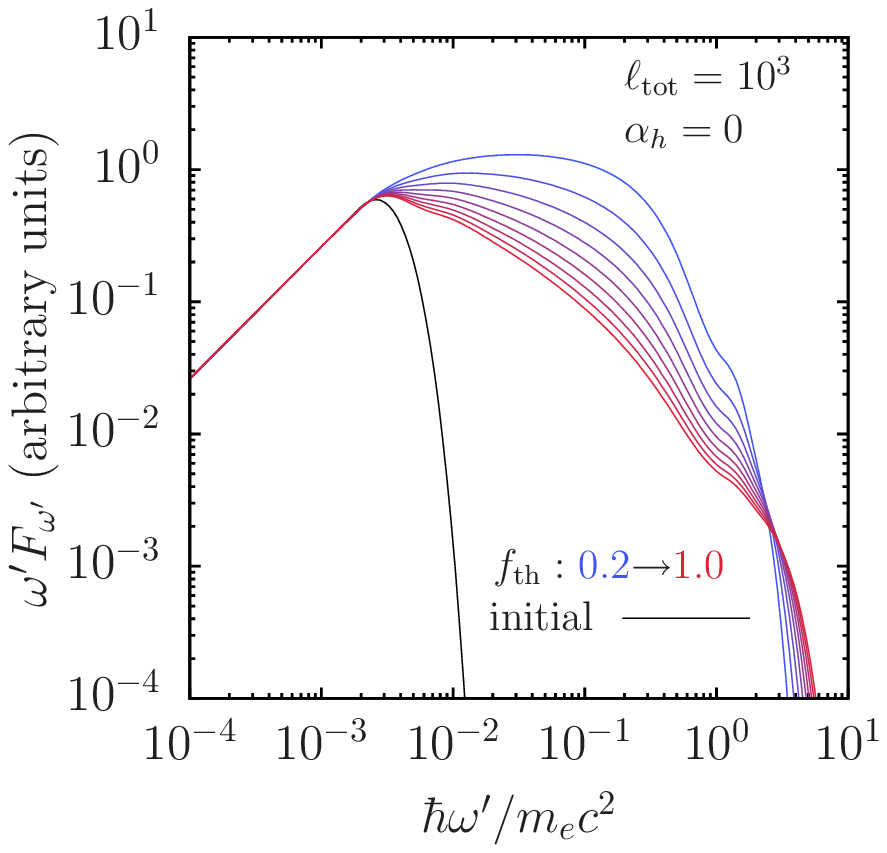}
\caption{Effect of raising seed photon temperature 
to $\theta_\gamma = 6\times 10^{-4}$ with $x_{\rm IC,max} = 10$.
Other quantities the same as in Figure \ref{fig:fhcompare}.}
\label{fig:ltot_1e3_3xpk}
\vskip .2in
\end{figure*}
\begin{figure*}
\epsscale{1.17}
\plottwo{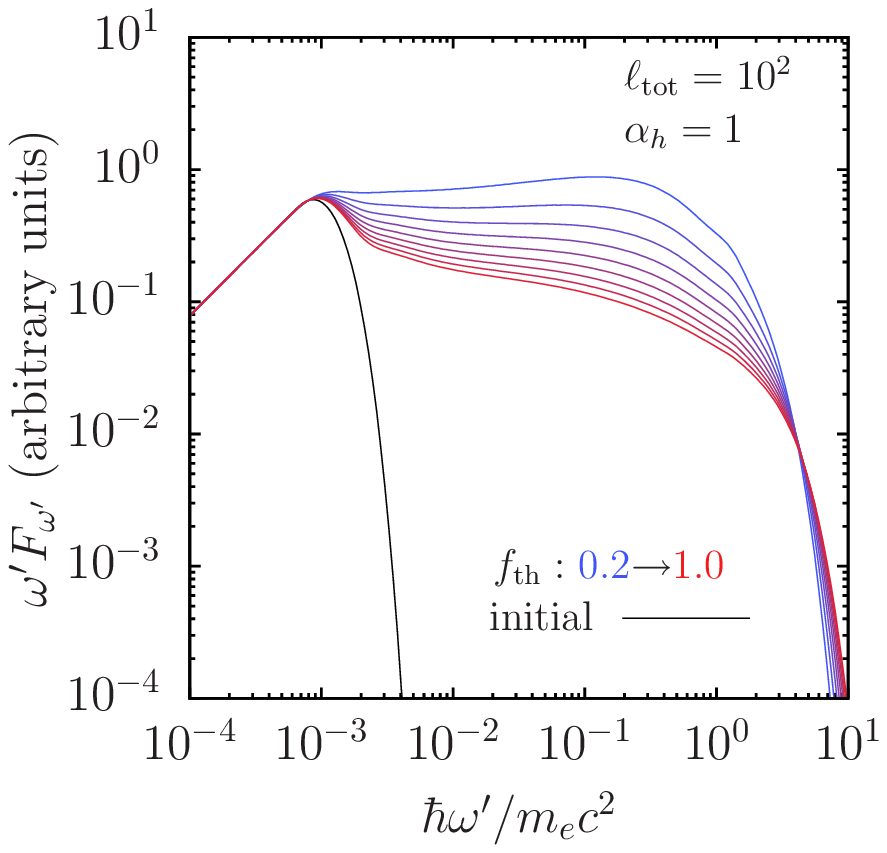}{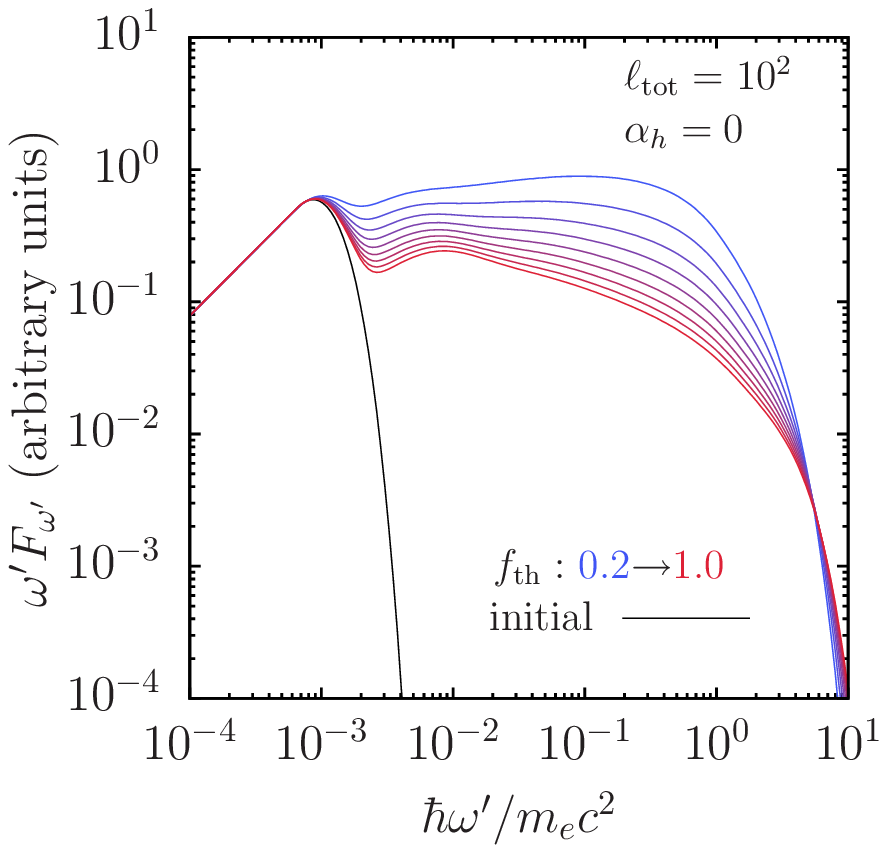}
\caption{Effect on the output spectrum of lowering the total compactness $\ell_{\rm th}+\ell_{\rm heat} = 10^2$. 
This gives a smaller pair yield (Figure \ref{fig:pairyield}), so that the pairs remain trans-relativistic at the end of heating.
Now the high-energy tail does not smoothly connect to the peak. Initial 
optical depth $\tau_{T,0}$ corresponds to $x_{\rm IC,max} = 5$ for photons drawn from the thermal peak.}
\label{fig:ltot_1e2}
\vskip .2in
\end{figure*}
\begin{figure*}
\epsscale{1.17}
\plottwo{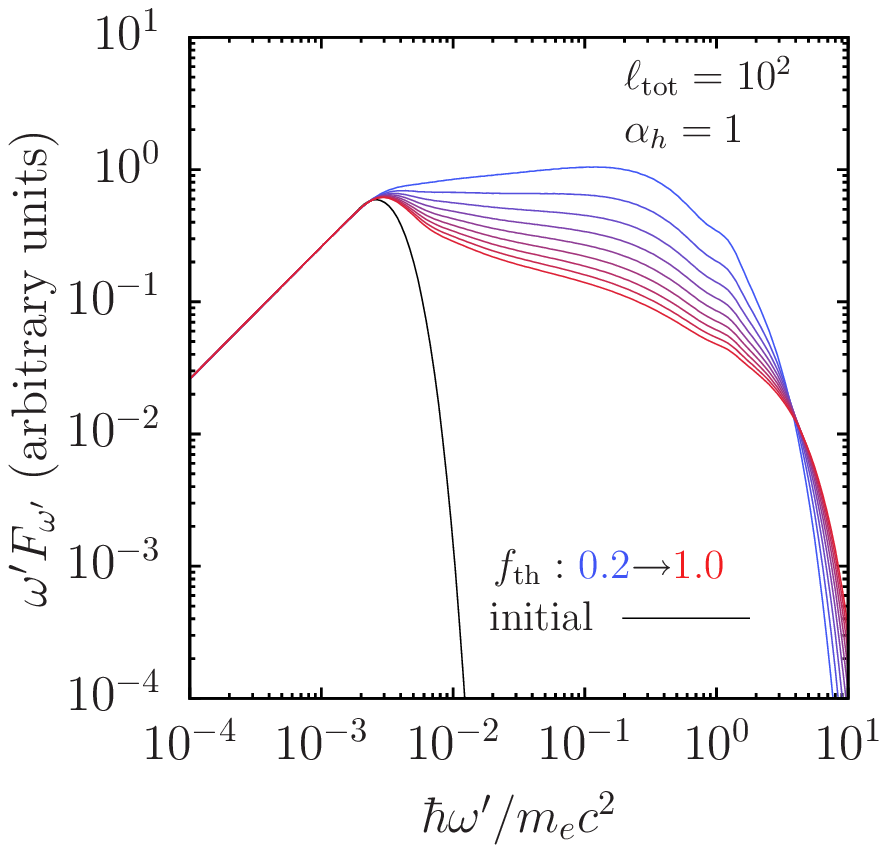}{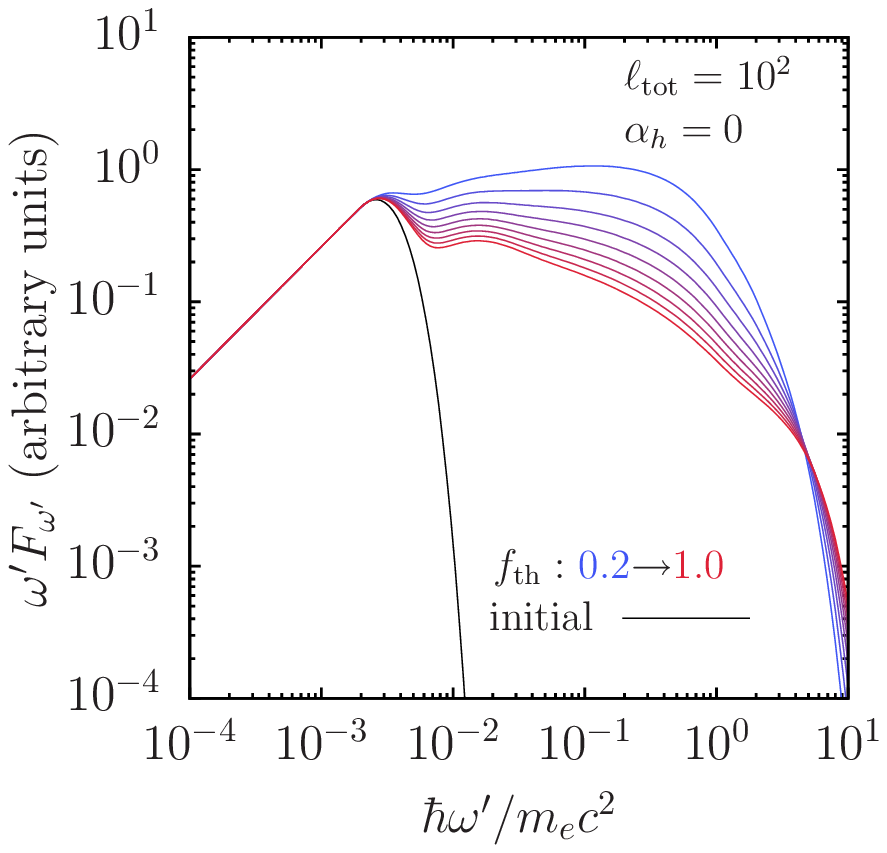}
\caption{Same as Figure \ref{fig:ltot_1e2} but now $\theta_\gamma = 6\times10^{-4}$
and $x_{\rm IC,max} = 10$.}
\label{fig:ltot_1e2_3xpk}
\vskip .2in
\end{figure*}
%\subfigure[$l_{\text{tot}}=10^3$, 
%$\alpha_h=1$, 
%$\tau_{T,0}^\star = 10~\tau_{T,0}$]{%
%\label{fig:ltot_1e3_fh_compare_10tau0}%
%\includegraphics[width=0.48\linewidth]{ltot_1e3_fh_compare_10tau0}
%}
%%%%%%%%%%%%%%% Figure - LTOT 1E2 FH COMPARE %%%%%%%%%%%%%%%%%%%
%\subfigure[$l_{\text{tot}}=10^2$, 
%$\alpha_h=1$]{%
%\label{fig:ltot_1e2_fh_compare}%
%\includegraphics[width=0.48\textwidth]{ltot_1e2_fh_compare}
%}
%\caption{Comparison of photon spectra 
%(comoving frame) for different levels of heat injected into pairs 
%where the heating compactness is controlled by the parameter 
%$f_{\text{th}} = l_{\text{th}}/l_{\text{heat}}$. 
%Larger heating compactness, 
%smaller $f_{\rm th}$, produces harder spectra. The initial optical depth 
%$\tau_{T,0}$ is determined from equation (\ref{eq:gmax}) where the peak 
%of the IC spectrum falls at $x^{\text{IC}}_{\text{pk}}\simeq 10$, 
%except for Figure $\text{c}^\star$ where the initial optical depth 
%was raised by an order of magnitude to see the effect of higher 
%optical depth.}
%\label{fig:ltot_fh_cases}
%\end{figure*}
%\clearpage
%}

\subsection{Dependence on Heating Compactness \\ and Pair Yield}

The spectral index above the peak is largely determined by the ratio $f_{\rm th}$
of the energy in seed thermal photons to that injected in heat (equation (\ref{eq:fth})).
A smooth connection of the spectrum to the peak is obtained for $\ell_{\rm tot} \gtrsim 10^3$,
but the efficiency of pair creation drops when $\ell_{\rm tot} \sim 10^2$ to the point
that a signficant thermal bump is preserved near the peak.  

At this juncture it is worth reviewing where in the outflow strong dissipation will
develop due to differential motion of baryons and magnetic field.  A net compactness
$\ell_{\rm tot} > 10^3$ is easily achieved, according to the following argument.

In the case of
a Wolf-Rayet progenitor, baryons derived from the stellar envelope can no longer be
accelerated outward by the anisotropic thermal photon pressure when $\ell_{\rm th}$
drops below $\sim 4\times 10^3$.  Baryon-free parts of the outflow can continue
to accelerate outward beyond this point, so that a Lorentz factor differential
$\Gamma/\Gamma_{\rm sat,ei} \sim 2$ develops at $\ell_{\rm th} \sim 2^{-4}(4\times 10^3)
\sim 250$.  The corresponding heating compactness is $\ell_{\rm heat}
= 500 (f_{\rm th}/0.5)^{-1}$, and $\ell_{\rm tot} \gtrsim 10^3$.  
The electron fraction in the wind emitted by a merged neutron star binary is only $Y_e \sim 0.05-0.1$,
meaning that the compactness during reheating is 5-10 times larger.

\begin{figure}
 \includegraphics[width=0.48\textwidth]{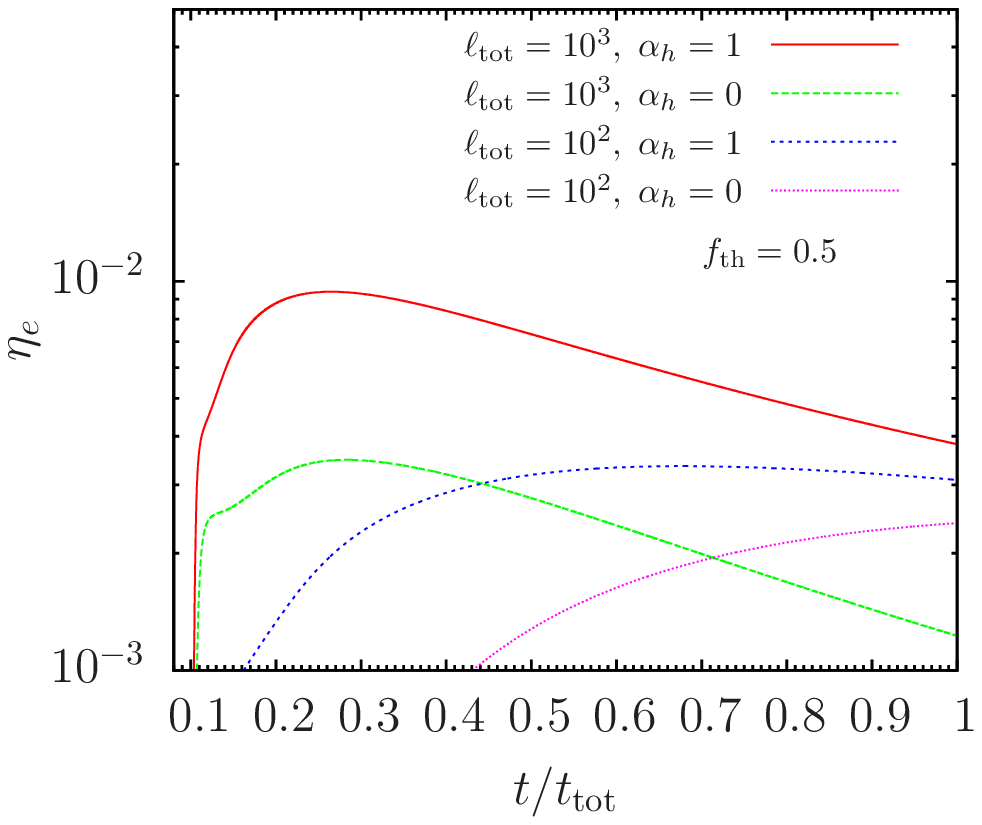}
 \caption{Comparison of the pair yield $\eta_e$ for various heating 
 profiles, and final compactness as plotted in Figures \ref{fig:fhcompare} and 
\ref{fig:ltot_1e2}. For all curves $f_{\text{th}}=0.5$. The initial optical depth for 
$\alpha_h=1~(\alpha_h = 0)$ is $\tau_{T,0}\simeq0.0034~(0.00087)$.}
 \vskip .2in
 \label{fig:pairyield}
 \end{figure}

We show the final comoving photon spectrum in Figure \ref{fig:fhcompare} for 
a range of $f_{\rm th}$ and the same total compactness
($\ell_{\rm tot} = 10^3$) as in Figures \ref{fig:timeprofile}-\ref{fig:ltot_1e3_fh_0.5_spec}.
Harder spectra result from larger proportional injections of heat (smaller $f_{\text{th}}$).
A shift to a constant heating rate $(\alpha_h=0)$ only produces subtle changes in the output spectra.
There is more of a difference in the pair yield $\eta_e$ (the rest energy in pairs created per 
$m_ec^2$ of injected heat),
\begin{equation}
\eta_e = \frac{d\tau_{\rm T}^+/dt}{d\ell_{\text{heat}}/dt}.
\end{equation}
We plot $\eta_e$ as a function of time in Figure 
\ref{fig:pairyield} for $f_{\text{th}} = 0.5$ and various heating profiles.
%used in Figures \ref{fig:fhcompare} and \ref{fig:ltot_1e2}.

We find a strong presence of the thermal bump for $\ell_{\rm tot} = 10^2$ (Figure \ref{fig:ltot_1e2}).  
A glance at the  pair yield corresponding to this case reveals insufficient pair production. This, consequently, 
leads to a higher average kinetic energy of pairs and failure of the high energy spectrum 
to connect smoothly with the thermal peak.

In Figures \ref{fig:ltot_1e3_3xpk} and \ref{fig:ltot_1e2_3xpk}, 
we show the effect on the photon spectrum of
 increasing the thermal seed photon temperature to $\theta_\gamma = 6\times10^{-4}$,
for the same range of compactness and heating profiles.
The corresponding results for the pair yield are in Figure \ref{fig:pairyield}.  

Concentrating the energy injection toward early times ($\alpha_h = 1$) allows
a greater number of hard, pair-creating photons to be upscattered before photon collisions
raise the pair density and force a drop in particle energy.  The net result is
a higher pair yield and a higher final optical depth (Figure \ref{fig:tauvsl}).
This also explains the stronger annihilation line as compared with the results
of constant heating.
 
%%%%%%%%%% Figure - Pair Yield %%%%%%%%%%%%%%%%%%%%%%%%%%%%%%%
%%%%%%%%%%%%%%%%%%%%%%%%%%%%%%%%%%%%%%%%%%%%%%%%%%%%%%%%%%%%%%%%%

\subsection{Buffering of the Scattering Depth}

\begin{figure}
\includegraphics[width=0.48\textwidth]{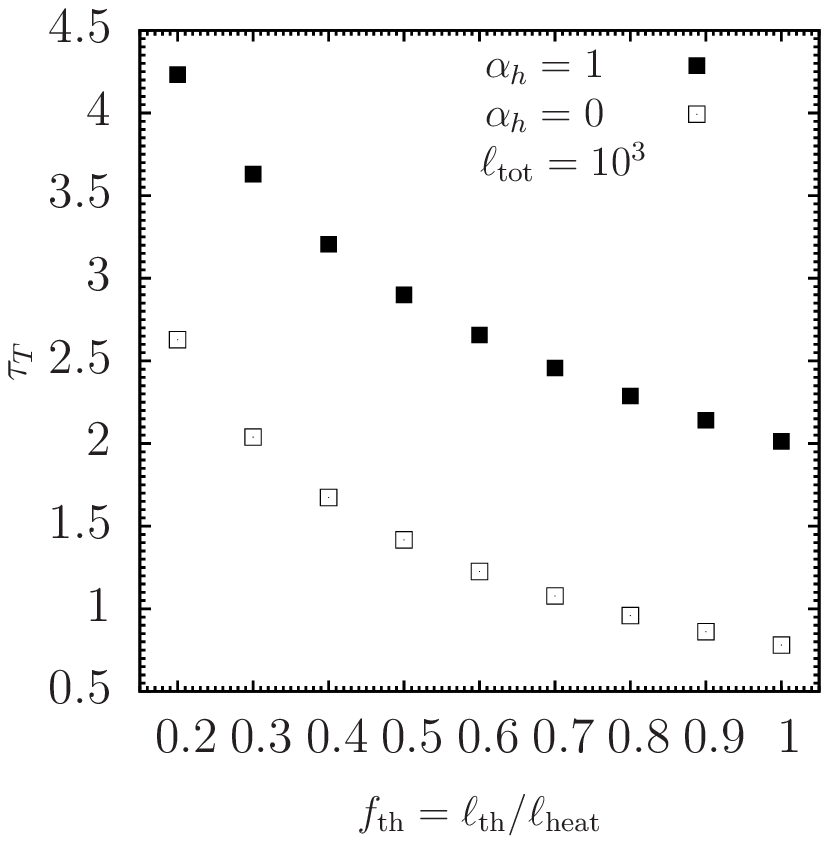}
\caption{Final scattering depth, approximated as Thomson, versus
heating compactness.  Total compactness (initial thermal $+$ injected
heat) $\ell = 10^3$.  A flat heating profile also leads to somewhat
lower scattering depth.}
\vskip .2in
\label{fig:tauvsl}
\end{figure}

Delayed heating in an optically thin outflow preserves a narrow thermal peak that
is generated before jet breakout.  The rate of pair creation is subject to a regulating
effect, due to the inverse relation (\ref{eq:gamsq}) between inverse-Compton energy and scattering
depth.  The scattering depth is a function of heating compactness, with lower
heating rates generating steeper spectra and lower optical depths
(Figure \ref{fig:tauvsl}).

More generally, the rate of pair creation in a compact and relativistically expanding outflow
is sensitive to the mechanism by which high-energy photons are produced.
A common starting assumption is that non-thermal electrons (and positrons)
are rapidly accelerated, so that radiative cooling follows acceleration.
Then a power-law high-energy photon spectrum depends on a power-law 
distribution of injected particles, such as might originate at a shock.
In this picture, the high-energy index (of charged particles or photons)
is not connected in a simple or obvious way with the compactness of the
outflow.  Rapid acceleration of a hard particle distribution of a high 
compactness would result in large scattering depths, $\tau_{\rm T} \sim \ell^{1/2}$ 
(e.g. \citealt{guilbert83}).  

%Continuous heating of thermal particles can generate a high-energy photon
%spectrum by multiple scattering near a scattering photosphere \citep{giannios06},
%when the heating starts at a scattering depth greater than unity.  This procedure 
%requires some fine tuning, however.   When scattering is dominated by the 
%electron-ion component, the heating must be concentrated near the scattering 
%photosphere.  Heating that is broadly distributed in radius might originate 
%from magnetic reconnection \citep{thompson94,drenkhahn02}, but the radial
%position of reconnection is still sensitive to the time profile of the magnetic
%field if the engine is a black hole ergosphere (\citealt{thompson06}, Paper III).

%Continuous heating in a pair-dominated, magnetized plasma generates a quasi-thermal 
%spectral peak if the compactness of the outflow is still high (Paper I).  This peak
%can be associated with the observed spectral peak in a GRB if the bulk Lorentz
%factor is still modest.  The scattering depth of the outflow regulates to a moderately
%high value, $\tau_{\rm T} \sim 5-10$, if the heating starts at $\ell \sim 10^5-10^8$
%as appropriate for breakout of a GRB jet.  If heating were to continue as the outflow 
%accelerated, then this thermal peak would be broadened by multiple scattering.

%%%%%%%%%%%%%% Figure %%%%%%%%%%%%%%%%%%%%%%%%%
%%%%%%%%%%%%%%%%%%%%%%%%%%%%%%%%%%%%%%%%%%%%%%%
%%%%%%%% CONTINUOUS HEATING IN AN ADIABATICALLY EXPANDING MEDIUM %%%%%%%%%%%%%%%%%%%%%
\section{Distributed Heating in an Expanding Medium}\label{s:expand}
\label{s:expand_results}
We now combine expansion with continued heating and explore their effect on the comoving photon spectrum.  
In this situation, radiation interacts with the magnetofluid in a complicated way.
The radiation field transfers energy to or from the expanding magnetofluid, depending on the 
magnetic field profile  (Section \ref{s:accel}, \citealt{russo13b}).    A simplification occurs when
$\Gamma \propto r$:  then the energy flux of an adiabatically evolved radiation field
is preserved in the frame of an external observer, and the differential flow of
radiation and matter is minimized \citep{beloborodov11}.  We focus on this case here.

%Although the medium is optically thin 
%($\tau_{T,0}\ll1$) to radiation when heating commences, the optical 
%depth quickly builds up and exceeds unity.  

We let the flow expand by a factor $\zeta$ during the heating episode, 
out to a final radius $R_h$, Lorentz factor $\Gamma_h$, and time $t_h$.  
The flow time in the comoving frame relates to the radial extent of the jet through
\be
dt = {dr\over \Gamma(r) c} = \left( \frac{R_{\rm h}}{\Gamma_{\rm h}  c} \right) \frac{d r}{r},
\ee
which gives an exponential expansion, $r(t) \propto e^{\Gamma_{\rm h}ct/R_{\rm h}}$.
A rescaled time coordinate is obtained by setting $\hat t = 0$~(1) at the start (end) of heating, 
\be
r(\hat t) = {R_h\over\zeta} \exp\left(\ln\zeta\,\hat t\right); \quad 
              \hat t = 1 + {\Gamma_h c\over R_h \ln\zeta} (t-t_h).
\ee
The initial time $t_0$ is found by setting $\hat t = 0$.

Adiabatic expansion of a photon gas is governed by the equation
\begin{equation}
\frac{\partial n_{\gamma} ( x )}{\partial t} = \frac{\partial}{\partial x}
  \left[ - \frac{d x}{d t}  n_{\gamma} ( x ) \right] + \frac{3}{x}   \frac{d
  x}{d t}  n_{\gamma} ( x ).
\end{equation}
The energies of individual photons (and all relativistic particles) evolve 
as $x \propto r^{-1}$, corresponding to
\begin{equation}
\frac{dx}{d\hat{t}} = -x\ln\zeta.
\end{equation}
The final expansion term in the photon evolution equation (\ref{eq:photon}) is,
\begin{equation}
\dot n_{\rm \gamma,exp}(x) = \frac{\partial}{\partial x}[x\ln\zeta~n_\gamma(x)]-3\ln\zeta~n_\gamma(x),
\end{equation}
and the corresponding term in equation (\ref{eq:particle}) for the pairs is,
\begin{equation}
\dot n_{\rm \pm,exp}(p) = \frac{\partial}{\partial p}\left[
\frac{\gamma^2}{p}\ln\zeta~n_\pm(p)\right]-3\ln\zeta~n_\pm(p).
\end{equation}

The corresponding advective coefficients that enter the FP equations are 
$\{A_\gamma,A_\pm\}_{\rm exp} = \{x\ln\zeta,\gamma\ln\zeta\}$, whereas 
the dilution of the number density enters the collision-integrals via 
the terms $\{\dot{n}_\gamma,\dot{n}_\pm\}_{\rm exp}^{\rm col} = 
\{-3\ln\zeta n_\gamma(x),-3\ln\zeta n_\pm(p)\}$. 

Expansion causes a rapid drop in the energy density in radiation,
\be
U_\gamma = m_ec^2 \int dx\, x n_\gamma(x) \propto r^{-4} \propto \exp\left(-4\ln\zeta\hat t\right),
\ee
corresponding to a net adiabatic dilution $\zeta^{-4}$ from the beginning to the end of heating.
Since the energy deposited in particles is rapidly transferred to the photons, it is useful
to subsume the effects of Compton scattering of the photons in a single heating term,
\be
{dU_\gamma\over d\hat t} + 4\ln\zeta U_\gamma = {dU_{\rm heat}\over d\hat t}.
\ee
The rate of heat deposition is scaled to the adiabatically evolved thermal energy density,
\be\label{eq:expandheat}
{dU_{\rm heat}\over d\hat t} \propto \hat t^{-\alpha_h} e^{-4\ln\zeta\hat t} U_{\rm th}(t_0). 
\ee

We define a compactness in terms of the adiabatically evolved energy
density at the end of heating.  From
\be
U_\gamma(t) = U_{\rm th}(t_0) e^{-4\ln\zeta\hat t}
 + \int_0^{\hat t} {dU_{\rm heat}\over d\hat t_2} e^{4\ln\zeta(\hat t_2-\hat t)} d\hat t_2,
\ee
the corresponding thermal and non-thermal compactness are
\be
\ell_{\rm th} = {\Gamma_h c\over R_h} {\sigma_T \over m_ec^2} {U_{\rm th}(t_0)\over \zeta^4}
\ee
and 
\be
\ell_{\rm heat} = {\sigma_T \Gamma_h \over R_h m_ec}
      \int_0^1 {dU_{\rm heat}\over d\hat t} e^{4\ln\zeta(1-\hat t)} d\hat t.
\ee
The total compactness can be written in terms of a thermal fraction 
$f_{\rm th} = \ell_{\rm th}/\ell_{\rm heat}$ as
%\begin{equation}
%\ell = \sigma_TU\left(\frac{R_{\rm br}}{\Gamma_{\rm br}}\right)
%\end{equation}
\be
\ell_{\rm tot} = \ell_{\rm heat}(1+f_{\rm th}).
\ee
In the case where heat is deposited at a uniform rate ($\alpha_h = 0$), the radiation compactness is
\begin{equation}
\ell_\gamma(\hat{t}) = \ell_{\rm heat}\left(\hat{t}+f_{\rm th}\right)e^{4\ln\zeta(1-\hat t)}.
\label{eq:l_that}
\end{equation}
One observes that the initial thermal compactness can significantly exceed the final
compactness with even a modest expansion.

\begin{figure}
\includegraphics[width=0.48\textwidth]{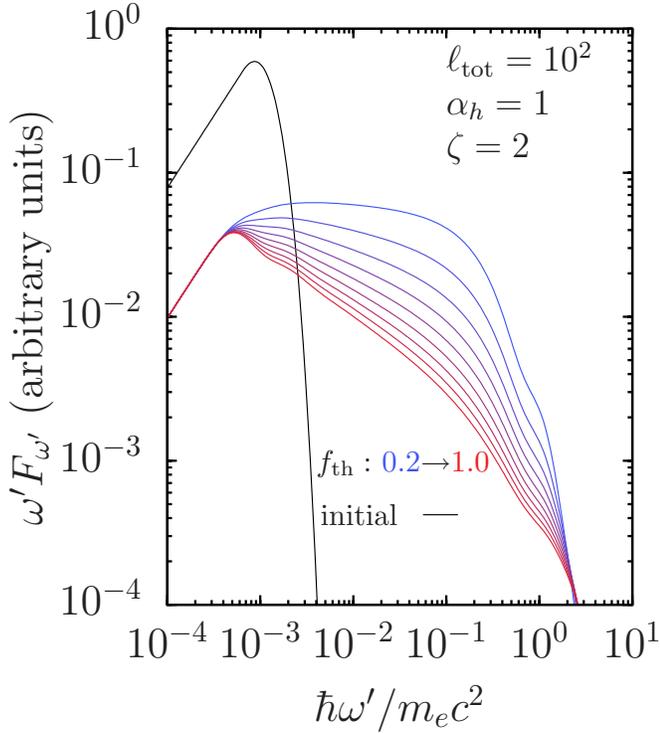}
\caption{Final comoving photon spectra in an expanding medium with
final compactness $\ell_{\rm tot} = 10^2$ and a range of $f_{\rm th} = 
\ell_{\rm th}/\ell_{\rm heat}$.  Net expansion factor $\zeta = 2$.
The injection of heat follows the profile (\ref{eq:expandheat}) with a $\hat t^{-1}$ envelope 
($\alpha_h = 1$) that is cut off at $\hat t = 0.1$.}
\vskip .2in
\label{fig:phot_spec_expand2}
\end{figure}
%%%%%%%%%%%%%%%%%%%%%%%%%%%%%%%%%%%%%%%%%%%%%%%%%%%%%%%%%%%%%%%%%%%%%%%%%%%%%%%%%%%%%%%

To compare with the static case, we align the final compactness $\ell_{\rm tot}$
and the comoving peak energy in the initial state.
Then the peak in the output comoving spectrum shows the effects of adiabatic cooling.
The initial optical depth is chosen in a similar way to the calculations of
Section \ref{s:results}, so that the maximum inverse Compton energy of photons drawn
from the thermal peak is $x_{\rm IC,max} \sim 5$ (equation (\ref{eq:xicmax})).

Although the compactness now has a much different time dependence than in the static case, 
the equilibrium Lorentz factor of the heated pairs depends mainly on the ratio of 
non-thermal and thermal compactness, and the scattering depth $\tau_{\rm T} \equiv
\sigma_T n_e (R_h/\Gamma_h)$,
\be
\frac{4}{3}(\gamma_e^2-1) = \frac{1}{\tau_{\rm T}\,U_\gamma\,\ln\zeta}\frac{dU_{\rm heat}}{d\hat{t}}
-\gamma_e.
\ee
Adiabatic expansion of the outflow adds an additional cooling term to equation (\ref{eq:gamsq}),
but it has a relatively small effect on $\gamma_e$ at large $\ell$.   The contribution of
the particles to the initial energy density is very small even after heating to relativistic
energies.
%eq. (\ref{eq:pair_heat_cool})
%\ee
%the solution of which defines an $\langle\gamma\rangle$ where most of 
%the pairs  are found, just like in the static case. Again, we ask that 
%the inverse  Compton scattered photon distribution has a peak 
%$\hbar\omega_{\rm IC}\simeq 10 m_ec^2$, such that 
%$\gamma^2 = \omega_{\rm IC,pk}/4\omega_{\rm pk}$. Then, at $\hat{t}=0$, 
%from the above equation, it follows that
% \be
% \tau_{T,0} = \ell_{\rm heat}e^{4\ln\zeta}
% \left[\ln\zeta\left\{\frac{4}{3}(\gamma^2-1)\ell_{\rm th,0}-\gamma\right\}
% \right]^{-1}
% \ee
% where $\tau_{T,0} = \sigma_T n_{e,0}R_{\rm br}/\Gamma_{\rm br}$ and 
% $\ell_{\rm th,0} = f_{\rm th}\ell_{\rm heat}$. 
 
% In order to compare the photon spectrum in both the static and expanding 
% cases, the final compactness $\ell_{\rm tot}$ must be the same. From 
% eq. (\ref{eq:le0} and \ref{eq:l_that}), we find 
% \be
% \ell_{\rm tot} = \ell_{\rm heat}+(\langle\gamma\rangle\tau_{T,0}+f_{\rm th}
% \ell_{\rm heat})e^{-4\ln\zeta}
% \ee
% where we solve for $\ell_{\rm heat}$ for a given $\ell_{\rm tot}$, 
% $f_{\rm th}$, and $\zeta$. The above expression can be simplified further by 
% noticing that $\ell_{e,0}\ll1$, which gives
% \be
% \ell_{\rm tot} \approx \ell_{\rm heat}(1+f_{\rm th}e^{-4\ln\zeta})
% \ee
% which reduces to the static case for $\zeta=0$.

Figure (\ref{fig:phot_spec_expand2}) shows output spectra corresponding to 
$\ell_{\rm tot} = 10^2$, with net expansion $\zeta = 2$ and a decaying envelope 
to the heating rate as defined in equation (\ref{eq:expandheat}) ($\alpha_h = 1$).
The initial compactness, which is closer to $\sim 10^3$,
plays a major role in determining the output spectrum:  a comparison with Figures
\ref{fig:fhcompare} and \ref{fig:ltot_1e2} shows a greater similarity with
the $\ell = 10^3$ fixed box runs than with the $\ell = 10^2$ case.  The effect
of adiabatic expansion on the energy of the spectral peak is clearly present.

\section{Effect of Residual Scattering on \\ Output Spectrum}\label{s:rescatt}

\begin{figure}
\includegraphics[width=0.48\textwidth]{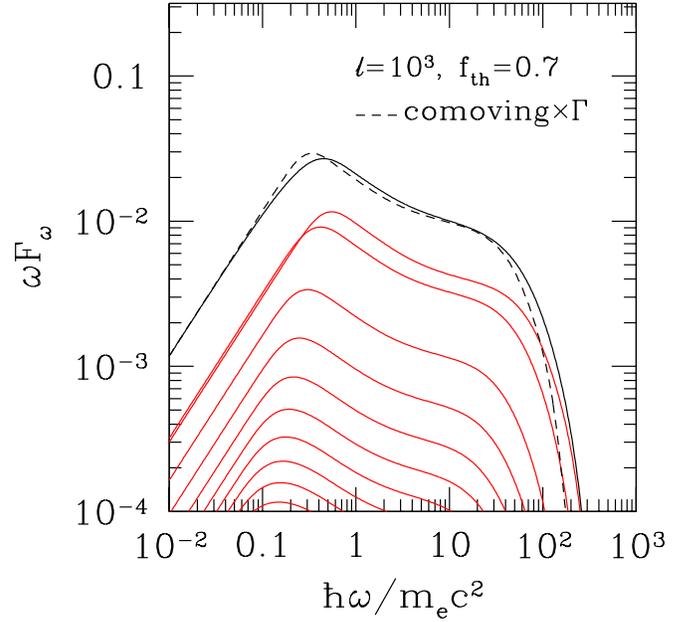}
%\vskip -0.5in
\caption{Output spectrum (solid black curve) resulting from multiple scattering off a passively expanding
pair gas with constant Lorentz factor $\Gamma = 100$ and scattering depth determined by the end of the 
kinetic calculation of Section \ref{s:results}.  Dashed black line:  source spectrum, boosted by a factor $\Gamma_0 = 100$ 
from the comoving one-box calculation, corresponding to $\ell = 10^3$ and $f_{\rm th} = 0.7$.  Red curves:  
time-resolved spectrum, plotted at intervals $\Delta t = 0.5 (R_0/2\Gamma_0^2c)$.}
\vskip .2in
\label{fig:scattspec}
\end{figure}
%\begin{figure}
%\includegraphics[width=0.48\textwidth]{pulse}
%\caption{Pulse profiles corresponding to the sequence of spectra in Figure \ref{fig:scattspec}.  Red curves
%now label a subset of the snapshots, separated in time by $\Delta\log(t) = 1$.}
%\vskip .2in
%\label{fig:pulse}
%\end{figure}
\begin{figure}
\includegraphics[width=0.48\textwidth]{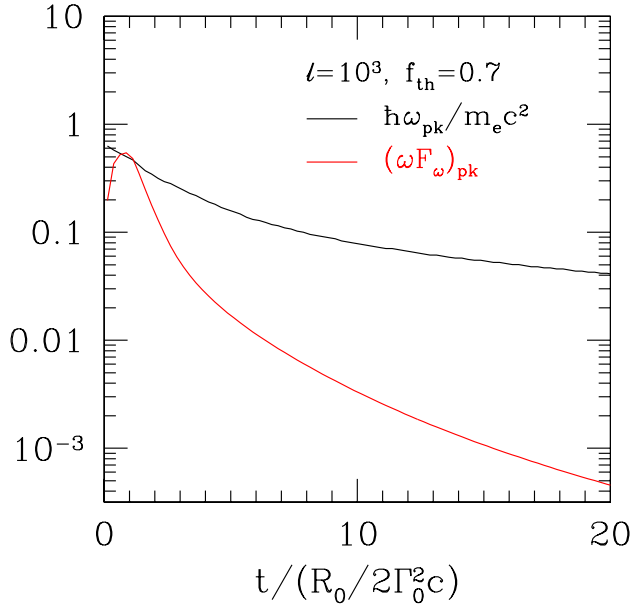}
%\vskip -0.5in
\caption{Variation with time of spectral peak energy and energy flux at the peak, in a burst
based on our one-zone spectral model with $f_{\rm th} = 0.7$, compactness $\ell = 10^3$, and $\Gamma = 10^2$
during the emission of the high-energy tail.}  
\vskip .2in
\label{fig:flux_peak}
\end{figure}
The photon spectra obtained from our static one-box calculations are now evolved by scattering
off the frozen, expanding pair gas.  We use the Monte Carlo code described in Paper III.  Heating is assumed to have turned off, so that the thermal
Compton parameter of the pair gas is low, $y_{\rm C} \sim \tau_{\rm T} T_e/m_ec^2 \ll 1$.  The spectrum
still evolves, but only modestly, by differential scattering off the bulk flow.  

In Figure \ref{fig:scattspec} we compare the output spectrum, averaged over an entire pulse, with the one-box
calculation boosted by a factor $\Gamma$ in energy.  The time-resolved spectrum shows
little evolution in shape, except for an overall reduction in energy due to side-ways
emission.   The strong hard-to-soft evolution is shown in Figure \ref{fig:flux_peak}.
This result is characteristic of optically thin and 
non-thermal emission from curved relativistic shells (\citealt{shenoy13} and references therein).  More
details of the pulse evolution are investigated in Paper III.

\subsection{Outflow Heating Continuously from a Large Scattering Depth}\label{s:shell}

A relativistic outflow that is heated continuously outward from a large scattering
depth develops an extended, high-energy spectral tail to a seed thermal radiation field
\citep{giannios06,beloborodov10,lazzati10}.  Here we compare the emergent spectrum and 
hardness evolution with that produced by pair breakdown.

Heating is assumed to continue from inside to outside the photosphere, as in the most recent
calculation of \cite{giannios08}.  The outflow starts at a certain initial scattering depth
(\ref{eq:tauT}) from an initial radius $R_0$, and the particle energy adjusts so that
\be\label{eq:yC}
{4\over 3}\left(\langle \gamma_e^2\rangle -1\right) n_e \sigma_T {r\over 2\Gamma^2} = {dy_{\rm C}\over d\ln t} = {\rm const}.
\ee
We choose a constant Lorentz factor $\Gamma$ and spherical geometry, 
and neglect any effect of pair creation or annihilation, as in the previous calculations.
Then $n_e(r) \propto r^{-2}$ when $\Gamma \gg 1$.

\begin{figure}
\includegraphics[width=0.48\textwidth]{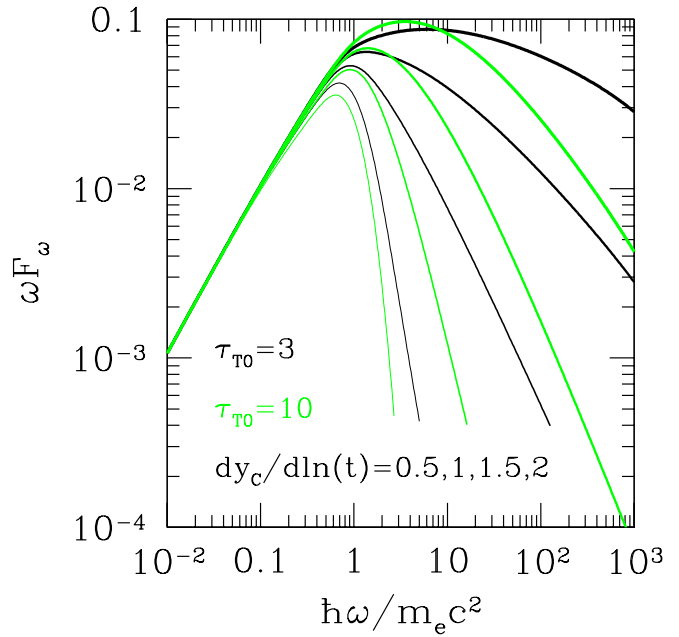}
%\vskip -0.5in
\caption{High-energy spectrum formed by continuous heating of an outflow that is making a transition
from large to small scattering depth, as discussed in the text.  Heating starts at a fixed scattering
depth (\ref{eq:tauT}):  $\tau_{T,0} = 3$ (black curves) and $\tau_{T,0} = 10$ (green curves).  Compton
parameter (\ref{eq:yC}) is $dy_{\rm C}/d\ln t = 2, 1.5, 1, 0.5$ as curves vary to from thickest to thinnest.}
\vskip .2in
\label{fig:spec_mult}
\end{figure}
\begin{figure}
\includegraphics[width=0.48\textwidth]{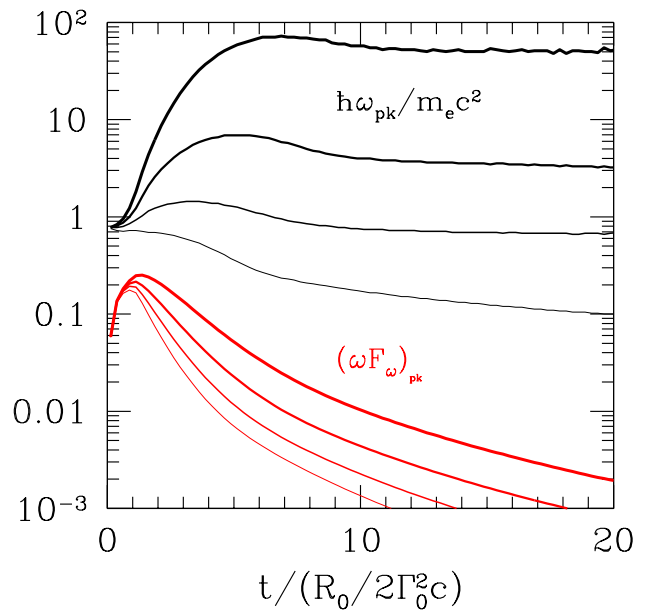}
%\vskip -0.5in
\caption{Variation with time of spectral peak energy and energy flux at the peak.
Spherical, relativistic outflow is continuously heated from inside its photosphere.
Curve colors and thickness label $\tau_{T,0}$ and $dy_{\rm C}/d\ln t$ as in Figure \ref{fig:spec_mult}.}
\vskip .2in
\label{fig:pulse_train_mult}
\end{figure}

The output spectrum, as shown in Figure \ref{fig:spec_mult}, confirms the formation of extended high-energy
tail, with an increasing hardness as $y_{\rm C}$ is raised.  However a hard high-energy spectrum is 
associated with a strong broadening of the spectral peak.  

Since harder photons are created by multiple scattering of softer photons, the pulses are broader 
at high energies, now in strong contradiction with observations (e.g. \citealt{fenimore95,norris96}).  
For example, the burst shows strong soft-to-hard evolution (Figure \ref{fig:pulse_train_mult}).
Further details of the pulse behavior in this model are discussed in Paper III.

\begin{figure}
\includegraphics[width=0.48\textwidth]{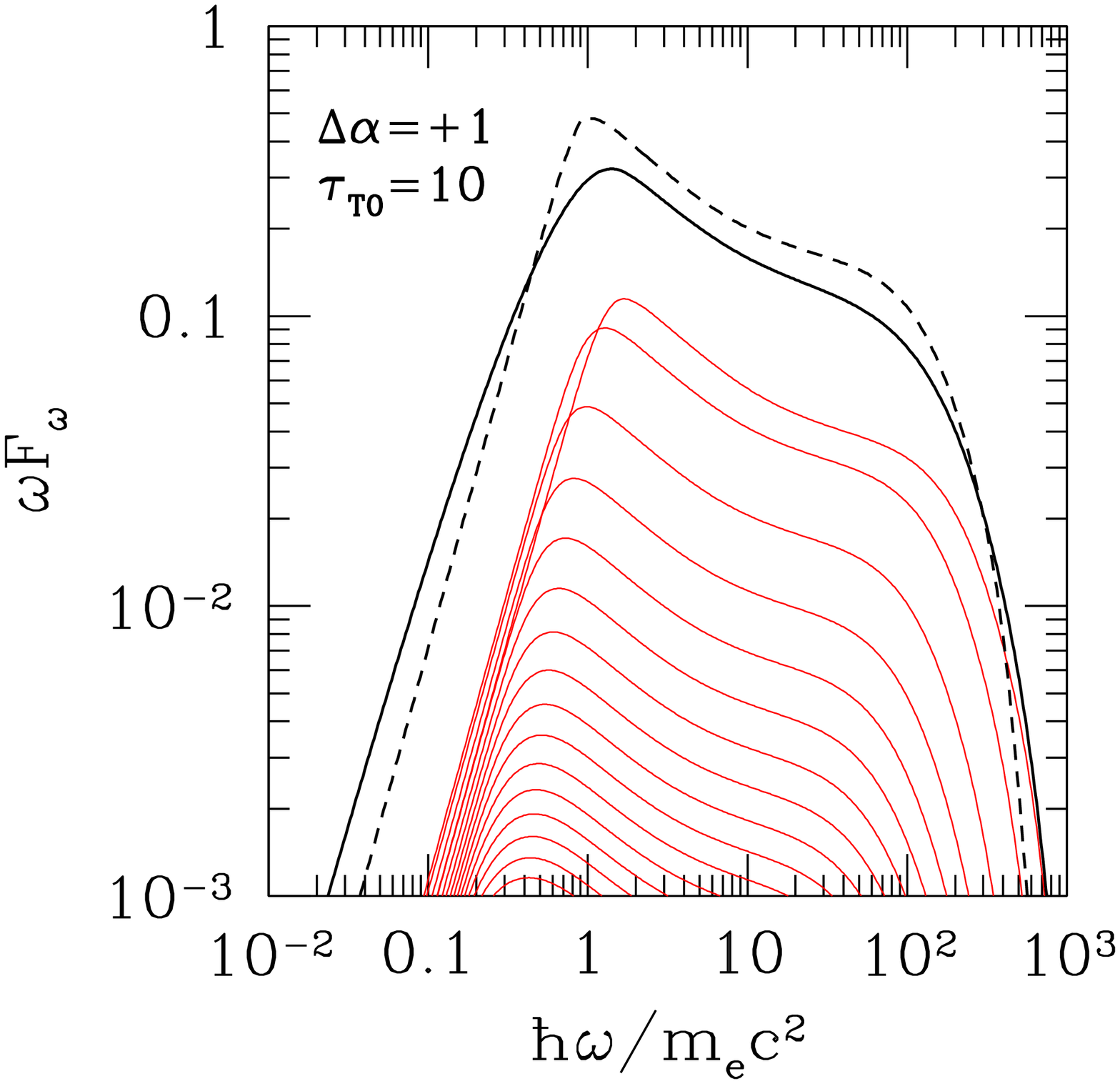}
%\vskip -0.5in
\includegraphics[width=0.48\textwidth]{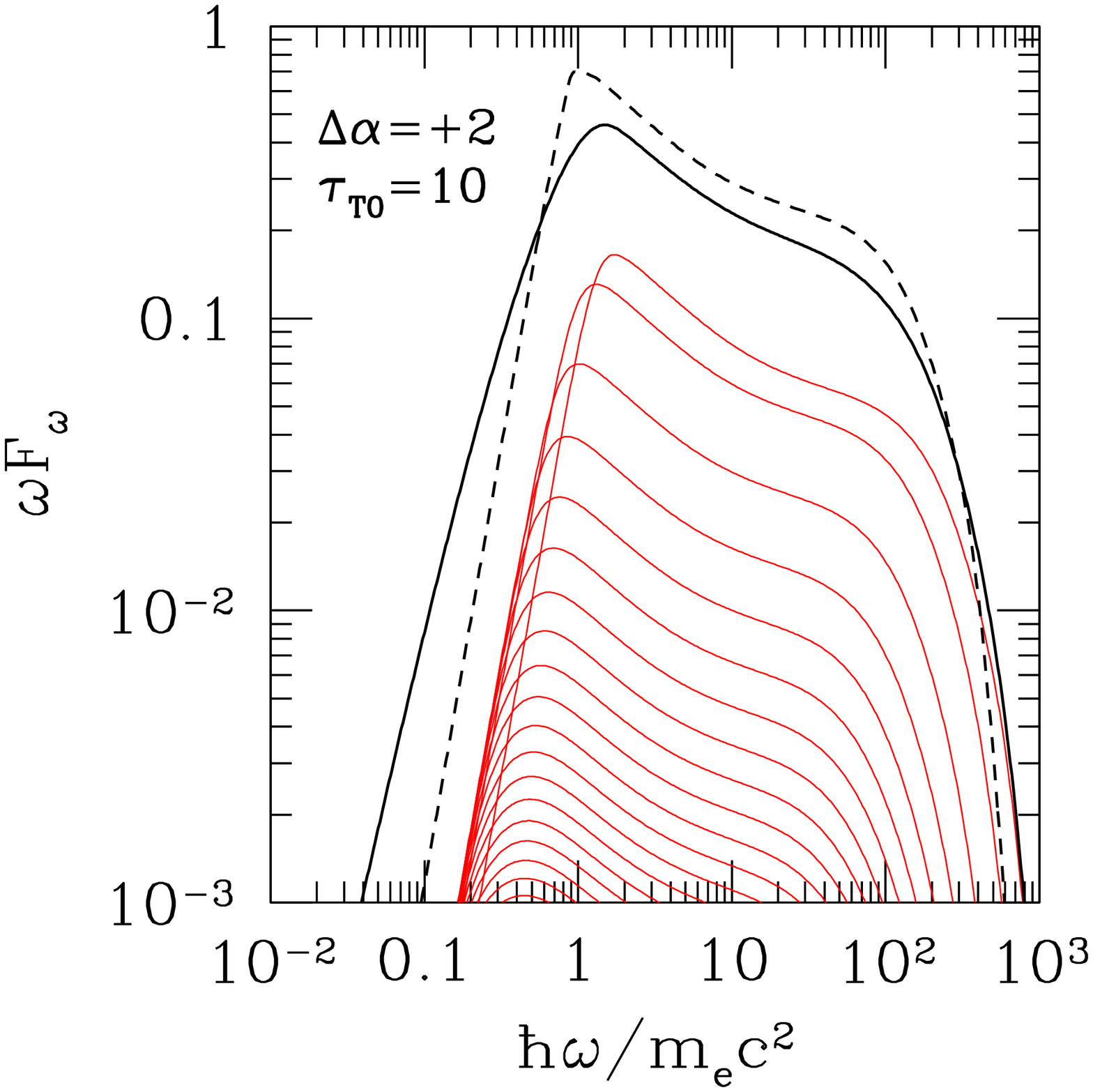}
%\vskip -0.5in
\caption{Effect of adjusting the input spectrum (dashed black curve, boosted by $\Gamma = 300$) on the 
output spectrum (solid black curve), processed by Compton scattering off a flow with radial scattering
depth $\tau_{T0}$ at the injection radius.  Top (bottom) panel:  $F_\omega \rightarrow 
(\omega/\omega_{\rm pk})^{\Delta\alpha} F_\omega$ for $\omega < \omega_{\rm pk}$ with 
$\Delta\alpha = 1$ (2).  Red curves:  time-resolved spectrum, plotted at intervals
$\Delta t = 0.25 (R_0/2\Gamma_0^2c)$.}
\label{fig:lowfreq}
\vskip .2in
\end{figure}

\subsection{Can Multiple Scattering Flatten \\ the Low-energy Spectrum?}

It has been suggested that a relatively hard (e.g. Planckian) source spectrum can, through multiple scattering
near a photosphere, be transformed into the flat low-energy spectrum that is characteristically
observed in a GRB (e.g. \citealt{asano13}).  Different results are obtained by \cite{beloborodov10}, using a Monte
Carlo approach, and \cite{deng14}, using semi-analytic methods.  They find that only a modest flattening of the 
spectrum is possible compared with Rayleigh-Jeans, unless possibly the emission is predominantly off-axis.

Here we revisit this problem in more detail, with the same Monte Carlo approach as above.
A spherical shell is assumed, and the low-energy spectral slope is modified
by hand.  (We use the output of the kinetic calculation with total compactness $\ell = 10^3$ and heating 
compactness $\ell_{\rm heat} = 2\ell_{\rm th}$ or $f_{\rm th} = 0.5$.)  

The output spectrum resulting from a sequence of low-energy spectra ($F_\omega \propto \omega^{1,2}$) is 
shown in Figure \ref{fig:lowfreq}.  Here the scattering depth has been adjusted upward (to $\tau_{\rm T} = 10$) to
accentuate the possible effect of a photosphere.   In spite of that, only a slight flattening of the low-energy spectrum
is observed.  Starting from a Planckian spectrum ($F_\omega \propto \omega^2$) does not lead to anything like a
GRB ($F_\omega \sim {\rm const}$).

%%%%%%%%%%%%%%% SECTION: DISCUSSION %%%%%%%%%%%%%%%%%%%%%%%%%%%%%%%
\section{Discussion}\label{s:discuss}

We have examined the observational imprint of a relativistic, ultraluminous, magnetized outflow emanating from a stellar mass black hole.
Such an outflow is naturally produced once the hole is able to acquire an ultra-strong ($\sim 10^{15}-10^{16}$ G) magnetic field 
from a massive, orbiting torus.  Our focus is on the spectral imprint of delayed dissipation after breakout from a confining medium.
We have considered i) acceleration of the magnetized jet following breakout; ii) reheating after it has expanded a factor $\sim 100$ 
and the magnetized component has become optically thin to scattering, leading to the creation of a secondary photosphere; 
iii) the non-thermal spectrum that arises in the simplest case of distributed heating; and iv) differences with the high-energy
emission from a photosphere that is continuously heated.  

The work presented here, in combination with \cite{russo13a,russo13b}
and Papers I, III, presents a detailed description of the dynamics of GRB outflows and the origin of the extremely bright
and variable gamma-ray pulses that are their defining signature.

{\it Energy source for high-energy emission.}  Differential motion between embedded baryon clouds and the lighter,
magnetized jet material produces strong disturbances in the magnetic field once the comoving photon compactness drops below
$\ell_{\rm th} \sim 4\times 10^3Y_{e\,0.5}^{-1}$.  This means that much of the high-energy emission originates in a fairly narrow zone,
less than a decade in radius after decoupling between the baryons and the photons.  

This differential motion has three features which
distinguish it from internal shocks:  i) it is actively {\it driven} within the dissipation zone
by the continuing acceleration of the lighter, magnetized jet fluid;  ii) it has significant {\it angular} as well as
radial structure; and iii) moderately relativistic differential motion is consistent with observed
GRB spectra.  In the case of internal shocks, high efficiencies demand relativistic collision speeds between shells \citep{beloborodov00},
but then the spectrum of accelerated electrons has a relativistic low-energy cutoff, which creates a cooling tail below
the spectral peak that is not observed \citep{ghisellini99}.

{\it Buffering of $\tau_{\rm T}$ during reheating.}  In strong contrast with emission models involving shock-accelerated
particles, the scattering depth is strongly buffered during reheating:  as the pair density increases, the energy of
the pairs drops, due to a decrease in the heating rate per particle.

{\it Above the spectral peak, harder photons are emitted before softer ones.}  A high-energy spectral tail, well approximating a power
law in many cases, originates from a sequence of relativistic particle states of diminishing mean energy.  The highest 
energy achieved by a Comptonized photon is $\hbar\omega_{\rm IC} \sim \hbar\omega_{\rm pk}/\tau_{\rm T,0}$, where $\tau_{\rm T,0}$ is the
radial scattering depth in the magnetized outflow at the onset of re-heating.  Within this tail, the highest energy
photons are emitted {\it before} those closer to the seed thermal peak.  

{\it Smooth high-energy spectral tail.}  The tail connects smoothly to the seed thermal peak if the total compactness 
$\ell_{\rm tot} \gtrsim 10^3$, and if the first photons upscattered during reheating reach an energy $> m_ec^2$ 
in the comoving frame.  For example, a photon index $-2.3$ above the peak corresponds to a ratio $f_{\rm th} \sim 0.5$ 
of seed thermal photon energy to injected energy.  The minimum thermal compactness is then $\sim 300$, which lies a factor
10-50 below the critical value $\ell_{\rm th} \sim 4\times 10^3(Y_e/0.5)^{-1}$ where embedded baryon clouds begin to move 
differentially with respect to the magnetofluid.  By the time the radiation compactness has dropped to this lower value,
the differential Lorentz factor has grown to $2-4$ and stronger heating is expected.

The second constraint requires a residual scattering depth $\tau_{\rm T,0} \lesssim 10^{-2}$ in the
magnetized jet at the onset of reheating.  We show that such a low optical depth is reached by a freely expanding
jet by the point that embedded baryons begin to decouple from the radiation field.

{\it Inconsistency between an extended, Comptonizing photosphere and observed GRB pulse behavior.}  A commonly explored
hypothesis is that the high-energy spectrum of a GRB forms in close analogy with an accretion disk corona, by
diffusive upscattering of softer thermal photons.  We confirm that a high-energy spectral tail can form by such
a mechanism, but point out two disagreements with observation:  the spectral peak tends to become very broad
if the high-energy tail is hard;  and, more seriously, the harder photons tend to lag softer ones.  We conclude
that multiple scattering at a photosphere in a relativistic outflow is not a viable explanation for high-energy
spectral tails in GRBs.  Residual scattering by the regenerated pairs is found to have a much milder effect on pulse widths. 

{\it How rapid is the bulk acceleration of very strongly magnetized jet?}  A baryon-free jet can reach transparency
while still very compact, e.g. $\ell_{\rm P} \sim 10^8$.  Then jet material formally has an enormous magnetization,  
$\sigma \sim \ell_{\rm P}/\tau_{\rm T} \sim \ell_{\rm P}$.  

This opens up the possibility that some portion of this material reaches
an enormous Lorentz factor in a short time.  For example, \cite{granot11} showed that a thin, outer layer of a planar
magnetofluid expanding into a vacuum reaches a Lorentz factor $\Gamma \sim 2\sigma$ almost instantaneously.  \cite{meszaros97} considered the 
possibility of very high Lorentz factors ($\Gamma > 10^5$) beyond the photosphere of a magnetized jet.  

Here we have shown that superposing a modest level of MHD turbulence onto the jet flow, with fractional energy density
$\varepsilon_t$ in effect limits the magnetization to $\sigma_{\rm eff} \sim 1/\varepsilon_t$.  This is demonstrated
in an analytic solution to the similarity problem posed by \cite{granot11}.  We also showed that including `turbulent inertia'
in a radiatively forced, transparent jet limits the growth of Lorentz factor to be only slightly faster than linear with
radius.

{\it Mapping of temperature at breakout to final spectral peak.}  We find that the observed spectral peak is the result 
of three steps.  First, as was investigated in Paper I, comptonization by a thermal pair gas with a cyclo-synchrotron 
source freezes out at a comoving peak energy $\hbar\omega_{\rm pk}' \sim 0.1m_ec^2$ when the compactness is high 
(e.g. $\ell \sim 10^7$) but the effective temperature is much less than $\sim 20$ keV.  The output spectrum 
(\ref{eq:grbthermal}) at this stage is quasi-thermal, with a Wien cutoff above the peak, but a flat spectrum 
($F_\omega = $ const) below the peak.  

Second, we have argued in Paper III for a modest delay between the emergence
of the outflow from the confining medium, and the decoupling of the relativistic magnetofluid from a forward
shell of baryonic material.  A limited amount of adiabatic softening is possible during this second step, corresponding 
to a maximum reduction $\sim {\cal R}_{\rm br}^{-2/3} \gtrsim 0.2$ in the peak energy.  (This softening would be negligible if the magnetofluid continues to be heated during its decoupling from the baryons.) The peak photon energy is
preserved after the outflow temporarily becomes transparent.

Third, we have in this paper considered the
rescattering of this GRB-thermal spectrum during a delayed pair breakdown that starts at $\tau_{\rm T} \sim 10^{-3}-10^{-2}$.  
Here $\omega_{\rm pk}$ increases by a factor $\sim 2$ as the high-energy spectral tail is generated.

Combining these steps, one finds that the observed peak energy remains within a factor $\sim 2$ of the seed
thermal peak (as seen Lorentz-boosted into the observer's frame).  In addition, the formation of individual
pulses by a causal process (a corrugation instability of the forward baryon shell) leads to a direct relation
between breakout Lorentz factor and opening angle of the outflow, $\Gamma_{\rm br} \sim \delta\theta$.  Both of
these ingredients are invoked in Paper I to `derive' the observed relation between $\hbar\omega_{\rm pk}$ and $E_{\rm \gamma,iso}$ \citep{amati02}.

\subsection{Outstanding issues}\label{s:outst}

{\it Maximum range of high-energy spectrum.}   This is an important test of the emission mechanism, and in principle can be used
to distinguish between models in which the source spectrum is a rigid power-law extending above $\sim m_ec^2$ in the comoving
frame (e.g. synchrotron radiation by internal shocks), or instead has a break around this energy, as in the mechanism developed here.
For example, an outflow producing a burst with $E_{\rm pk} \sim 300$ keV and reaching $\Gamma \sim 300$ in the high-energy emission
zone would, in the second case, have a high-energy spectrum extending to $\sim 10^2$ MeV.  

Fermi measurements of GRBs have revealed the onset of extended tails of emission above
$\sim 100$ MeV that appear to be powered by a forward shock \citep{ackermann13}.  The pulse structure generally simplifies into one or
two pulses at these energies, and there is evidence for the emergence of a rising high-energy spectrum during the later stages of
the burst  (\citealt{ackermann14}, as well as the earlier GRB 941017: \citealt{gonzalez03}).  

In several bright bursts there 
is a significant deficit in the $> 100$ MeV emission relative to an upward extension of the high-energy power-law measured 
near $\sim 1$ MeV \citep{guetta11}.  This is generally consistent with our spectral calculations.
Suppression of the high-energy emission could also be due to pair conversion if $\Gamma \lesssim 300$ in the 
prompt emission region, and the emission zone is compact enough.  

To decide between these possibilities, an important clue comes from the delay in the onset of 
the $\sim 100$ MeV emission, with respect to the $\sim$ MeV band, that is frequently observed \citep{ackermann13}.  
If this delay {\it and} the deficit in overall $> 100$ MeV emission were due to pair conversion during the first
part of the burst, then the deficits in fluence and duty cycle would be roughly proportional.
But, in fact, the $> 100 $ MeV emission is significantly extended in time.

The delayed onset of the hard emission is consistent with expectations based on the pair loading and radiative acceleration
of a Wolf-Rayet wind \citep{thompson06}.  Early on in a burst, the gamma-ray flux
across the forward shock is high enough that the wind medium reaches a comparable Lorentz factor to that of the gamma-ray emitting
material.  The transition from a weak forward shock to a relativistic shock moving into a nearly static medium occurs {\it during}
the prompt phase, leading to the emergence of a high-energy spectral tail with a cooling spectrum $F_\omega \propto \omega^{-1/2}$
(see Sections 3 and 8.2 of \citealt{thompson06}).  Detailed calculations of the emergent high-energy spectrum have recently
been performed of the interaction between pre-acceleration and high-energy emission including a range of emission channels \citep{beloborodov14}
but making a simplifying assumption about the dynamics of the GRB ejecta (that they have already collected into 
a thin shell that is in dynamical contact with the forward shock). 

We conclude that there is evidence for a high-energy spectral tail extending above the peak over a range of $\sim 10^2$, but
not necessarily for more extended prompt gamma-ray emission that is not associated with the forward shock.  High-energy emission
by delayed pair breakdown tentatively passes this test, although more detailed comparisons of different spectral components along
the lines of \cite{guetta11} are in order.

{\it Particle energy diffusion.}
Cascading Alfv\'enic turbulence in a very strongly magnetized pair plasma can become charge starved at a high wavenumber,
because wavepackets become very elongated along the background magnetic field.
The fluctuating current density $(c/4\pi) k_\perp \delta B$ can exceed the maximum conduction current $n_e e c$ that the
ambient pairs can supply.  Equating these two quantities, and making use of the wave power spectrum
$\delta B(k_\perp)$ fixes the transverse wavenumber $k_\perp^{\rm starve}$.  Some recent numerical experiments
\citep{maron01,boldyrev06} find $\delta B \sim k_\perp^{-1/4}$, somewhat flatter than the Kolmogorov scaling 
$\delta B \sim k_\perp^{-1/3}$  that was initially suggested in the theory of strong Alfv\'enic turbulence developed
by \cite{gs95}.  

The charge-starvation scale can be compared with 
the wavenumber $k_\perp \sim \omega_{\rm Pe}/c$ at which Alfv\'en waves Landau damp on the thermal motion of 
sub-relativistic pairs.  For the adopted wave power-spectrum, this is (\citealt{thompson06}, Paper I)
\be
\left({k_\perp^{\rm starve} c\over\omega_{\rm Pe}}\right)^2 =
{\bar\lambda_c\over r/\Gamma}\left({B\over B_{\rm Q}}\right)^{-8/3}\left({3\tau_{\rm T}\over 2\alpha_{\rm em}}\right)^{5/3}
{B^2\over \delta B_0^2}.
\ee
Here all quantities refer to the comoving frame, $\delta B_0$ is the wave amplitude at the outer scale, 
$\omega_{\rm Pe} = (4\pi n_e e^2/m_e)^{1/2}$ is the plasma frequency, $\bar\lambda_c = \hbar/m_ec$, and
$k_{\perp,0} \sim (\delta B_0/B)^{-1} \Gamma/r$ is assumed.  
Then at breakout at a radius $R_{\rm br} \gtrsim 2\Gamma^2 ct_{\rm eng} \sim 10^{12}$ cm, one finds
\be\label{eq:kperpbr}
\left({k_\perp^{\rm starve} c\over\omega_{\rm Pe}}\right)^2_{\rm br} =
0.02\,{R_{\rm br,12}^{5/3} (\tau_{\rm T}/3)^{5/3} (\Gamma_{\rm br}/3)^{11/3}\over L_{\rm P\,iso,51}^{4/3} (\delta B/B_0)^2}.
\ee
One sees that charge starvation can set in at a shallower depth in the cascade than Landau damping.  The scattering depth outside 
breakout in a spherically diverging flow scales as $\tau_{\rm T} \propto r^{-1} \Gamma^2$, hence
\be
\left({k_\perp^{\rm starve} c\over\omega_{\rm Pe}}\right)^2 = \left({\Gamma\over \Gamma_{\rm br}}\right)^{1/3}\,
\left({k_\perp^{\rm starve} c\over\omega_{\rm Pe}}\right)^2_{\rm br} \quad (r > R_{\rm br}).
\ee
Combining this with equation (\ref{eq:kperpbr}) one see that an Alfv\'enic cascade still becomes charge starved 
during the first stages of a delayed pair breakdown, occurring at a Lorentz factor $\sim 10^2\Gamma_{\rm br}$.

It is straightforward to show that the wave energy density $\delta B^2/8\pi$
at $k_\perp \sim k_\perp^{\rm starve}$ is much less than the rest energy density of the background charges, so that its
dissipation in one waveperiod is consistent with gradual heating.  

More intermittent heating
associated with stronger magnetic field gradients remains an interesting possibility, which could be encapsulated in 
a particle energy diffusivity that is a strong function of energy, e.g. ${\cal D} \propto \gamma_e^{2-3}$.  This would have the effect
of generating a high-energy power-law tail to the particle spectrum, even while the low-energy cutoff remained fixed
by a near balance between global heating and cooling.  
Spectral calculations based on more elaborate models for particle energy diffusion will be explored elsewhere.
We are nonetheless encouraged that a high-energy power-law
spectrum emerges without invoking such strong particle scattering, by starting from the simplest prescription for uniform, 
distributed heating.

%%%%%%%%%%%%%%% SECTION: APPENDIX %%%%%%%%%%%%%%%%%%%%%%%%%%%%%%%%%

\begin{appendix}

\section{A. Details of the Kinetic Calculations}\label{s:kinetic}

%%%%%%%%%%%%%%% COMPTON SCATTERING %%%%%%%%%%%%%%%%%%%%%%%%%%%%%%%%
 \subsection{Compton Scattering}
In describing Compton scattering, we follow the treatment of 
\citet{Belmont2009}.  In particular, we make use of i) the 
prescription given there for switching from the continuous F-P formalism 
to a discrete collision integral; and ii) the method for calculating the exact Compton 
scattering cross-section $d\sigma/dx$ over a wide range of photon
and particle energies (see equations (27)-(39) of \citealt{Belmont2009}).  We summarize
the approach here for convenience.

Compton scattering can be described exactly as a collision process 
using the following two collision integrals:
%\begin{widetext}
\begin{eqnarray}
&&\dot{n}_{\pm,\text{cs}}^{\text{col}}(p) = \int dp_0~n_\pm(p_0)
\int_{x_c(p_0)}^\infty dx_0~n_\gamma(x_0)~c\frac{d\sigma}{dp}
(p_0,x_0\rightarrow x(p)) - n_\pm(p)\int_{x_c(p)}^\infty dx_0~
n_\gamma(x_0)c\sigma(p,x_0) \\
&&\dot{n}_{\gamma,\text{cs}}^{\text{col}}(x) = \int dx_0~
n_\gamma(x_0)
\int_{p_c(x_0)}^\infty dp_0~n_e(p_0)~c\frac{d\sigma}{dx}
(p_0,x_0\rightarrow x)-n_\gamma(x)\int_{p_c(x)}^\infty dp_0~
n_e(p_0)~c\sigma(p_0,x).
\end{eqnarray}
%\end{widetext}
Here $x(p) = x_0+\gamma_0(p_0)-\gamma(p)$ is obtained from 
energy conservation.  The first and second terms represent 
scattering into and out of a given energy/momentum bin.

%The differential cross-section $d\sigma/dx$ is 
%given in equations (27)-(39) of \citet{Belmont2009}. 

The collision integral cannot be evaluated accurately when
$|x-x_0| \ll x_0$:  then energy distribution of scattered photons
is narrowly peaked around the initial energy, and very high grid resolution is needed to 
resolve it. A workaround is to Taylor expand the collision integrals 
to second order over the width of the scattered 
distribution, which yields F-P equations (\ref{eq:fpeq1}, 
\ref{eq:fpeq2}). The advection and diffusion coefficients are 
obtained by taking moments of the differential cross section,
\begin{eqnarray}
\{A_\gamma(x),D_\gamma(x)\} &&= \int_0^{p_c(x)}dp~n_e(p)~
c\sigma_{\{1,2\}}(p,x) \\
\{A_\pm(p),D_\pm(p)\} &&= \int_0^{x_c(p)}dx~n_\gamma(x)~
c\sigma_{\{1,2\}}(p,x)
\end{eqnarray}
where
\begin{eqnarray}
&&\sigma(p_0,x_0) = \int dx~\frac{d\sigma}{dx}
(p_0,x_0\rightarrow x) \\
&&\sigma_1(p_0,x_0) = \int dx~(x-x_0)\frac{d\sigma}{dx}
(p_0,x_0\rightarrow x) \\
&&\sigma_2(p_0,x_0) = \int dx~(x-x_0)^2\frac{d\sigma}{dx}
(p_0,x_0\rightarrow x)
\end{eqnarray}
The limits of integration $x_c(p)$ and $p_c(x)$ depend on the grid 
resolution, and are shown in Figures 5 and 6 of \citet{Belmont2009} 
for different resolutions.
%%%%%%%%%%%%%%% COULOMB SCATTERING %%%%%%%%%%%%%%%%%%%%%%%%%%%%%%%%
\subsection{Coulomb Scattering}
The F-P treatment of Coulomb scattering for arbitrary distribution 
of particles is developed in the study of 
\citet{NayakshinMelia1998}. Here we use the expressions for the 
diffusion and advection coefficients given in 
equations (F1)-(F5) of \citet{VurmPoutanen2009}
\begin{eqnarray}
A_{\pm,\text{coul}}(p) = \int dp'~N_\mp(p')a_{\text{coul}}(p,p') \\
D_{\pm,\text{coul}}(p) = \int dp'~N_\mp(p')d_{\text{coul}}(p,p')
\end{eqnarray}
%%%%%%%%%%%%%%% PAIR PRODUCTION/ANNIHILATION %%%%%%%%%%%%%%%%%%%%%%
\subsection{Pair Production and Annihilation}
The processes of pair production and annihilation can be exactly 
described by collision integrals, but their evaluation is typically
not limited by grid resolution:
%\begin{widetext}
\begin{eqnarray}
&&\dot{n}_{\pm,\text{pp}}^{\text{col}}(p) = \int dx~n_\gamma(x)
\int dx'~n_\gamma(x')R_{\gamma\gamma}(x,x'\rightarrow p); \\
&&\dot{n}_{\pm,\text{pa}}^{\text{col}}(p) = - n_\pm(p)\int dp'~
n_\mp(p')R_\pm(p,p'); \\
&&\dot{n}_{\gamma,\text{pa}}^{\text{col}}(x) = \int dp_+~n_+(p_+)
\int dp_-~n_-(p_-)R_\pm(p_+,p_-\rightarrow x); \\
&&\dot{n}_{\gamma,\text{pp}}^{\text{col}}(x) = - n_\gamma(x)\int dx'~
n_\gamma(x')R_{\gamma\gamma}(x,x'),
\end{eqnarray}
%\end{widetext}
where
\begin{eqnarray}
&&R_{\gamma\gamma}(x,x') = 2\int dp~R_{\gamma\gamma}
(x,x'\rightarrow p); \\
&&R_\pm(p,p') = \frac{1}{2}\int dx~R_\pm(p,p'\rightarrow x).
\end{eqnarray}
The rate of producing a lepton with momentum $p$ upon the 
annihilation of two photons with energies $x$ and $x'$, 
$R_{\gamma\gamma}(x,x'\rightarrow p)$, is given in the work of 
\citet[][equations (24)-(29)]{BottcherSchlickeiser1997}. 
An analytical expression for the rate 
$R_\pm(p,p'\rightarrow x)$ of producing a photon with 
energy $x$ when two leptons of momenta $p$ and $p'$ annihilate 
each other are provided by \citet[][equations (23), (24), (55)-(58)]
{Svensson1982}.  The factors of 1/2 (2) account for the fact that
two photons are emitted (absorbed) during pair annihilation (creation).

\section{B. Drag Force On Relativistic Pairs Heated Parallel to ${\bf B}$.}\label{s:drag}

We now consider the drag force imparted to longitudinally heated electrons and positrons.  In addition to
non-resonant Compton drag off thermal photons, there is a contribution from cyclo-synchrotron absorption, 
which we show to be negligible.

In the rest frame of its guiding center, an electron preferentially emits and absorbs soft photons of a similar frequency
if the ambient radiation field has a Rayleigh-Jeans slope.  After boosting to the lab frame, the absorbed photons 
then have a characteristic frequency
\be\label{eq:omres}
\omega \sim \left[1 + \left({p_\perp\over m_ec}\right)^2\right] {\omega_{\rm ce}\over\gamma_{e\,\parallel}}.
\ee
The frequency (\ref{eq:omres}) does indeed sit in the low-frequency Rayleigh-Jeans tail of the ambient thermal radiation field,
and below the flat portion of the spectrum that connects to the spectral peak (Paper I).  

Therefore we 
can define the Rayleigh-Jeans temperature $T_c$ in terms of the pair temperature at breakout, where free-free processes are strong enough to
fill in the low-frequency spectrum.   Including the adiabatic cooling associated with jet collimation and acceleration outside
breakout, one has
\be
T_c(r) \sim T_e(R_{\rm br})\left({r\over R_{\rm br}}\right)^{-1}.
\ee
The temperature at breakout is $\kB T_e(R_{\rm br}) \sim E_{\rm pk}/3 \sim 0.04\,m_ec^2$, as was found for
a magnetized pair plasma of a compactness $\sim 10^4-10^6$ (Paper I).  Therefore $\kB T_c \sim 1$ keV 
in the reheating zone concentrated at a distance $r \gtrsim 10\,R_{\rm br}$ from the engine.  

The drag force that is felt by pairs in relativistic motion along the magnetic field depends on 
their gyrational energy.  A simple close-formed expression for this force is obtained by assuming
a two-dimensional thermal distribution with temperature $T_\perp$ in the plane perpendicular to ${\bf B}$,
boosted by a Lorentz factor $\gamma_{e\,\parallel}$ along ${\bf B}$.  As we now derive,
\be\label{eq:synchdrag}
{dp_\parallel\over dt}\biggr|_{\rm synch~abs} = {5\over 2\gamma_{e\,\parallel}}{T_c\over T_\perp}\sigma_T {B^2\over 8\pi}.
\ee
This drag force is suppressed compared with non-resonant Compton drag off the thermal peak,
\be
{dp_\parallel/dt|_{\rm synch~abs}\over dp_\parallel/dt|_{\rm IC}} = {5\over 16\gamma_{e\,\parallel}^3} {T_c\over T_\perp}
\left({T_\perp\over m_ec^2}\right)^{-2}{B^2\over 8\pi U_\gamma}
\ee
when $\kB T_\perp \gg m_ec^2$.  

An analogous expression is obtained assuming sub-relativistic gyrations, with most of the absorption at the
cyclotron fundamental.  Then the ambient photons are aberrated nearly into a direction parallel to ${\bf B}$,
and the corresponding absorption cross section is $\sigma(\omega') = 2\pi^2 (e/B) \omega_{\rm ce}\delta(\omega'-\omega_{\rm ce})$.
Given the same Rayleigh-Jeans distribution of soft target photons, one obtains
\be
{dp_\parallel\over dt} = {dp'_\parallel\over dt'} = {12\over\gamma_{e\,\parallel}}\ln(2\gamma_{\rm e\,\parallel})
\left({T_c\over m_ec^2}\right)\sigma_T {B^2\over 8\pi}.
\ee

To derive expression (\ref{eq:synchdrag}), consider a planar distribution of relativistic $e^\pm$ (the sign of the absorbing charge
is immaterial) with a temperature $T_\perp \gg m_ec^2/\kB$.  Superposed on this thermal
distribution of gyrations is a uniform motion of the particle guiding centers along ${\bf B}$ at a Lorentz
factor $\gamma_{e\,\parallel}$, independent of gyrational energy.

Ambient soft photons of frequency $\omega$ and intensity $I_\omega = T_c \omega^2/4\pi^3 c^2$
have a frequency $\omega' = \omega/{\cal D}$ and intensity $I'_{\omega'}
= I_\omega/{\cal D}^3$ in the guiding center rest frame, 
where ${\cal D}(\gamma_{e\,\parallel},\mu') = \gamma_{e\,\parallel}(1+\beta_{e\,\parallel}\mu')$
is the Doppler factor.  The absorption coefficient in this frame is
\be
\alpha_{\omega'}' = {4\pi^3 c^2\over T_\perp {\omega'}^2}
\int d\gamma_\perp {\partial^2 P'\over \partial\omega'\partial\Omega'} {dn_e'\over d\gamma_\perp},
\ee
where $\partial^2P'/\partial\omega'\partial\Omega'$ is the synchrotron power radiated at frequency
$\omega'$ and direction cosine $\mu'$ with respect to ${\bf B}$.  An electron or positron feels a net force
\be
{dp_\parallel'\over dt'} = {1\over n_e'}\int 2\pi d\mu' \int d\omega' (\mu')^2 \alpha'_{\omega'}{I'_{\omega'}\over c}
\ee
in the direction of the magnetic field.  This force is invariant under the parallel boost,
$dp_\parallel/dt = dp_\parallel'/dt'$.  After performing the integral over $\omega'$, we can substitute
\be
{dP'\over d\mu'} = {7\over 16}{\gamma_\perp^3 e^4 B^2\over m_e^2c^3}\left[{1\over (1+{\theta'}^2\gamma_\perp^2)^{5/2}}
+ {5\over 7}{{\theta'}^2\gamma_\perp^2\over (1+{\theta'}^2\gamma_\perp^2)^{7/2}}\right],
\ee
with radiation concentrated at an angle $\theta' \simeq \mu' \sim 1/\gamma_\perp$ from the gyrational plane,
to obtain equation (\ref{eq:synchdrag}).

\end{appendix}

%%%%%%%%%%%%%%% BIBLIOGRAPHY %%%%%%%%%%%%%%%%%%%%%%%%%%%%%%%%

\end{document}